\documentclass[aps,prl,preprint,groupedaddress]{revtex4-2}

\usepackage{amsmath}
\usepackage{palatino,graphicx,wrapfig}
\usepackage[small,it]{caption}
\usepackage{amsmath}
\usepackage{amssymb,bm}
\usepackage{parskip}

\usepackage{booktabs}

\usepackage{multirow}
\usepackage[
linesnumbered,ruled,vlined]{algorithm2e}
\usepackage {algpseudocode}
\usepackage{algorithmicx}
\usepackage{algcompatible}
\usepackage{pythonhighlight}
\usepackage{float}
\usepackage{hyperref}
\usepackage{enumitem}
\usepackage{caption}
\usepackage{ragged2e}

\begin{document}


\title{Generalizable data-driven turbulence closure modeling on unstructured grids with differentiable physics}



\author{Hojin Kim}
 \affiliation{School of Mechanical Engineering, \\ Purdue University, West Lafayette, 47907, IN, USA}
\author{Varun Shankar}%
\affiliation{%
 Mechanical Engineering Department, Carnegie Mellon University, Pittsburgh, 15213, PA, USA}%

\author{Venkatasubramanian Viswanathan}
\affiliation{
 Department of Aerospace Engineering, University of Michigan, Ann Arbor, 48109, MI, USA}%
\author{Romit Maulik}
\affiliation{%
 College of Information Sciences and Technology, Pennsylvania State University, University Park, 16802, PA, USA
}%
\affiliation{%
 Mathematics and Computer Science Division, Argonne National Laboratory, Lemont, 60439, IL, USA}%


\date{\today}

\begin{abstract}
Differentiable physical simulators are proving to be valuable tools for developing data-driven models for computational fluid dynamics (CFD). In particular, these simulators enable end-to-end training of machine learning (ML) models embedded within CFD solvers. This paradigm enables novel algorithms which combine the generalization power and low cost of physics-based simulations with the flexibility and automation of deep learning methods.  In this study, we introduce a framework for embedding deep learning models within a finite element solver for incompressible Navier-Stokes equations, specifically applying this approach to learn a subgrid-scale (SGS) closure with a graph neural network (GNN). We first demonstrate the feasibility of the approach on flow over a two-dimensional backward-facing step, using it as a proof of concept to show that solver-consistent training produces stable and physically meaningful closures. Then, we extend this to a turbulent flow over a three-dimensional backward-facing step. In this setting, the GNN-based closure not only attains low prediction errors, but also recovers key turbulence statistics and preserves multiscale turbulent structures. We further demonstrate that the closure can be identified in data-limited learning scenarios as well. Overall, the proposed end-to-end learning paradigm offers a viable pathway toward physically consistent and generalizable data-driven SGS modeling on complex and unstructured domains.
\end{abstract}


\maketitle

\section{Introduction}
\label{sec:Introduction}


Accurate computational modeling of turbulent flows represents a critical requirement for the analysis of physical systems in a variety of engineering and scientific disciplines \cite{anderson1995computational}. Although the equations of fluid motion are well-understood, accurate numerical simulation of the governing Navier-Stokes equations with computational fluid dynamics (CFD) methods remains a formidable task \cite{Tennekes2018-mf} when significant turbulent behavior is observed. In such cases, solutions to the Navier-Stokes equations are characterized by chaotic and multi-scale behavior, which makes resolving the entire flow field through direct numerical simulation (DNS) often impractical \cite{Moin1998}. The number of grid points required for DNS can be shown to scale with $Re^{9/4}$, where the Reynolds number $Re$ represents the ratio of inertial to viscous forces in the flow \cite{Pope2000}. High Reynolds number flows are often of practical interest, but simultaneously pose a challenging modeling endeavor. 

Developments in CFD have largely centered on reducing the computational complexity of attaining numerical solutions to computationally expensive flow configurations. Historically, major efforts have been directed towards progressing reduced-order models (ROMs) \cite{Raveh2004,Mannarino2015} and turbulence closure models \cite{Spalart2000,Hoffman2006,Defraeye2010}. In such cases, the system complexity is decreased by modeling only a subset of the dominant modes of the system, as in the case of ROMs, or the mean flow or large scales of the flow, in the case of Reynolds-Averaged Navier-Stokes (RANS) \cite{Gatski2001} or large eddy simulation (LES) \cite{Sagaut2006-ln} respectively.  Recently, more attention has been turned towards machine learning (ML) as a potential pathway for reducing the cost of CFD methods \cite{Kutz2017}. The success of deep neural networks (DNNs), in particular, has provided new strategies for interacting with flow data, for tasks such as regression, optimization, or classification. DNNs have demonstrated enormous potential to recognize and extract complex relationships in existing data, which may be used for the prediction or control of new flows. This flexibility opens up numerous avenues for novel model design and the integration of ML in CFD, with a variety of approaches explored in the past few years \cite{Vinuesa2022}. The development of ROMs using ML was among the first efforts to tackle dynamical modeling of fluid systems with this new wave of data-driven learning \cite{Vinuesa2022}. Previous works have used deep learning architectures like convolutional neural networks (CNNs) \cite{Guo2016,Shankar2022} and recurrent neural networks (RNNs) \cite{Rajendran2018} to predict the spatiotemporal evolution of flow fields, or to learn nonlinear encodings of low dimensional flow features \cite{Milano2002}. More recent advances in ML architectures have also been applied to fluid modeling, such as using generative adversarial networks (GANs) for super-resolution \cite{Xie2018}, long short-term memory networks (LSTMs) for improved dynamic prediction \cite{mohan2018deep}.
Notably, graph neural networks (GNNs) have demonstrated a remarkable ability to model spatiotemporal data for fluid-flow prediction tasks \cite{Chen2021,pfaff2021learning, Yang2022, barwey2023multiscale, barwey2025Interpretable, barwey2025mesh, raut2025fignn}, in addition to applications from different domains including molecular dynamics \cite{Batzner2022}, traffic forecasting \cite{Bui2021}, and structural mechanics \cite{pmlr-v119-sanchez-gonzalez20a}. Specifically, GNNs have exhibited their capability to generalize to out-of-sample datasets (for example different flow geometries) due to their ability to perfectly represent nodes and edges in unstructured meshes, and their formulation which identifies local and non-local correlations across an unstructured dataset.



Many of these DNN-based modeling strategies for direct flow-field prediction, however, suffer from a fundamental drawback, which is that ML is ultimately interpolative and can fail catastrophically on out-of-distribution tasks. Without an interpretable framework, it can be difficult to determine if or how a model will fail when confronted with novel data. In contrast, hybrid learning approaches combine first-principles modeling built on fundamental physical assumptions with ML. By training ML models under more guaranteed physics constraints, hybrid methods can secure stronger guarantees on generalizability and stability. For example, physical symmetries and constraints can be embedded into ML models using invariant and equivariant networks \cite{Berrone2022,suk2021equivariant}. Turbulence closure modeling with ML represents one example of hybrid learning for CFD research. Closure models are typically partially data-driven by nature, offering a pathway to integrating ML with more fundamental first-principles knowledge. More importantly, closure models merely augment known governing laws (here the Navier Stokes equations) which must continue to be solved to machine precision albeit on coarser spatial and temporal length scales. 

There has been significant interest in using DNNs for turbulence modeling in the past few years \cite{Beck2021,Duraisamy2019,Pandey2020}. Given their extensive use in engineering disciplines, RANS models have received considerable attention. ML has been used to identify uncertain model predictions \cite{Ling2015}, to augment existing turbulence models \cite{Singh2017,Wu2018, heo2024simulation}, and to correct discrepancies in the Reynolds stress with respect to high-fidelity simulations \cite{Ling2016,Jiang2021}. DNNs have also been used in LES, for predicting the subgrid-scale stress that encompasses the effect of unresolved scales on the resolved scales \cite{Zhou2019,Maulik2018,Stoffer2021,Wang2018}. Common in DNN-based modeling approaches for subgrid-scale closure is the use of multi-layer perceptron architectures, which predict the closure term locally using resolved quantities. However, non-local models such as CNNs, GNNs and Fourier Neural Operators (FNOs) have also been explored, showing improvement over these simpler architectures \cite{Guan2022,Subel2023, quattromini2024graph, kurz2025harnessing, shankar2025differentiable}.

{From the perspective of LES, there are two pathways to building a closure model using data-driven techniques. The first is the so-called \textit{a-priori} learning approach, where the closure models are trained prior to integration within a partial differential equation (PDE) solution scheme. This requires supervised data for training that must come from, typically, DNS of a ground truth flow configuration. Furthermore, performance in an \textit{a-posteriori} setting is not guaranteed by good \textit{a-priori} performance, as the solver introduces new artifacts, such as numerical errors and temporal effects, that cannot be captured during training \cite{Maulik2018}. Even worse, solutions from turbulence modeling can become unstable, and these issues may arise despite achieving low \textit{a-priori} errors, which is due to a problem called model-data inconsistency \cite{sanderse2024review}. These disadvantages have thus generally restricted the applicability of these approaches to modeling more canonical flows, where DNS is feasible to obtain.}


Fortunately, the accuracy and stability in \textit{a-posteriori} settings can be guaranteed by employing an \textit{a-posteriori} learning approach \cite{shankar2025differentiable}, where the training procedure includes solving PDE and the solution from the PDE solver is used to train a closure model. Recent studies have demonstrated the effectiveness of deep reinforcement learning-based \textit{a-posteriori} training frameworks, where the closure parameters are optimized online through direct interaction with the LES environment\cite{bae2022scientific, beck2023toward}. The utilization of a differentiable PDE solver further enables an end-to-end training of ML models with backpropagation, which can lead to substantially improved \textit{a-posteriori} accuracy and use insight via numerical analysis \cite{chakraborty2024note}. However, differentiable CFD solvers are not widely available due to the nontrivial nature of changing existing CFD solvers to differentiable forms, which has hindered the investigation of this paradigm. Generally, differentiability can be realized via source-to-source transformations of existing code bases \cite{Bischof2008}, implementation of a solver in an automatic differentiation (AD) framework, or exploitation of existing adjoint-capable PDE solvers \cite{Mitusch2019}. Several previous efforts have leveraged these methods for improving CFD schemes. Kochkov et al. accelerated DNS computations through learned interpolations of the numerical scheme \cite{Kochkov2021}. Ho et al. use the adjoint-enabled field inversion technique for improving turbulence modeling of separated flows \cite{Ho2021}. Bezgin et al. have developed a differentiable CFD solver for two-phase flows in JAX \cite{BEZGIN2022108527}. Fan et al. explore accelerating turbulent fluid-structure interaction (FSI) problems using a JAX-based differentiable solver framework\cite{fan2026diff}. Several other works have explored ML modeling for subgrid closure in LES using differentiable simulations \cite{Sirignano2020,MacArt2021,List2022, sirignano2023deep}. List et al. utilize a differentiable simulator with a loss function based on turbulence physics to model closure for various evolving turbulent flow scenarios \cite{List2022}. Sirignano et al. deploy a deep learning-based closure model in differentiable simulation for flows around bluff bodies and evaluate their model's \textit{a-posteriori} accuracy and stability \cite{sirignano2023deep}. \textcolor{black}{A more recent study by Quattromini et al. utilizes GNNs to model the Reynolds stress tensor in RANS simulations for the reconstruction of the mean flow around a cylinder at low Reynolds numbers \cite{quattromini2024graph}.} While many studies on RANS modeling have employed an \textit{a-posteriori} approach \cite{zhao2020rans, wu2018physics, singh2017augmentation}, turbulence modeling for LES has often taken an \textit{a-priori} learning approach, where the closure models are trained prior to integration within a PDE solution scheme. The reader is directed to \cite{sanderse2024review} for a recent review on closure modeling using ML. We also remark that while this review is focused on data-driven closures, similar approaches are also showing promise in the direction of wall-modeling \cite{ma2025machine,bae2022scientific,vadrot2023survey}


Building on these earlier studies, the primary goal of this work is to introduce a subgrid-scale turbulence closure for unsteady flows on unstructured grids, embedded within a differentiable LES solver. Unlike the authors’ previous work \cite{shankar2025differentiable}, which employed the finite volume method (FVM) on structured grid simulations, the present study utilizes the finite element method (FEM). FEM offers greater flexibility in handling complex geometries and unstructured meshes, which are often necessary in real-world flow applications. Its variational formulation also naturally integrates with differentiable programming frameworks, making it a suitable choice for gradient-based optimization and learning tasks. Furthermore, a graph neural network (GNN) is adopted as the base architecture for the machine-learning-based subgrid closure model. GNNs are particularly well-suited for unstructured mesh data, as they can effectively handle non-Cartesian geometries and irregular topological structures. This makes them ideal for modeling subgrid-scale interactions across irregular grid connectivity. To evaluate the efficacy of this GNN-based closure model and differentiable solver on complex flow scenarios, we apply our framework to the canonical backward-facing step (BFS) case in both two- and three-dimensions, which features near-wall turbulence, flow separation, and the advection of shed vortices. An overview of the proposed framework is illustrated in Fig.~\ref{fig:flowchart}. The main contributions of this study are summarized as follows:

\begin{figure}[t]
  \centering
  \includegraphics[width=\linewidth]{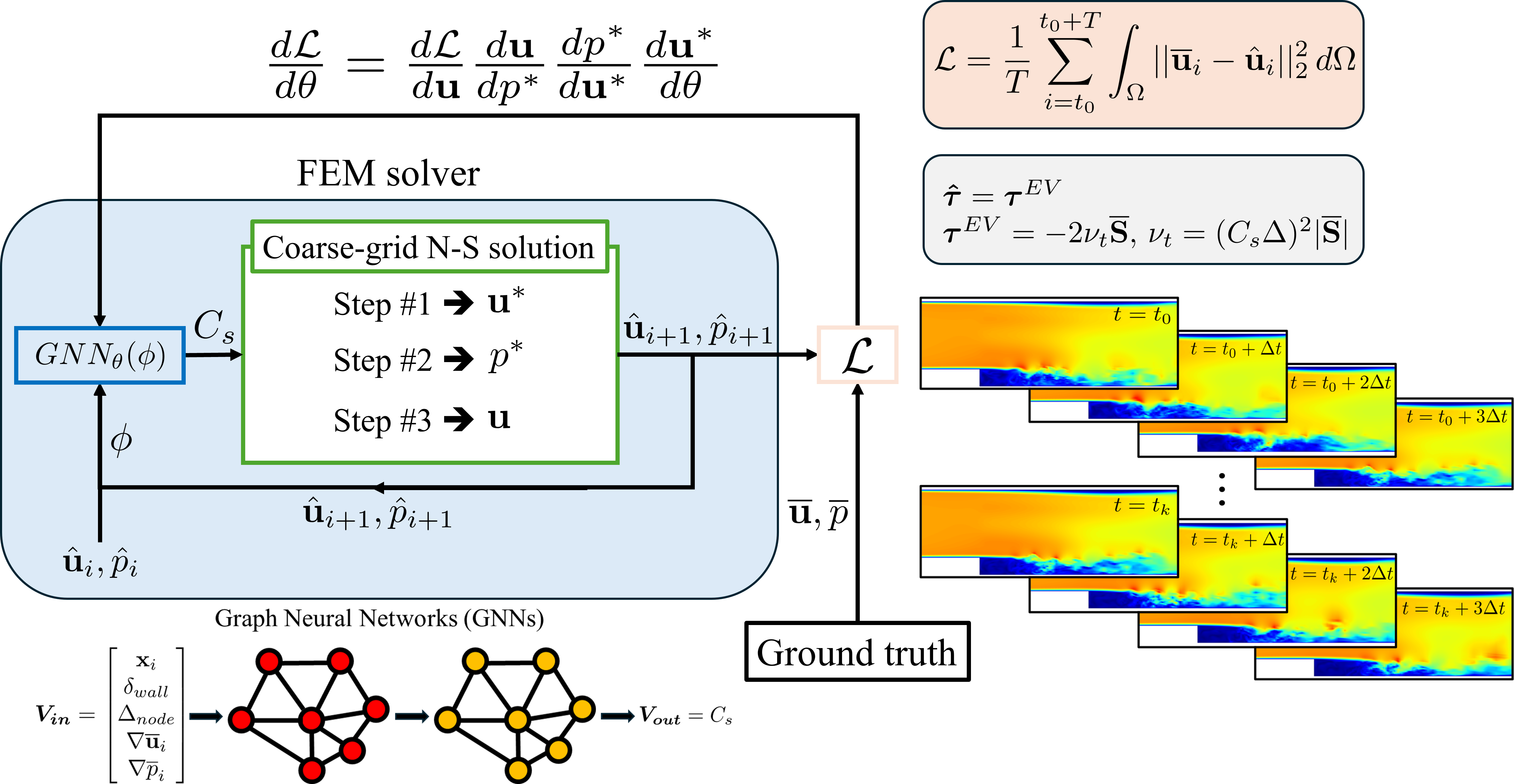}
  \captionsetup{singlelinecheck=false} 
  \caption{\justifying
A schematic of the differentiable FEM-based LES solver for training a GNN-based closure model. $\overline{\mathbf{u}}_i$ and $\overline{p}_i$ represent velocity and pressure data from the filtered ground truth solutions, respectively. $\hat{\mathbf{u}}_i$ and $\hat{p}_i$ represent predicted velocity and pressure data. The LES equations are solved using the IPCS algorithm \cite{goda1979multistep}. The GNN's inputs are the node features, $\boldsymbol{V_{in}}$, and the edge features, $\boldsymbol{E_{in}}$, and the output $C_s$ is a spatiotemporally variable Smagorinsky coefficient for subgrid-scale closure. The loss $\mathcal{L}$ is calculated over multiple snapshots, and gradients are backpropagated across the PDE solver to update GNN parameters $\theta$ using the discrete adjoint method. Further details about each component are explained in the Sec. \ref{sec:Methods}.
  }
  \label{fig:flowchart}
\end{figure}

\begin{itemize}[topsep=0pt, itemsep=5pt, parsep=0pt, partopsep=0pt]
    \item \textbf{Differentiable FEM-based LES solver:} A differentiable FEM-based LES solver compatible with unstructured meshes is developed. This differentiable solver is used to generate datasets including the ground truth data and a solution from a trained model. We implement the solver in the open-source FEniCS/FEniCSx finite-element framework \cite{fenics, baratta2023dolfinx}: the two-dimensional preliminary case is built on legacy FEniCS, while the three-dimensional LES experiments use FEniCSx. The use of FEniCSx for the three-dimensional case is motivated by its native support for distributed-memory parallelism, improved scalability on modern HPC systems, and data-oriented architecture that facilitates integration with machine learning models, which are not fully supported in the legacy version \cite{baratta2023dolfinx}. In both implementations, differentiability is instantiated by using the discrete adjoint method.
    
    \item \textbf{Generalizable subgrid closure modeling using GNN:} A subgrid-scale closure model based on a GNN is trained within the proposed differentiable algorithm. In this setup, the GNN-based closure is learned in an \textit{a-posteriori} manner: each training iteration solves the coarse-grained Navier-Stokes equations with the current closure, and the resulting PDE solution is then used to update the model parameters. We first apply this framework to a two-dimensional BFS as a preliminary proof of concept. We then train and evaluate the same framework on a three-dimensional BFS and further test it on an unseen geometry to assess its generalization capability.
 
    \item \textbf{Training under incomplete data availability:} We also investigate whether the proposed GNN-based closure can be learned when only partial observations are available. We first demonstrate this in the two-dimensional BFS setting, where only downstream flowfield data are available. We then extend the same sparse-data setup to the three-dimensional BFS case to evaluate its performance under constrained data availability.


\end{itemize}

\color{black}


The remainder of this paper is organized as follows. The LES governing equations, the definitions of graph and its structures, and implementation of the adjoint method are explained in Sec. \ref{sec:Methods}. Section~\ref{sec:Results and Discussion}. then details the generation of the training dataset, the training procedure, the training results, the generalization experiments, and the experiments conducted under limited-data conditions. Within this section, Sec.~\ref{subsec:Preliminary 2D validation}. presents the two-dimensional preliminary validation results, while Sec.~\ref{subsec:3D turbulence results: main contribution}. provides the main three-dimensional turbulence results. Finally, Section~\ref{sec:Conclusions}. concludes the paper.

\section{Methods}
\label{sec:Methods} 

\subsection{Governing equations}

We seek to numerically approximate solutions to the incompressible Navier-Stokes equations, given in their non-dimensional form as:
\begin{align}
\frac{\partial \mathbf{u}}{\partial t}+\nabla\cdot(\mathbf{u}\otimes\mathbf{u})&=\frac{1}{Re}\nabla^2\mathbf{u}-\frac{1}{\rho}\nabla p+\mathbf{f} \\
\nabla\cdot\mathbf{u}&=0,
\label{eq:ns}
\end{align}
where $\mathbf{u}$ is the velocity vector, defined $\{ u,v\}$, $p$ is the pressure, $\rho$ is the density, $Re$ is the Reynolds number, and $\mathbf{f}$ represents any external forces.
The nonlinear nature of the convective term $\nabla\cdot(\mathbf{u}\otimes\mathbf{u})$ is a particularly problematic aspect of the system, resulting in chaotic solutions that are multi-scale -- spanning several orders of magnitude in space and time. Accurate direct numerical simulations of the governing equations must resolve the smallest scales of the flow, which is infeasible with our current computational capabilities for all but simplified, idealized, canonical flows. The number of grid points required scales with $Re^{9/4}$ \cite{Pope2000}, which makes DNS for many high Reynolds number flows intractable. Fortunately, it is often sufficient in scientific and engineering applications to obtain only coarse-grained approximations of the true solution fields. The vast majority of energy in the flow is contained within the large scales, and consequently relevant quantities, e.g. lift or drag, are often mainly influenced by the scales encompassed in coarse-grained approximations.

Coarse-grained solution fields, however, cannot be recovered simply by simulating the governing equations on a coarser computational mesh. This can be shown mathematically by modeling the coarse-graining operation as a low-pass filtering operation, denoted with an overbar $\overline{\boldsymbol{\;\cdot\;}}$. Filtering, as a result of coarse-graining, leads to the introduction of additional terms in the equation due to the nonlinear convective term:
\begin{equation}
    \frac{\partial \overline{\mathbf{u}}}{\partial t}+\nabla\cdot(\overline{\mathbf{u}}\otimes\overline{\mathbf{u}})=\frac{1}{Re}\nabla^2\overline{\mathbf{u}}-\frac{1}{\rho}\nabla \overline{p}+\overline{\mathbf{f}} + \nabla\cdot \boldsymbol{\tau},
    \label{eqn:les}
\end{equation}
where $\boldsymbol{\tau}$ represents the effects of the unresolved velocity components on the resolved field and is defined as
\begin{equation}
    \tau_{ij} = \overline{u_iu_j}-\overline{u}_i\overline{u}_j.
    \label{eqn:t_dns}
\end{equation}
The additional terms are representative of the ``closure'' problem that has plagued numerical analysis of the Navier-Stokes equations for many years, and must be modeled to effectively solve these filtered equations. The equations above represent the filtered Navier-Stokes equations used in LES, where the unknown term is treated as an additional stress, known as the subgrid-scale (SGS) stress. We remark here, that the nature of the filtering induced by the coarse-graining is typically unknown and the rest of this manuscript will utilize filtering and coarse-graining interchangeably. Furthermore, since we are limited by no assumptions of an effective filter, we do not consider comparisons with benchmark closure modeling strategies which require approximations of low-pass filters such as dynamic Smagorinsky \cite{germano1991dynamic}, which requires a test filter, and approximate deconvolution-based methods \cite{layton2012approximate}, which iteratively invert a Gaussian filter.

Numerous methods to model the SGS stress have been proposed over the last few decades, using various theoretical assumptions and empirical observations. The most well-known and perhaps widely used SGS model \cite{SMAGORINSKY1963} employs a linear eddy viscosity, which hypothesizes that the SGS stress is proportional to the filtered rate-of-strain in the fluid. The Smagorinsky model computes an effective eddy viscosity,
\begin{align}
    \nu_t &= (C_s \Delta)^2|\overline{\mathbf{S}}| \\
    \boldsymbol{\tau}^{EV} &= -2\nu_t\overline{\mathbf{S}},
    \label{eqn:smag}
\end{align}
where $C_s$ is a dimensionless empirical coefficient, $\Delta$ is a characteristic length scale, and $|\overline{\mathbf{S}}| = (2\overline{S}_{ij}\overline{S}_{ji})^{1/2}$ is the magnitude of the rate-of-strain tensor. In our model, $C_s$ adaptively varies across the domain and over time as a function of both local and neighboring flow features, adjusting the SGS tensor to better match the ground truth flowfield.

\subsection{Graph neural networks}
\label{sec:Graph neural networks}

From the ansatz presented in the prior section, the SGS stress can be estimated using a deep learning model. We set a few requirements to determine a choice of DNN architecture. We wish for $C_s$ to be spatiotemporally varying and a function of the instantaneous resolved velocity and pressure fields $\overline{\mathbf{u}}, \overline{p}$. In addition, it is understood that the SGS stress is non-local \cite{ClarkDiLeoni2021}, in that the stress at a point in space is influenced not only by the local flow, but also the surrounding flow as well. Lastly, the network architecture must be amenable to the choice of discretization of the coarse-grained resolved solution field, which, to be most generalizable, can be assumed to be an unstructured computational mesh. Based on these requirements, we propose to model the desired turbulence quantity, $C_s$, using graph neural networks (GNNs). Here, we use the Encode-Process-Decode paradigm that is commonly found in deep GNN models. We provide an overview of the GNN architecture in this section.



\subsubsection{Graph representation of flowfield}
\label{sec:Graph representation of flowfield} 
The input graph is defined by $G_0=(\mathbf{V_0}, \mathbf{E_0}, \mathbf{A})$, where $\mathbf{V_0}, \mathbf{E_0}$ and $\mathbf{A}$ represent the node features, edge features and red{adjacency matrix} between nodes, respectively. Specifically, $\mathbf{V_0}$ contains the feature vectors for each node, while $\mathbf{E_0}$ contains the feature vectors for each edge. The adjacency matrix $\mathbf{A}$ is a binary (one-hot) matrix that defines the connections between nodes, where $\mathbf{A}_{ij}=1$ if there is an edge between node $i$ and node $j$, and $\mathbf{A}_{ij}=0$ otherwise.

\begin{figure}
  \begin{center}
      \includegraphics[width=0.65\linewidth]{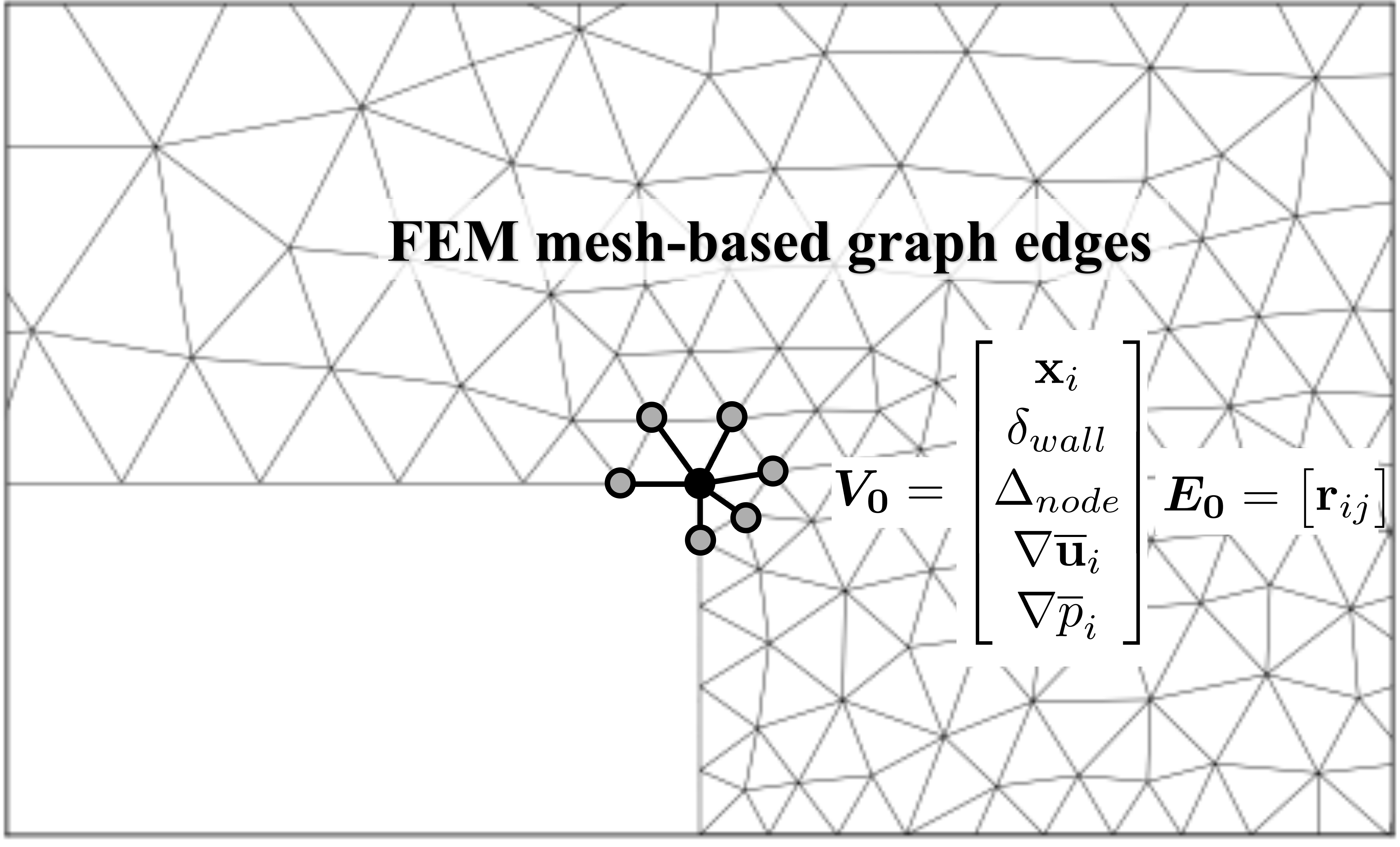}
  \end{center}
  \caption{Schematic of GNN nodes and edges representing the FEM-based computational mesh for the two-dimensional BFS case.}
  \label{fig:GNN_mesh}
\end{figure}

To represent flowfield data from an FEM discretization as a graph, all nodes in the finite element mesh are mapped to graph nodes, $\mathbf{V_0}$, while the edges of finite elements between corresponding nodes are represented by graph edges, $\mathbf{E_0}$, as in Fig. \ref{fig:GNN_mesh}. The node features, $\mathbf{V_0}$, are composed of both fixed and time-varying components. The fixed node features include the position vector, the binary wall indicator, and the characteristic length scale associated with each node. The position vector of each node, $\mathbf{x}_i$, defines the spatial arrangement of the computational mesh, providing a point cloud on which the required SGS stress model parameters are computed. This feature is time-invariant because the mesh coordinates remain fixed. The binary wall indicator, $\delta_{\text{wall}}$, identifies whether a node is located on a no-slip wall ($\delta_{\text{wall}} = 1$) or not ($\delta_{\text{wall}} = 0$). This feature is crucial for near-wall modeling, as SGS stress is significantly influenced by boundary conditions at walls. The characteristic length scale, $\Delta_{\text{node}}$, is the same quantity used to compute the eddy viscosity contribution in the SGS stress model. It is calculated as the minimum edge length of each element at the cell center, and then interpolated to the nodes to define $\Delta_{\text{node}}$. The time-varying node features consist of the local resolved velocity gradient tensor, $\nabla \overline{\mathbf{u}}_i$, and the resolved pressure gradient vector, $\nabla \overline{p}_i$. A basis transformation is applied to the velocity gradient tensor to decompose it into its isotropic, symmetric-traceless, and skew-symmetric components. The edge features, $\mathbf{E_0}$, include the relative position vector between nodes, $\mathbf{r}_{ij} = \mathbf{x}_i - \mathbf{x}_j$, which is time-invariant since the node positions are fixed. As a result, for the two-dimensional BFS case, the node and edge feature dimensions are 10 and 2; for the three-dimensional case, they are 17 and 3, respectively. The node and edge features for the two-dimensional BFS case are briefly described in Fig. \ref{fig:GNN_mesh}. We remark that time-varying FEM discretizations are also compatible with graph representations but are not implemented in this study.

\subsubsection{Encode-Process-Decode}
We adopt a simple Encode-Process-Decode architecture for our GNN implementation as follows. After the graph is generated from the raw dataset, the node features in the input graph, $\mathbf{V_0}$, are mapped to a higher-dimensional latent node space, $\mathbf{h}_i$, through a linear projection. Similarly, the latent edge features, $\mathbf{h}_{ij}$, are obtained by applying a linear projection to the edge features from the input graph. Here, we set the dimension of latent node and edge spaces to 128 for the two-dimensional case. For the three-dimensional case, we increase the latent dimension to 256 to accommodate the larger node/edge feature dimensionality (17/3 vs. 10/2) and the greater complexity of three-dimensional turbulence, which requires higher representational capacity. This enrichment of node and edge feature representations enhances the expressiveness of the architecture, enabling subsequent GNN layers to more effectively capture nonlinear relationships within local graph neighborhoods. The encoding process is formally defined as follows:
\begin{align}
    \mathbf{h}_i &= \mathcal{E}_v(\mathbf{V_{0}}) \\
    \mathbf{h}_{ij} &= \mathcal{E}_e(\mathbf{E_{0}}),  
    \label{eq:encoder}
\end{align}
where $\mathcal{E}_v$ and $\mathcal{E}_e$ are linear projection-based encoders for node and edge features with the structure of $[17, 256]$ and $[3,256]$ for the three-dimensional case ($[10, 128]$ and $[2,128]$ for 2D), respectively, and are applied node/edge-wise. This reflects the input feature sizes (see Sec. \ref{sec:Graph representation of flowfield}.) and the latent widths used (256 in 3D; 128 in 2D). Once the input graph's node and edge features are encoded into latent node and edge representations, the processor performs multiple rounds of message passing on these latent features. In our setup, we use four message-passing layers within the processor. Each layer consists of arithmetic operations and multi-layer perceptrons (MLPs).

After processing, the node features are decoded using a linear projection to predict the target quantity, $C_s$, for the SGS stress model. 
Then, the decoding process is defined as follows:
\begin{align}
    C_s &= \mathcal{D}_v(\mathbf{h}_{i}), 
    \label{eq:decoder}
\end{align}
where $\mathcal{D}_v$ is a linear projection-based decoder for the processed node features $\mathbf{h}_{i}$, with the structure of $[256, 1]$ for the three-dimensional case ($[128, 1]$ for 2D).

\subsubsection{Message-passing}
The processor in the GNN model consists of sequential nonlinear message-passing layers that propagate information throughout the graph. The latent node and edge features, $\mathbf{h}_{i}$ and $\mathbf{h}_{ij}$,  are used as input to the successive message-passing layers. Similar to standard GNN frameworks, we apply a residual update mechanism to edge features. Specifically, the edge features are updated as $\mathbf{h}_{ij} \leftarrow \mathbf{h}_{ij} + \mathbf{h}_{ij}'$, where the new edge features are aggregated along edges. Following this, a residual update is applied to node features as $\mathbf{h}_{i} \leftarrow \mathbf{h}_{i} + \mathbf{h}_{i}'$. The edge update is defined as follows:
\begin{align}
    \mathbf{h}_{ij}' &= f_e(\mathbf{h}_{ij},\mathbf{h}_{i},\mathbf{h}_{j}) \\
    \mathbf{h}_{ij} &= \mathbf{h}_{ij} + \mathbf{h}_{ij}',
    \label{eq:edge}
\end{align}
where $f_e$ is an edge update function containing MLPs with the structure of $[768, 256, 256]$ for the three-dimensional case ($[384, 128, 128]$ for 2D). The sigmoid linear unit (SiLU) serves as the activation function at the hidden layer.

The node update is performed similarly:
\begin{align}
    \mathbf{h}_{i}' &= f_v\left(\frac{1}{|\mathcal{N}|}\sum_\mathcal{N}\mathbf{h}_{ij},\mathbf{h}_{i}\right) \\
    \mathbf{h}_{i} &= \mathbf{h}_{i} + \mathbf{h}_{i}' ,
\end{align}
where $\mathcal{N}$ represents a neighborhood of connectivity and $f_v$ is a node update function containing trainable MLPs with the structure of $[768, 256, 256]$ for the three-dimensional case ($[256, 128, 128]$ for 2D) and SiLU serving as the activation function at the hidden layer.

\subsection{Implementation of the adjoint method}
\label{subsec:Implementation of the adjoint method}

The objective of differentiable physics based turbulence closure modeling is to pair the deep learning model with a differentiable solution algorithm for the governing equations, such that the parameters of the neural network predicting a closure model can be trained in an end-to-end fashion. Automatic differentiation (AD) frameworks have become a necessary tool for training deep neural networks using gradient-based optimization. In AD, compositions of differentiable functions can be exactly computed using the chain rule. Most standard DNNs are composed of straightforward feed-forward operations, such as linear operations and pointwise nonlinearities, which can easily be written in high-level scientific programming languages. However, the integration of PDE solution algorithms, particularly using unstructured computational meshes, can add enormous complexity that makes writing these algorithms in an AD-enabled framework impractical. While the development of PDE solvers in AD-enabled languages is underway \cite{bezgin2023jax}, there currently exists a plethora of efficient and robust non-differentiable PDE solvers equipped with discrete adjoint solvers, that can interface with high-level languages and be called as external routines. If these can be integrated within an AD framework such as PyTorch \cite{paszke2019pytorch}, we can leverage the advantages of a differentiable solution scheme without rewriting complex PDE solvers from scratch.

Fortunately, AD toolchains provide a pathway to integrating this functionality through the use of custom gradients. In reverse-mode AD, we leverage the chain rule as follows:
\begin{align}
    x_1 &= f_0(x_0) \label{eqn:ch0}\\
    x_2 &= f_1(x_1) \label{eqn:ch1}\\
    &\vdots \notag \\
    y &= f_n(x_n) \label{eqn:chn}\\ 
    \frac{dy}{dx_i} &= \frac{dy}{dx_{i+1}}\frac{dx_{i+1}}{dx_i} = \frac{dy}{dx_{i+1}}\frac{df_i(x_i)}{dx_i} .\label{eqn:chain}
\end{align}
The iterative nature of Eq.~\ref{eqn:chain} allows gradients of $y$ to be propagated backwards starting from $\frac{dy}{dx_n}$. For each step in the forward pass, Eqs.~\ref{eqn:ch0}-\ref{eqn:chn}, a corresponding reverse function should be implemented, also known as a vector-Jacobian product (VJP):
\begin{equation}
    \frac{dy}{dx_i} = f_i'\left(\frac{dy}{dx_{i+1}},x_i\right).
\end{equation}
The backpropagation of gradients then constitutes a composition of these VJP functions. In an AD framework, VJPs are supplied for the majority of $f_i$'s that are encountered, e.g. addition, multiplication, and other simple operations. However, if $f_i$ represents a call to an external routine not known by the framework, a custom VJP must be supplied.

We first look at an outline of the implicit pressure correction scheme (IPCS) \cite{goda1979multistep} for solving the unsteady Navier-Stokes equations, shown in Algorithm \ref{alg:ipcs}. The algorithm is initialized from a choice of discretization and flow configuration. At each timestep, the GNN is used to predict the turbulence model parameters, denoted $\alpha_{\boldsymbol{\tau}}$. Since the GNN is written in the AD framework, no further work is required to enable gradient backpropagation. The next three functions, \texttt{tentative\_vel}, \texttt{pres\_correct}, and \texttt{vel\_correct}, represent solutions to three PDEs that are obtained from external routines outside of the AD framework. For this reason, we must implement a custom VJP for each of the PDE solution steps.

\begin{algorithm}
\caption{IPCS method for Navier-Stokes equations}
\label{alg:ipcs}

\textbf{Initialize:} \\ 
\qquad \texttt{GNN}$_\theta$ \\ 
\qquad \texttt{pres\_correct} \\ 
\qquad \texttt{vel\_correct} \\
\qquad $\mathbf{u},p$ \\
\qquad $t \gets 0$ \\
\For{$t<T$}{
$\alpha_{\boldsymbol{\tau}} \gets \texttt{GNN}_\theta(\mathbf{u},p)$\\
$\mathbf{u}^* \gets \texttt{tentative\_vel}(\mathbf{u},p,\alpha_{\boldsymbol{\tau}})$ \\
$p^* \gets \texttt{pres\_correct}(\mathbf{u}^*,p)$ \\
$\mathbf{u} \gets \texttt{vel\_correct}(\mathbf{u}^*,p^*,p)$ \\
$p \gets p^*$ \\
$t \gets t + \Delta t$} 

\end{algorithm}

The three PDE-solving functions can be represented in the FEniCS/FEniCSx formalism as:
\begin{equation}
    \mathbf{x} = \mathrm{\texttt{PDE\_solve}}(\mathbf{m}),
\end{equation}
where $\mathbf{m}$ represents any time-varying inputs to each function, and $\mathbf{x}$ denotes the solutions to the corresponding PDEs. For example, for the function \texttt{tentative\_vel}, $\alpha_{\boldsymbol{\tau}}$ serves as time-varying input, and $\mathbf{u}^*$ is the solution to the PDE, while $\mathbf{u}^*$ represents time-varying input, and  $p^*$ is the solution to the PDE for the function \texttt{pres\_correct}. A custom reverse-mode gradient computation would implement:
\begin{equation}
    y_{\mathbf{m}} = \mathrm{\texttt{PDE\_solve\_vjp}}\left(y_{\mathbf{x}}, \mathbf{m}\right),
\end{equation}
where $y$ represents the objective function, and we indicate partial derivatives with a subscript.
The forward \texttt{PDE\_solve} can be further expanded as the solution to the linear system:
\begin{equation}
    \mathbf{A}(\mathbf{m})\mathbf{x} = \mathbf{b}(\mathbf{m}),
\end{equation}
where $\mathbf{A}$ and $\mathbf{b}$ represent the discretized left hand side (LHS) and right hand side (RHS) of the PDE respectively, which in general, are functions of the input $\mathbf{m}$. Given this description, we can implement \texttt{PDE\_solve\_vjp} using the discrete adjoint method. The discrete adjoint method defines
\begin{equation}
    y_{\mathbf{m}} = \boldsymbol{\lambda}^T\left(\mathbf{A}_\mathbf{m}\mathbf{x} - \mathbf{b}_\mathbf{m}\right),
\end{equation}
where $\boldsymbol{\lambda}$ is the solution to the adjoint equation given by
\begin{equation}
    \mathbf{-A}^T\boldsymbol{\lambda} = y_{\mathbf{x}}^T.
\end{equation}

A PDE solver that exposes $\mathbf{A}$, as well as enables, importantly, the computation of $\mathbf{A}_\mathbf{m}$ and $\mathbf{b}_\mathbf{m}$, offers sufficient functionality to implement \texttt{PDE\_solve\_vjp}.

Our choice of FEniCS/FEniCSx \cite{fenics, baratta2023dolfinx} for this study is motivated by this criteria. FEniCS/FEniCSx offers a general-purpose PDE solving framework using the finite element method. FEniCS/FEniCSx provides the ability to define PDEs symbolically using unified form language (UFL) \cite{alnaes2014unified} and automatic differentiation functionality that is useful for computing $\mathbf{A}_\mathbf{m}$ and $\mathbf{b}_\mathbf{m}$. In addition, FEniCS/FEniCSx incorporates an integrated PETSc library \cite{petsc-user-ref}, which facilitates the efficient computation of linear systems during the solution of PDEs and the discrete adjoint method. PDEs are defined using their variational form. For example, a Poisson system can be defined as:
\begin{python}
    F = inner(nu*grad(u), grad(v))*dx
\end{python}
where \texttt{u} is a \texttt{TrialFunction}, \texttt{v} is a \texttt{TestFunction}, and \texttt{nu} is a given \texttt{Function}. The matrix $\mathbf{A}$ at the left-hand side and vector $\mathbf{b}$ at the right-hand side of the discretized linear system can be produced by calling the \texttt{assemble} routine on \texttt{F}. To compute $(\mathbf{A}-\mathbf{b})_{\mathrm{\texttt{nu}}}$, one need only \texttt{assemble} a new form: 
\begin{python}
    F_nu = derivative(F, nu)
\end{python}
With these procedures in hand, a custom VJP for \texttt{PDE\_solve} can be supplied, described in Algorithm \ref{alg:adj}.

\begin{algorithm}
\caption{Custom gradient for PDE solve}
\label{alg:adj}
\textbf{function} \texttt{PDE\_solve}($\mathbf{m};$ F) \\ 
\qquad $\mathbf{A},\mathbf{b} =$ assemble(F(\textbf{m})) : LHS/RHS of form F \\
\qquad $\mathbf{x} =$ solve($\mathbf{A},\mathbf{b}$) : Linear solve \\ 
\qquad \textbf{return x} \\
\textbf{end function} \\

\textbf{function} \texttt{PDE\_solve\_vjp}($y_\mathbf{x},\mathbf{m};$ F)\\ 
\qquad $\boldsymbol{\lambda} =$ solve($-\mathbf{A}^T,y_\mathbf{x}^T$)   :  Adjoint solution \\
\qquad $F_\mathbf{m} =$ derivative(F(\textbf{m}; \textbf{x}), \textbf{m})  : Residual gradient \\ 
\qquad $y_{\mathbf{m}} =$ assemble($-\boldsymbol{\lambda}^TF_\mathbf{m}$)  : Assembled gradient \\ 
\qquad \textbf{return} $y_{\mathbf{m}}$ \\
\textbf{end function} 

\end{algorithm}

Note that for brevity, we have included only the major steps of the procedure and removed many of the specific details of the implementation, such as conversion of PyTorch tensors to FEniCS/FEniCSx variables, the application of boundary conditions, and certain FEniCS/FEniCSx syntax. With \texttt{PDE\_solve\_vjp} implemented, \texttt{PDE\_solve} can be used as any other differentiable PyTorch layer, and the entire IPCS algorithm can be written using PyTorch variables, which naturally record the gradient tape required for automatic backwards differentiation.

\section{Results and Discussion}
\label{sec:Results and Discussion}

\subsection{Preliminary 2D validation}
\label{subsec:Preliminary 2D validation}
The purpose of this section is to demonstrate the feasibility of our framework using two-dimensional flow cases.
\subsubsection{Dataset generation for 2D flows}
\label{subsubsec:Dataset generation for 2D flows}
Channel flows with obstacles positioned along one or either side of the channel wall are prevalent across various scientific and engineering applications. Examples include applications in environmental science, biomedical engineering, and industry-scale fluid dynamics \textcolor{black}{\cite{pandey2023flow, sarwar2024modeling, ma2021thermal}}. Depending on the shape and configuration of these obstacles, the flow can exhibit a complex mix of phenomena such as stagnation, separation, recirculation, reattachment, and vortex shedding. These behaviors are highly sensitive to obstacle geometry and flow conditions. Consequently, accurately predicting the intricate flow behavior around these obstacles often demands significant computational resources, particularly when dealing with turbulent or unsteady flows. Advanced modeling techniques or high-resolution simulations are typically required to capture the full dynamics, making efficient computation an ongoing challenge in the field.

\begin{figure}
  \begin{center}
      \includegraphics[width=0.7\linewidth]{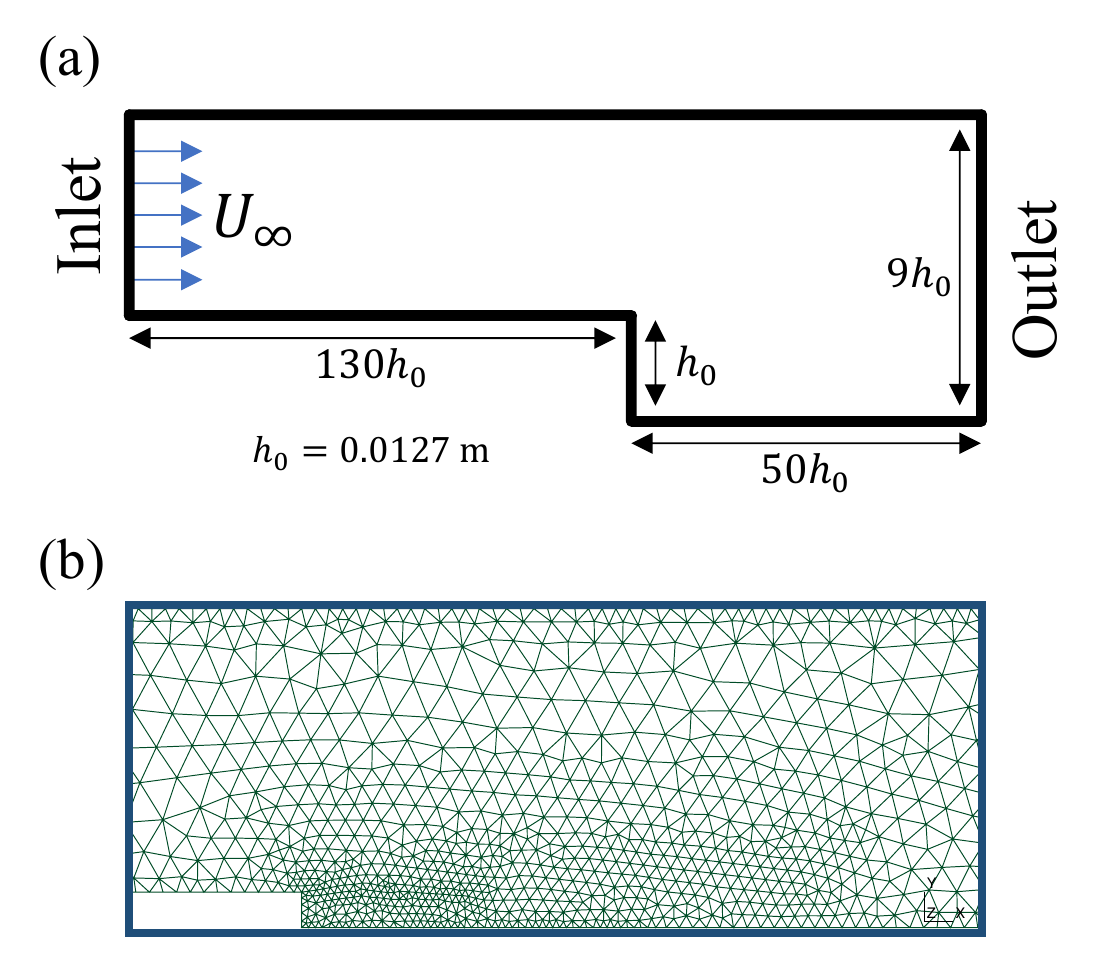}
  \end{center}
  \caption{(a) Schematic of the two-dimensional backwards-facing step domain with inlet, outlet, and dimensions labeled. (b) A subsection of the ``5k'' mesh used to evaluate the ML closure. Greater density of elements is placed in front of the step.}
  \label{fig:2d_dom}
\end{figure}

We first describe our training configuration given by flow over a backwards-facing step (BFS). This is a common benchmark problem in fluid dynamics that has been widely studied in both experimental \cite{armaly1983experimental} and computational \cite{wee2004self} settings. BFS flow is characterized by flow separation at the step and a subsequent reattachment region further downstream, which introduces a recirculation region near the step and transient vortex shedding. These phenomena are found frequently in more complex non-idealized flows and can be challenging to model accurately, making the BFS a valuable case study for evaluating modeling approaches. The BFS configuration can be described as flow in a channel with a sudden expansion. An inlet is prescribed on the left with a uniform velocity in the $x$-direction, entering a channel of fixed width. Once the flow has fully developed, the channel width is increased, creating a step at the lower wall. This expansion produces a shear layer at the step height, which leads to turbulent mixing and vortex development. The flow reattaches downstream and eventually exits the domain through a pressure outlet. A schematic of the computational domain for the two-dimensional BFS is presented in Fig. \ref{fig:2d_dom}(a).

Ground truth data is taken from a high-resolution two-dimensional LES simulation of the BFS flow under the Reynolds number, $Re=U_{\infty}h_0/\nu=26,051$, where $U_{\infty}=32m/s$ and $\nu=1.56\times10^{-5}$. We use an unstructured triangular mesh of on average 80,000 Taylor-Hood elements, which is refined near all walls and particularly around and downstream from the step. The elements with the highest density reside in the shear layer at the step, with generally decreasing density further away. The LES equations using the standard Smagorinsky turbulence model with $C_s=0.16$ are solved using an implicit pressure correction scheme and Crank-Nicolson timestepping. We employ only the least-squares incompressibility constraint (LSIC) stabilization in the finite element formulation, while the streamline-upwind Petrov-Galerkin (SUPG) term is deactivated.
For consistency, we set the ground truth timestep $\Delta t$ to $5\times10^{-6}s$, which is sufficient to keep the Courant-Friedrichs-Lewy (CFL) number approximately equal to or less than 1, where $\mathrm{CFL}<1$ is not strictly required due to the implicit timestepping. Snapshots of the velocity and pressure fields are saved every $10^{-4}$ seconds, and a total of 1024 snapshots are generated. We denote results from this ground truth simulation as ``LES-80k''. 
More comprehensive efforts to align the modeling with experimental or DNS studies under specific operating conditions are considered beyond the scope of this work, because the primary objective of this work is to demonstrate the capabilities of the graph neural network and the differentiable framework.

We generate a significantly coarser mesh for the two-dimensional BFS flow, consisting of 5,253 triangular cells with Taylor-Hood elements, on which we apply our ML-based closure and evaluate it against baseline comparisons. Fig. \ref{fig:2d_dom}(b) shows a portion of this mesh in the area around the step. The scaling factor for element sizes was set to be approximately 4.5 times of the size of the high-resolution mesh. Simulations on this coarse grid use nearly the same solution scheme as the ground truth, except the timestep size is increased to $\Delta t=10^{-4}s$. We test two baseline comparisons on this grid, no model (``NM-5k''), where no closure model is used, and the standard Smagorinsky closure (``LES-5k''), where we set $C_s=0.06$ based on optimizing \textit{a-posteriori} results with respect to the ground truth. The ML-closure is denoted ``ML-5k''.

A third mesh, consisting of 23,502 elements, is also created, positioned between the coarse ML mesh and the fine ground truth mesh in terms of size. This mesh serves as an additional baseline comparison (referred to as 'LES-20k') for the model in relation to a more well-resolved LES simulation. The Smagorinsky model is used with $C_s=0.07$, and the timestep is reduced to $\Delta t=5\times10^{-5}s$.

\subsubsection{Training procedure and loss formulation}
\label{subsubsec:Training procedure and loss formulation}
We optimize the ML model by minimizing the $L^2$ norm error of the resulting \textit{a-posteriori} velocity field with respect to the training dataset from the ground truth velocity field. During training, a predicted solution trajectory $\hat{\mathbf{u}}$ is obtained by simulating the LES equations using the GNN-based SGS closure model. The training dataset is generated using only the first 256 snapshots of the LES-80k simulation. We set multiple trajectories for training dataset, where each of the trajectories consists of 32 consecutive snapshots with different initial conditions separated by 16 timesteps. This is based on our own empirical analysis and suggestions from previous work to avoid gradient explosion \cite{shankar2025differentiable}. Ultimately, this leads to a total of 15 trajectories for training. The velocity field in each trajectory {$\{ \mathbf{u}_k \}_{k=0}^{n}$} is projected onto the coarse mesh used in the ML-LES simulation to generate the target velocity field {$\{ \overline{\mathbf{u}}_k \}_{k=0}^{n}$}. The projection is performed using the Galerkin projection method as follows. In a FEM-based CFD solver, the flow field data is represented as a function of the $x$, $y$ and $z$ coordinates within a specified function space. In this section, the two-dimensional ground truth velocity field, $\mathbf{u}$, defined on the dense grid, is represented within each cell as a function of $x$ and $y$ and belongs to a second-order continuous Galerkin function space. Likewise, the ground truth velocity data projected onto the coarse grid, $\overline{\mathbf{u}}$, is defined on each cell of the coarse grid and belongs to a second-order continuous Galerkin function space based on the coarse mesh. The weak form equation of the projection from $\mathbf{u}$ to $\overline{\mathbf{u}}$ is expressed as follows:

{
\begin{equation}
    \int_{\Omega_e} \overline{\mathbf{u}} \cdot v \; dx = \int_{\Omega_e} \mathbf{u} \cdot v \; dx, \quad \forall v \in{V_c},
\end{equation}
}
where $\Omega_e$ represents each element in the computational domain, $V_c$ is a second-order continuous Galerkin function space defined on the coarse grid, and $v$ is a test function within the function space. For simplicity, this weak form equation can be written as follows:
{
\begin{equation}
    a_e(\overline{\mathbf{u}},v) = L_e(v), \quad \forall v \in{V_c},
\end{equation}
}
where $a_e$ and $L_e$ represent bilinear and linear forms, respectively. Since this simplified weak form equation is satisfied for all test functions within $V_c$, by choosing basis functions, $\{\phi_i\}_{i=1}^6$, as a test function, a linear system can be configured as follows: 
\begin{equation}
\begin{split}
    \mathbf{A_e}\mathbf{c_e} &= \mathbf{L_e}, \\
    \text{where} \quad (\mathbf{A_e})_{ji} &= a_e(\phi_i, \phi_j), \; (i,j=1,2,...,6)\\
    \mathbf{c_e} &= [c_1\, c_2\, ...\, c_6]^T, \\
    \mathbf{L_e} &= [L_e(\phi_1)\, L_e(\phi_2)\, ...\, L_e(\phi_6)]^T
\end{split}
\label{eqn:weakform}
\end{equation}

In FEniCS/FEniCSx, a second-order Lagrange function is chosen as the basis function within the second-order continuous Galerkin function space, where each component of $\mathbf{c_e}$ represents the coefficient of a basis function for $\overline{\mathbf{u}}$. Since the linear system in Eq.~\ref{eqn:weakform} is defined locally on each element, the global linear system is assembled to ensure continuity between elements as follows. 
{
\begin{equation}
    \mathbf{A}\mathbf{c}=\mathbf{L}
\end{equation}
}
This assembly process combines the contributions of each element, resulting in a system that enforces the continuity of the solution across shared boundaries between adjacent elements. The pressure field is also projected for the initial condition. The initial condition is then integrated over 32 steps using the GNN-based turbulence model. The overall loss is computed as
\begin{equation}
    L = 0.9\times\frac{1}{T}\sum_{t=0}^T \int_\Omega ||\overline{\mathbf{u}}_t - \hat{\mathbf{u}}_t||_2^2 \; d\Omega+0.1\times\int_\Omega ||\overline{\mathbf{u}}_{mean} - \hat{\mathbf{u}}_{mean}||_2^2 \; d\Omega,
\end{equation}
where both the instantaneous and the time-averaged accuracy are considered, and the integral is computed using the FEM discretization. This formulation of the loss function enables the model to learn subgrid closures that not only match the instantaneous flow field values but also maintain consistency with the overall trend of actual values. Since the integral is computed in FEniCS an adjoint can be obtained using the same method described in Sec. \ref{subsec:Implementation of the adjoint method}. The loss is backpropagated in an end-to-end fashion, through all PDE evaluations and timesteps, to update the parameters of the GNN. The model is trained for 50 epochs using the Adam optimizer \cite{kingma2014adam} with a learning rate of $10^{-4}$. Training took about 6 hours on 2 NVIDIA A100 GPUs and 8 AMD EPYC 7543P CPU-cores. Much of the inefficiency in the model can be attributed to data transfer between the CPU and GPU, where the CPU is used to solve the PDE and the GPU is used to evaluate the GNN. Training history is shown in the Fig.~\ref{fig:Training_history} in Appendix. \ref{app:Training history}.

\subsubsection{Training results for 2D flows}
\label{subsubsec:Training results for 2D flows}
We evaluate the performance of the hybrid ML-LES solution algorithm by examining several \textit{a-posteriori} metrics from the flow. For each generalization case, the trained ML closure is used in a simulation of the flow beginning with zero initial conditions. The solution is integrated over a total of 1024 timesteps (we remind the reader that error computation and backpropagation during training are performed over the first 256 timesteps in batches of 32). We analyze the efficacy of the model with respect to several baselines: ``NM-5k'', no closure on the coarsest mesh, ``LES-5k'', Smagorinsky model on the coarsest mesh, and ``LES-20k'', Smagorinsky model on a finer mesh. 

For all predicted velocity fields, we compute the normalized time-varying $L^2$ error at each time step with respect to the ground truth velocity, given by:
\begin{equation}
    \frac{\int_\Omega ||\overline{\mathbf{u}}_t - \hat{\mathbf{u}}_t||_2^2 \;d\Omega}{U_\infty^2},
\end{equation}

where $U_\infty^2$ is the inlet velocity, and $\Omega$ is the computational domain. Additionally, we provide the velocity sampled from a probe in front of the step at the location $(h,h)$. This location is based on the region where the error between the LES-5k and the LES-80k models is the largest. For these transient plots, we report values from the first 1024 timesteps, extending well beyond the training dataset, which consists of the first 256 timesteps. This allows us to evaluate the accuracy and stability of pointwise predictions over a longer prediction horizon. We also examine the time-averaged statistics of the flow, as these are often the most relevant metrics of interest. We present the time-averaged x-velocity field $U_x$ along with the corresponding error field to assess each model's ability to accurately capture the recirculation zone and identify areas of discrepancy within the flow. This series of \textit{a-posteriori} metrics provides a comprehensive evaluation of model performance, ensuring that the predicted flow aligns closely with the ground truth.

\begin{figure}
  \begin{center}
      \includegraphics[width=0.7\linewidth]{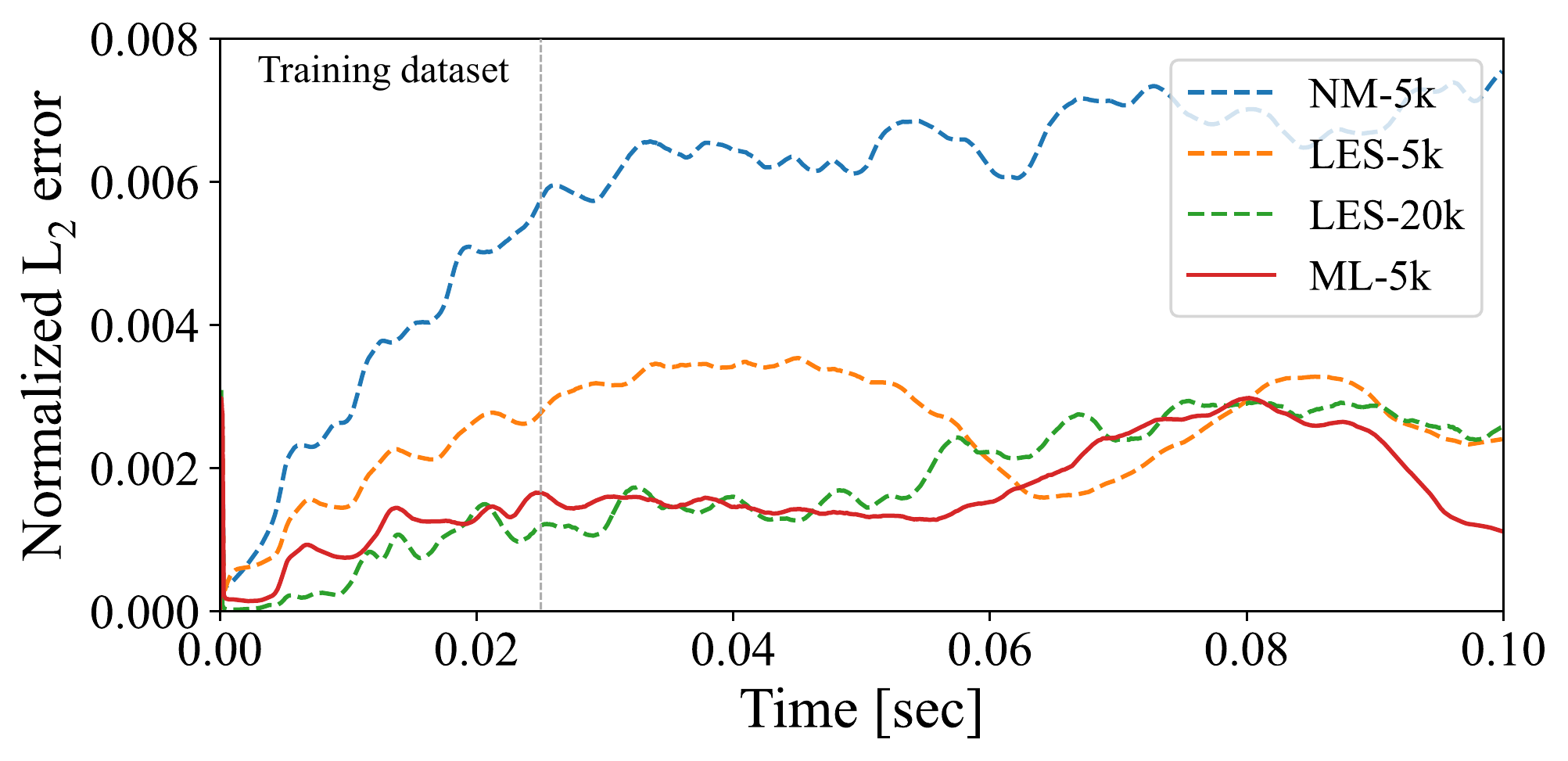}
  \end{center}
  \caption{Normalized $L_2$ error on the two-dimensional BFS case. The gray dashed line represents the extent of training data collected.}
  \label{fig:BFS_error}
\end{figure}

\begin{figure}
  \begin{center}
      \includegraphics[width=\linewidth]{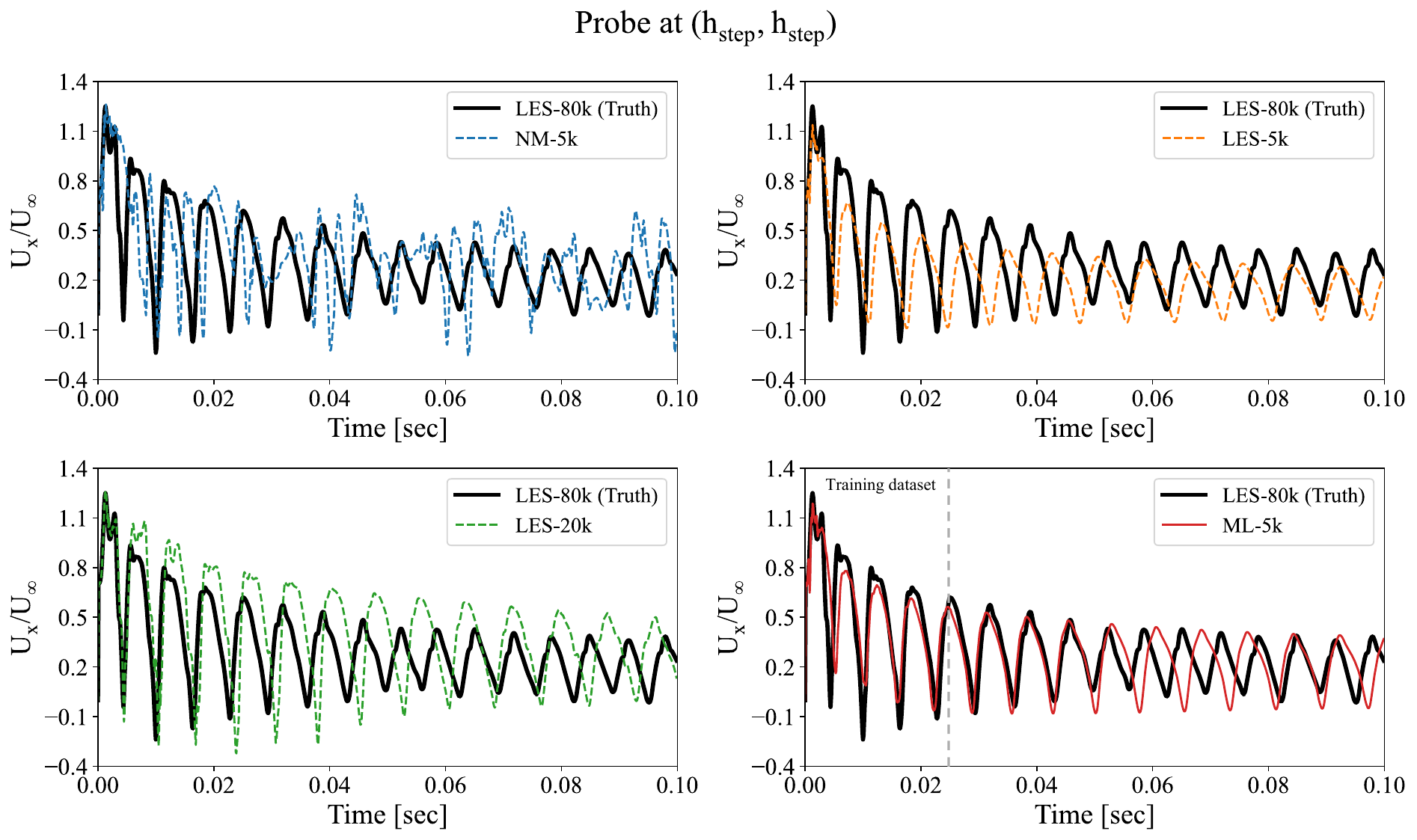}
  \end{center}
  \caption{$x$-direction Velocity measurement from a probe in front of the step at $(x, y)=(h, h)$.}
  \label{fig:BFS_probe}
\end{figure}

Figure \ref{fig:BFS_error} shows the normalized $L^2$ error over time for the four models tested (with the gray dashed line representing the extent of training data used). We observe that NM-5k shows the steepest growth and highest overall error, indicating the need for a closure model. LES-5k shows a smaller initial rate of error increase and fluctuates around a significantly lower error than the NM-5k model, while the LES-20k model results in the smallest initial rate of error increase and smaller error than LES-5k. We see that the learned model ML-5k generally results in the lowest overall $L^2$ error, and given that this metric is reflective of the training cost function, we determine that the training process is functioning as intended to minimize the $L^2$ error. Furthermore, the ML-5k model is stable for more time steps than those used during training ($t=0.0256s$). However, the transient $L^2$ error is just one metric we should use to evaluate the trained model's performance. To further examine the transient behavior of the model, pointwise accuracy of each model near the wall is compared. Fig. \ref{fig:BFS_probe} illustrates pointwise accuracy of the models from a probe at $(x,y)=(h,h)$ using the normalized $x$-direction velocity. It is observed that the ML-5k model is more accurate in capturing periodic fluctuations of velocity than other models and remains stable for a longer time horizon.

\begin{figure*}
  \begin{center}
      \includegraphics[width=\linewidth]{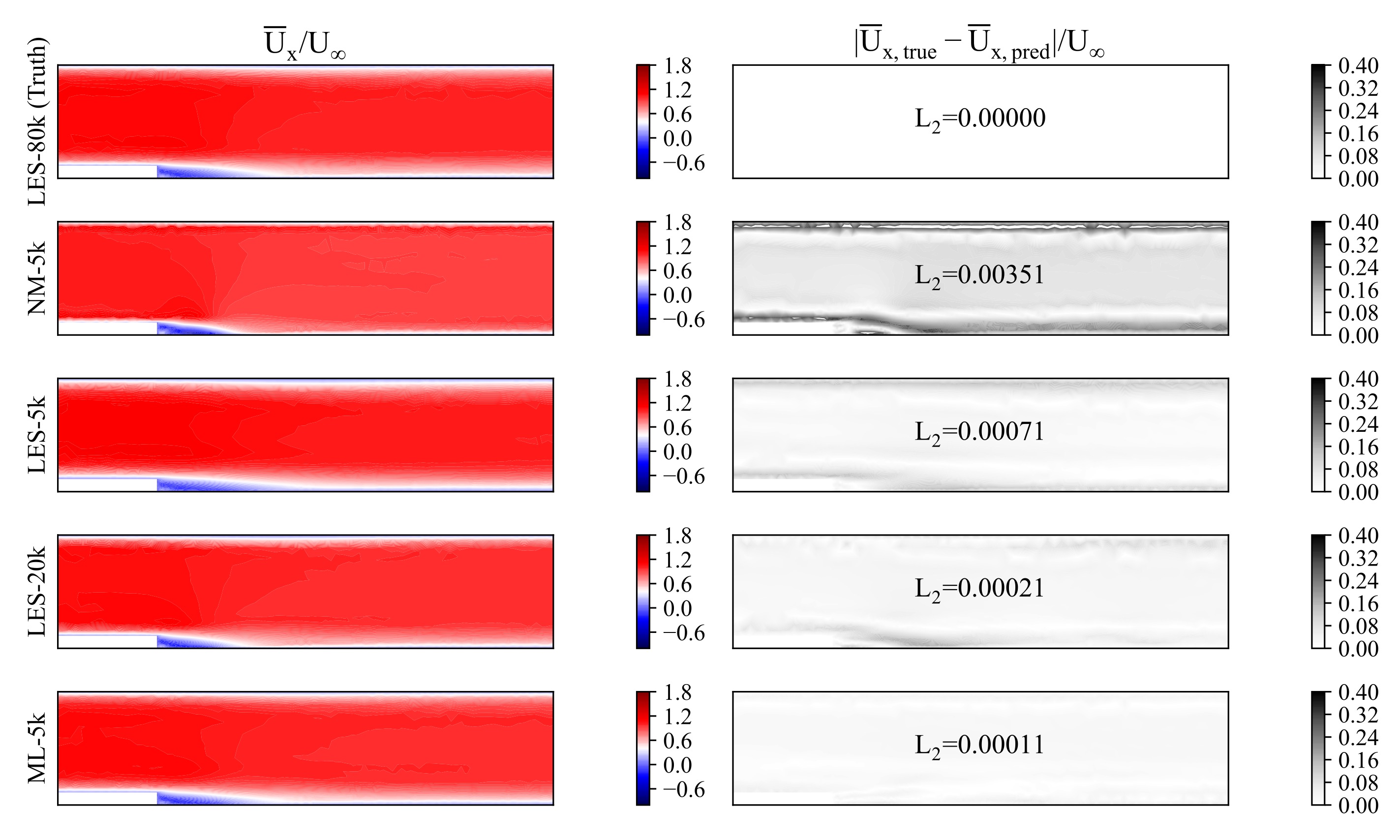}
  \end{center}
  \caption{Plots of the mean flow from two-dimensional BFS case for various models. \textbf{(Left column)} The normalized mean x-velocity and \textbf{(Right column)} The error with respect to the truth in space, as well as the integrated total.}
  \label{fig:Flowfield_BFS}
\end{figure*}

The mean flow is a particularly useful measure that decouples the transient effects of the model from flow features of interest. Fig. \ref{fig:Flowfield_BFS} depicts the normalized x-velocity of the flow and the resulting error field with respect to the LES-80k model. We also report the integrated total error for each model. We can observe the recirculation region characterized by the negative (blue) x-velocity that is present in each model. the NM-5k model contains discrepancies throughout the portion of the domain shown, including prior to the step, in front of the step, and at the boundary layer. The LES-5k model reduces the error by a factor of over 5, but still contains deviations, notably at the boundary layer, while the LES-20k model achieves smaller error than the LES-5k model by reducing the error in this region. The ML-5k model achieves the smallest errors among models by reducing the error within the region downstream of the step.

\begin{figure}
  \begin{center}
      \includegraphics[width=0.5\linewidth]{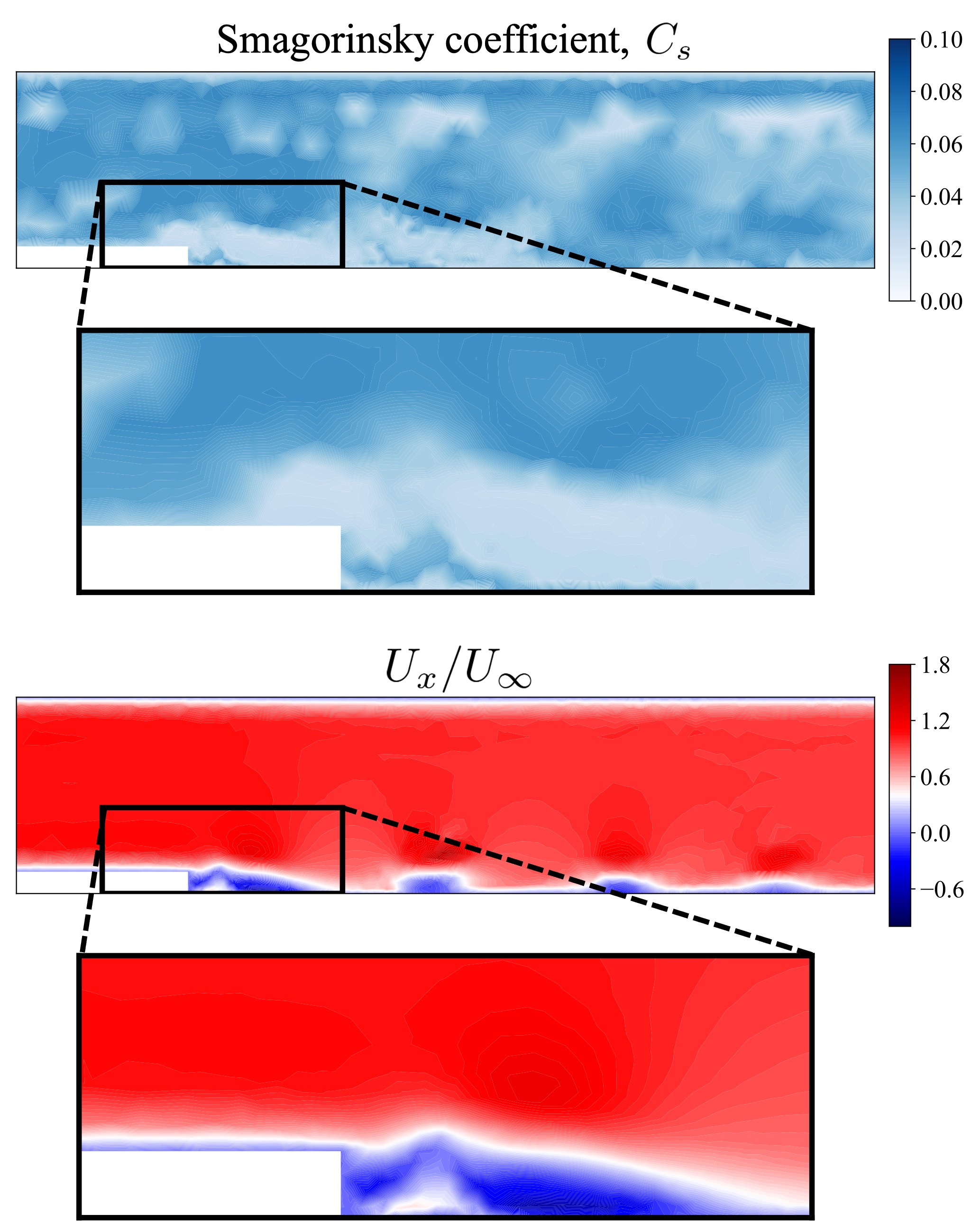}
  \end{center}
  \caption{\textbf{(Top row)} Smagorinsky coefficient and \textbf{(Bottom row)} Normalized $x$-direction velocity predicted by trained model across domain near the two-dimensional BFS and their zoomed-in views at t=0.02} 
  \label{fig:Smagorinsky_BFS}
\end{figure}

Figure \ref{fig:Smagorinsky_BFS} illustrates the visualization of the normalized $x$-direction velocity and the Smagorinsky coefficient predicted by the subgrid closure model at $t= 0.02s$. In LES simulation with the conventional Smagorinsky model, $C_s$ is predetermined before the simulation to model the viscous effect of subgrid-scale eddies, and every mesh point in the computational domain has the same $C_s$ value, which remains fixed throughout the simulation. This often leads to excessive dissipation in the viscous sub-layer region near walls where wall viscosity is already strong enough to damp the eddies. Furthermore, flow state predictions near vortices with high local shear can be inaccurate due to this over-dissipation. In contrast, the ML-5k adaptively adjusts $C_s$ during the simulation to tune the subgrid dissipation. This is possible because our ML subgrid closure model can learn the optimal amount of local subgrid dissipation to minimize the ground truth error. Specifically, the ML-5k model predicts smaller $C_s$ values in the near-wall regions compared to the areas farther from the walls. In addition, the model predicts lower $C_s$ value near the vortices where local shear is high due to rapid velocity changes around the vortex cores. This adaptive adjustment of $C_s$ reduces the dissipation caused by the subgrid eddies, thereby decreasing errors within the near-wall region and the region upstream of the step as shown in Fig. \ref{fig:Flowfield_BFS}.
It is notable that such complex subgrid learning is extracted despite our model being exposed to a training dataset solely comprised of sampled velocity time-series across the domain.

\subsubsection{Generalization to new geometries}
\label{subsubsec:Generalization to new geometries}
We apply our GNN-based subgrid closure model, trained on the two-dimensional BFS case, to different geometries to evaluate its generalization ability. These new geometries include two-dimensional ramp and wall-mounted cube (WMC) cases with the same height of $h=0.0127m$ as the BFS case and are tested under the same Reynolds number, $Re=26,501$. They present a more challenging extrapolation task for the trained model as these configurations have distinct geometric differences near the separation regions and exhibit different separation physics. Visualizations of the discretizations of these geometries are provided in Fig.\ref{fig:Grid_BFR_WMC} in Appendix. \ref{app:Grids used for data generation}

\begin{figure}
  \begin{center}
      \includegraphics[width=\linewidth]{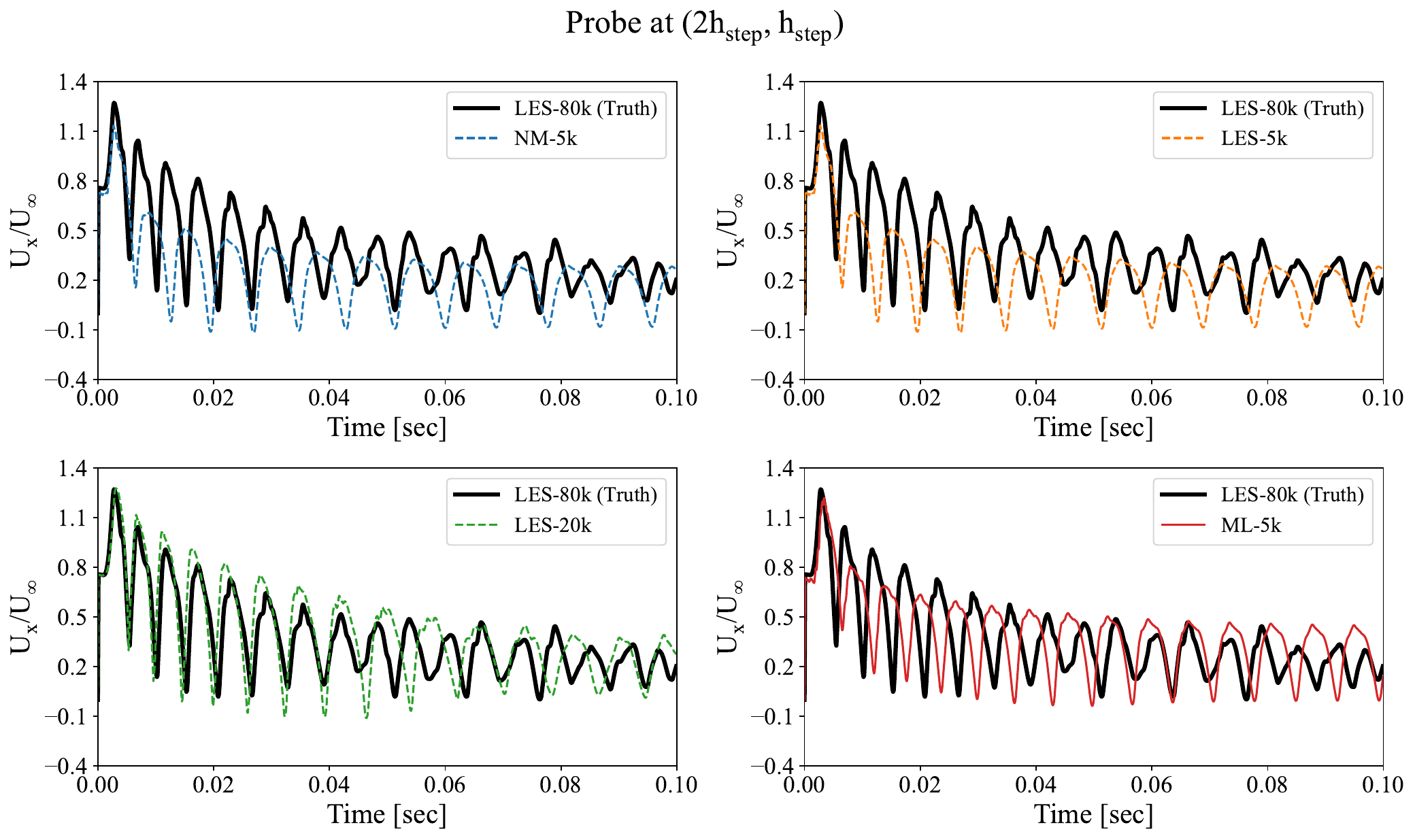}
  \end{center}
  \caption{$x$-direction Velocity measurement taken from a probe in front of the two-dimensional ramp at $(x, y)=(2h, h)$}
  \label{fig:BFR_probe}
\end{figure}

Fig. \ref{fig:BFR_error} shows the normalized $L^2$ error over time for the two-dimensional ramp case, where we compare the performance of the trained model with three other baseline predictions. We observe that NM-5k exhibits the highest overall error (due to the lack of any closure), while LES-5k has lower error than NM-5k, highlighting the need for a closure model. Furthermore, ML-5k shows a lower error than LES-5k, whereas LES-20k achieves the lowest error. This suggests that even when tested on the ramp case, which has a smoother separation due to a smaller flow turning angle, our model is improved compared to LES-5k model. Furthermore, the proposed model exhibits stability beyond the time steps used during training, suggesting its robustness in extended time-step integration. Fig. \ref{fig:BFR_probe} presents the pointwise accuracy of the models based on normalized $x$-direction velocity data obtained from a probe located at $(x,y)=(2h,h)$. The trends observed in the pointwise analysis are consistent with the overall error profiles, demonstrating that the ML model not only improves upon the performance of LES-5k but also remains stable for longer prediction horizons than those encountered during training.

\begin{figure}
  \begin{center}
      \includegraphics[width=0.7\linewidth]{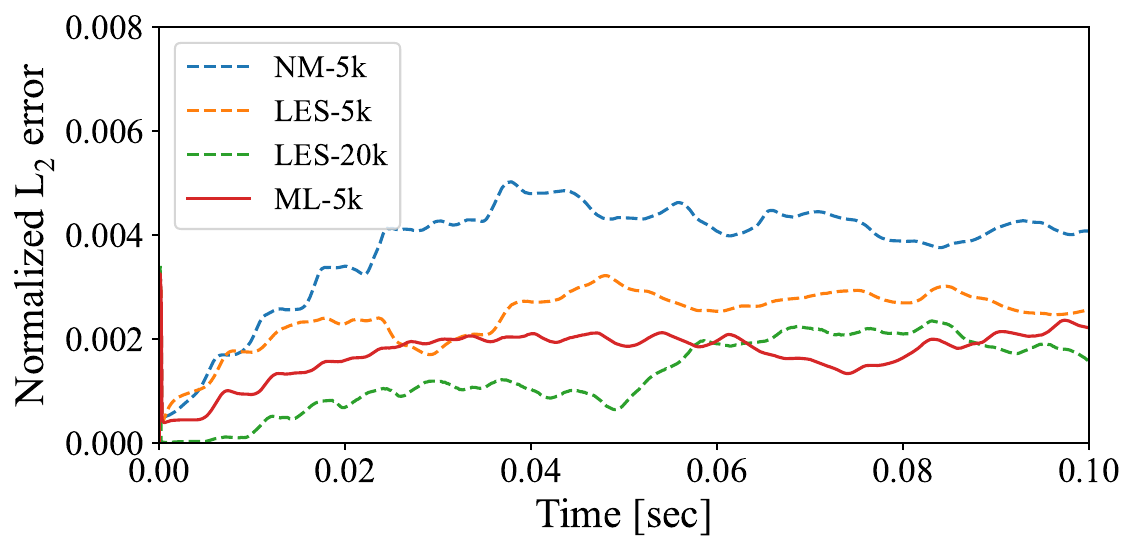}
  \end{center}
  \caption{Normalized $L_2$ error on the two-dimensional ramp case. This geometry was unseen during training.}
  \label{fig:BFR_error}
\end{figure}

\begin{figure*}
  \begin{center}
      \includegraphics[width=\linewidth]{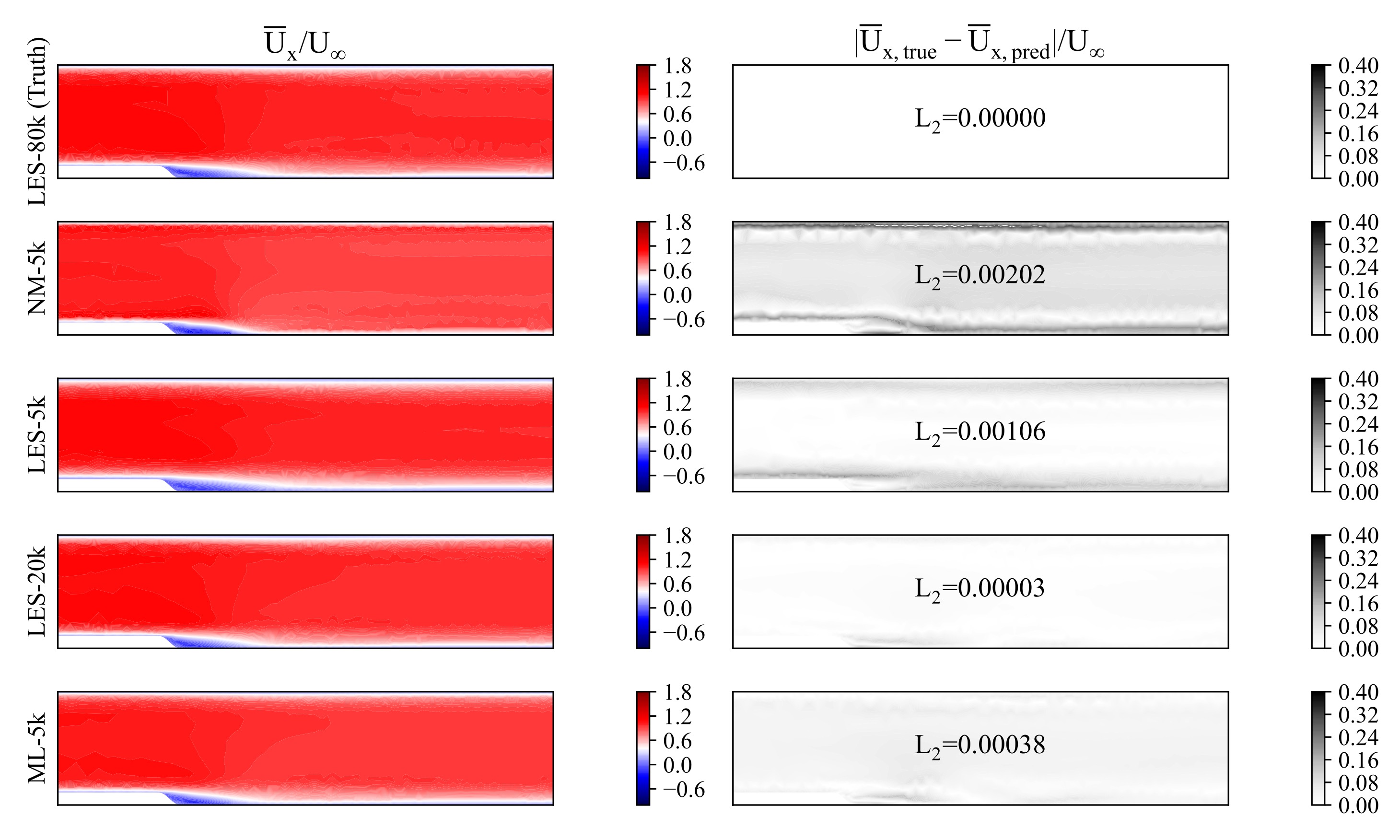}
  \end{center}
  \caption{Plots of the mean flow from the two-dimensional ramp case for various models. \textbf{(Left column)} The normalized mean x-velocity and \textbf{(Right column)} The error with respect to the truth in space, as well as the integrated total.}
  \label{fig:Flowfield_BFR}
\end{figure*}

\begin{figure}
  \begin{center}
      \includegraphics[width=0.5\linewidth]{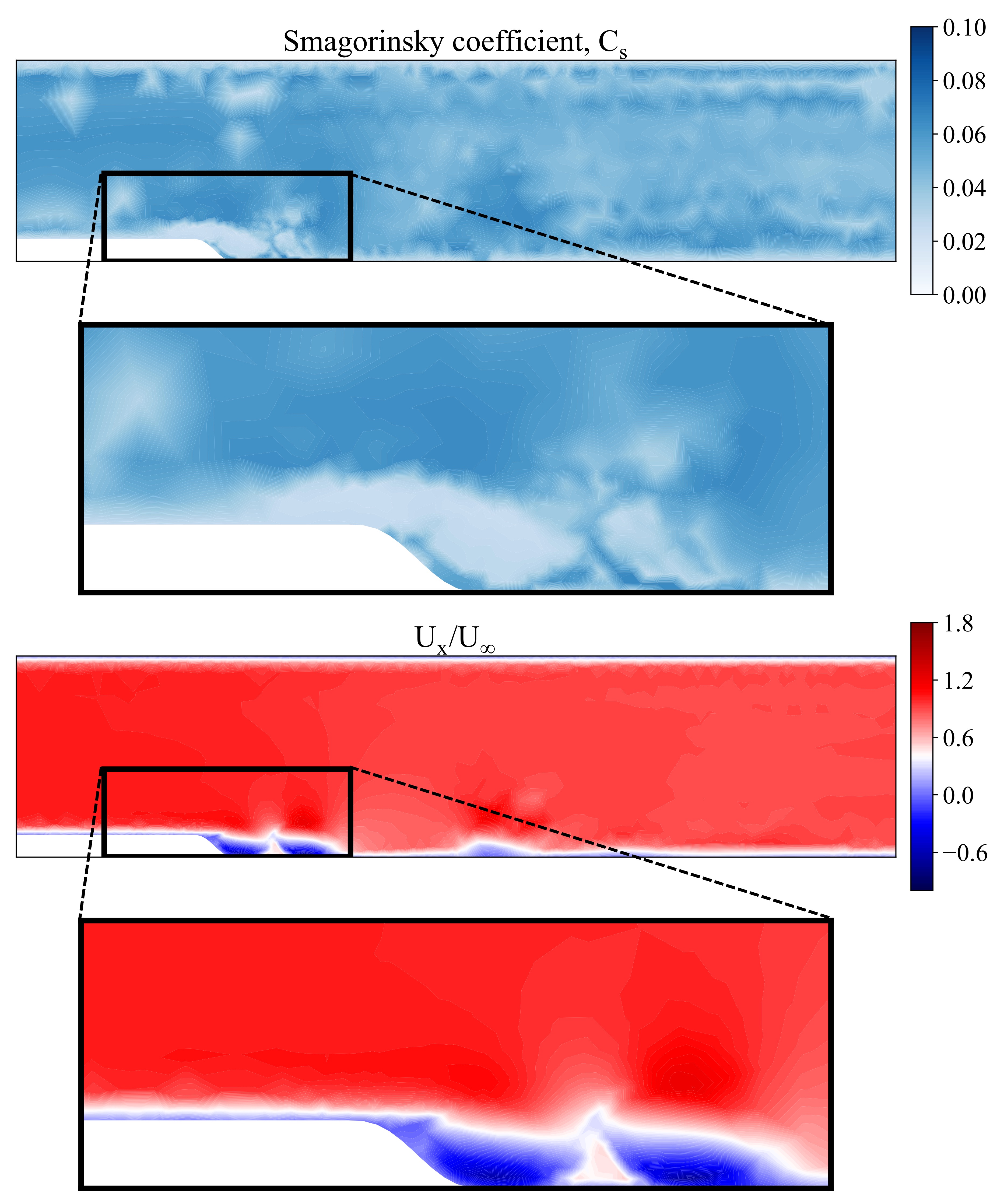}
  \end{center}
  \caption{\textbf{(Top row)} Smagorinsky coefficient and \textbf{(Bottom row)} Normalized $x$-direction velocity predicted by trained model across domain near the two-dimensional ramp and their zoomed-in views at t=0.02}
  \label{fig:Smagorinsky_BFR}
\end{figure}

Figure \ref{fig:Flowfield_BFR} presents a contour plot of the normalized mean $x$-direction velocity around the two-dimensional ramp, along with the error distribution for each model when compared to the LES-80k reference solution. Additionally, the integrated error values for each model are reported. Across all models, the recirculation zones immediately downstream of the ramp are well captured. However, the NM-5K and LES-5k models exhibit the largest errors near these recirculation zones and along the wall boundary. In contrast, the LES-20k model shows relatively smaller errors in these areas, achieving the lowest error among the models. The ML-5k model effectively reduces errors in the high shear flow and near-wall turbulence regions by adaptively modulating the Smagorinsky coefficient. Fig. \ref{fig:Smagorinsky_BFR} shows a visualization of the Smagorinsky coefficient and the normalized $x$-direction velocity across the domain at $t=0.02s$. It is observed that our model changes $C_s$ dynamically depending on localized flow quantities, thereby reducing errors near the wall boundary and vortices. 

\begin{figure}
  \begin{center}
      \includegraphics[width=0.7\linewidth]{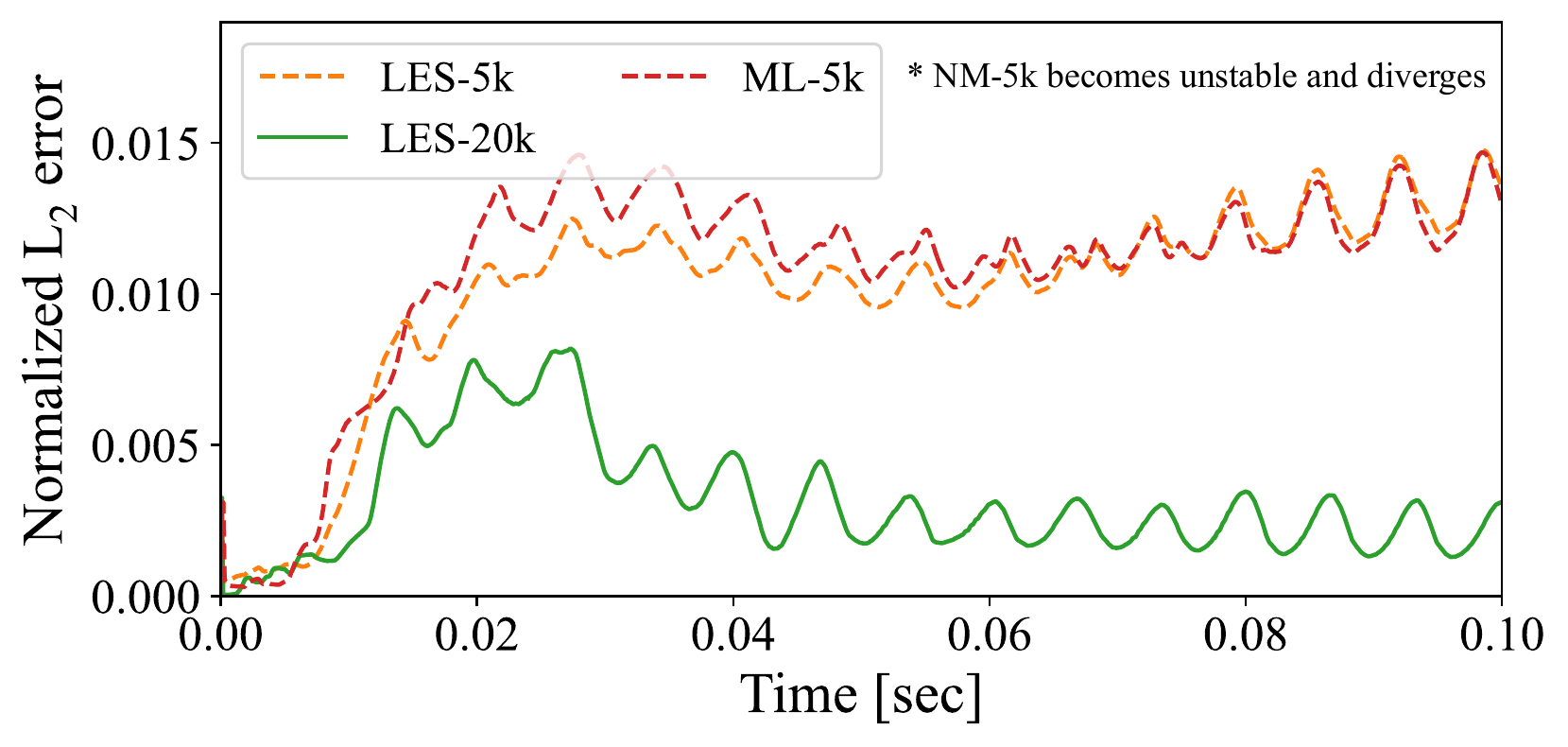}
  \end{center}
  \caption{Normalized $L_2$ error on the two-dimensional WMC case. This geometry was unseen during training.}
  \label{fig:WMC_error}
\end{figure}

\begin{figure}
  \begin{center}
      \includegraphics[width=\linewidth]{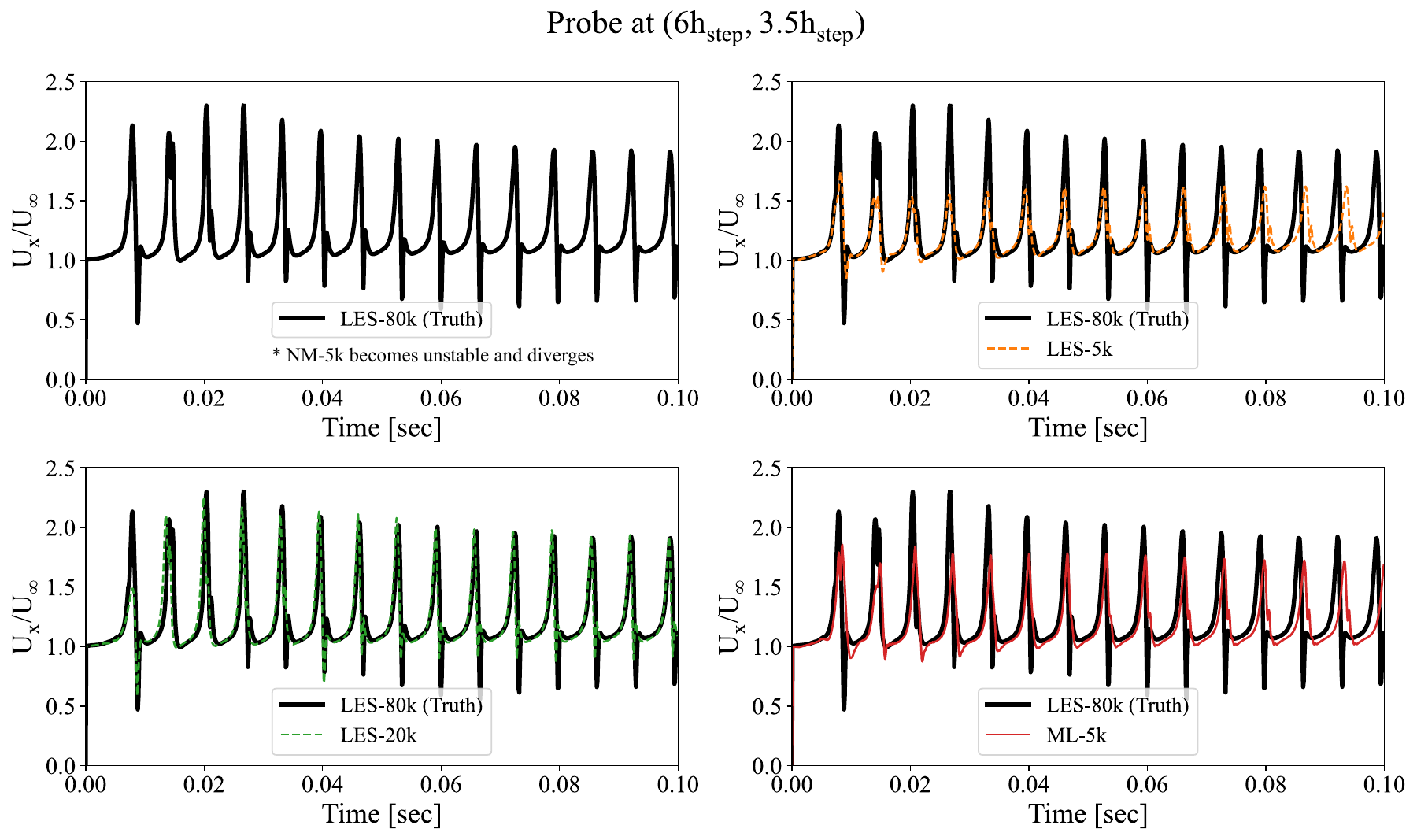}
  \end{center}
  \caption{$x$-direction Velocity measurement taken from a probe in front of the two-dimensional cube at $(x, y)=(6h, 3.5h)$}
  \label{fig:WMC_probe}
\end{figure}

Figure \ref{fig:WMC_error} compares the predictability of the trained model for the two-dimensional WMC case with three other models using the $L^2$ error over time. In the WMC case, a simulation without a turbulence model becomes unstable and diverges after several time steps, indicating the necessity of incorporating a closure model. While the LES-20k achieves the lowest error among the models, using the ML-5k model does not result in significant performance improvement over the LES-5k model unlike the ramp case. This can be attributed to the difference in separation mechanisms between the two cases. In contrast to the ramp case where flow originates from inflow and simply separates near the slope, the flow upstream of the cube encounters a stagnation point with high pressure and recirculation before separation---a fundamentally different separation mechanism from what the trained model is exposed to during training. Nevertheless, the trained model recovers the performance of the LES-5k model and remains stable in extended time-step integration above time span used during training. Figure \ref{fig:WMC_probe} compares the pointwise accuracy of models from a probe at $(x,y)=(6h,3.5h)$ using the $x$-direction velocity normalized by the freestream velocity. At this location, the LES-20k model can capture peaks of the periodic fluctuation of the true data, while the LES-5k model predicts lower peaks than the actual values. Although using the trained model does not provide a marked improvement in peak prediction, ML-5k remains stable for longer prediction horizons than those used during training.

\begin{figure*}
  \begin{center}
      \includegraphics[width=\linewidth]{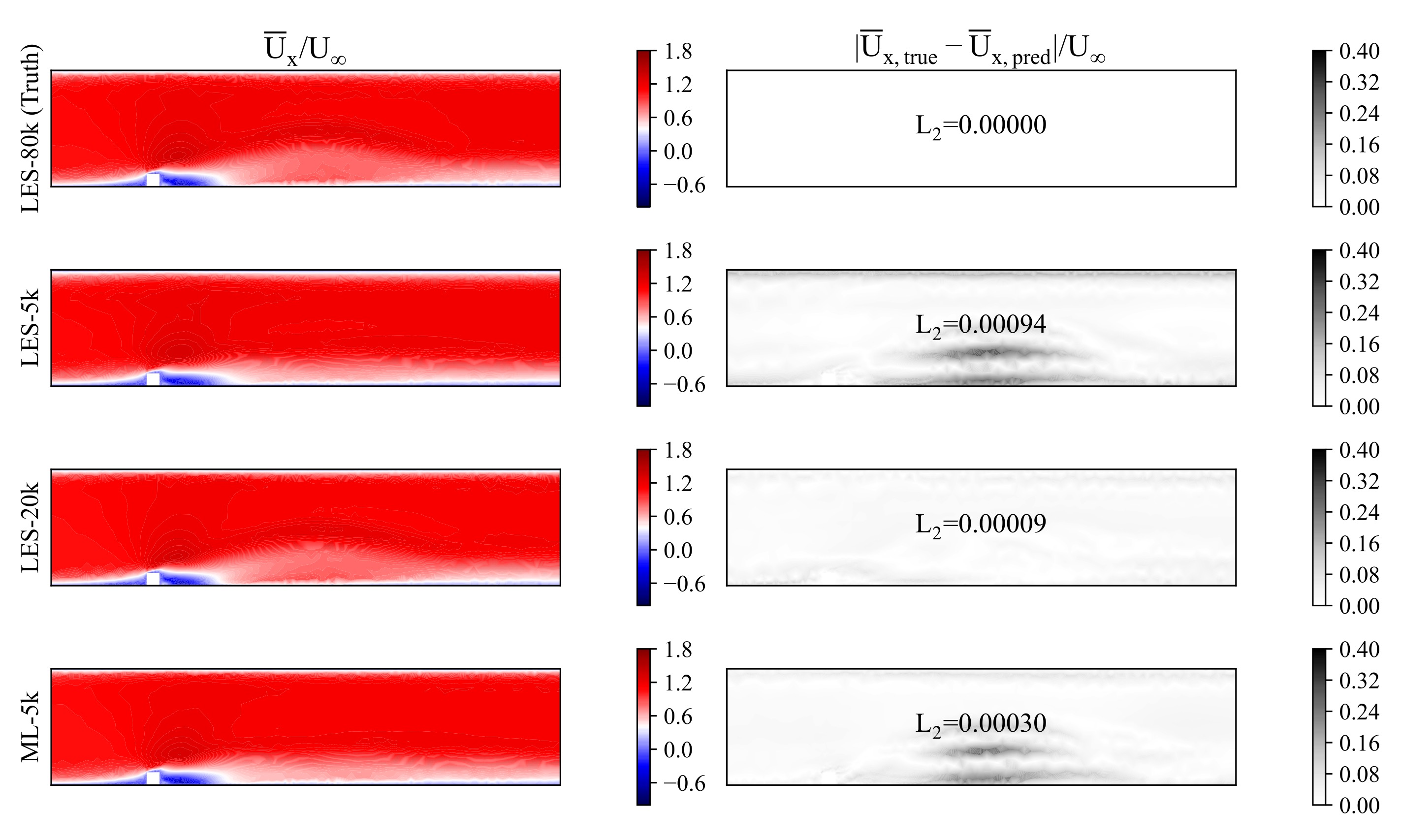}
  \end{center}
  \caption{Plots of the mean flow from the two-dimensional WMC case for various models. \textbf{(Left column)} The normalized mean x-velocity and \textbf{(Right column)} The error with respect to the truth in space, as well as the integrated total}
  \label{fig:Flowfield_WMC}
\end{figure*}

\begin{figure}
  \begin{center}
      \includegraphics[width=0.5\linewidth]{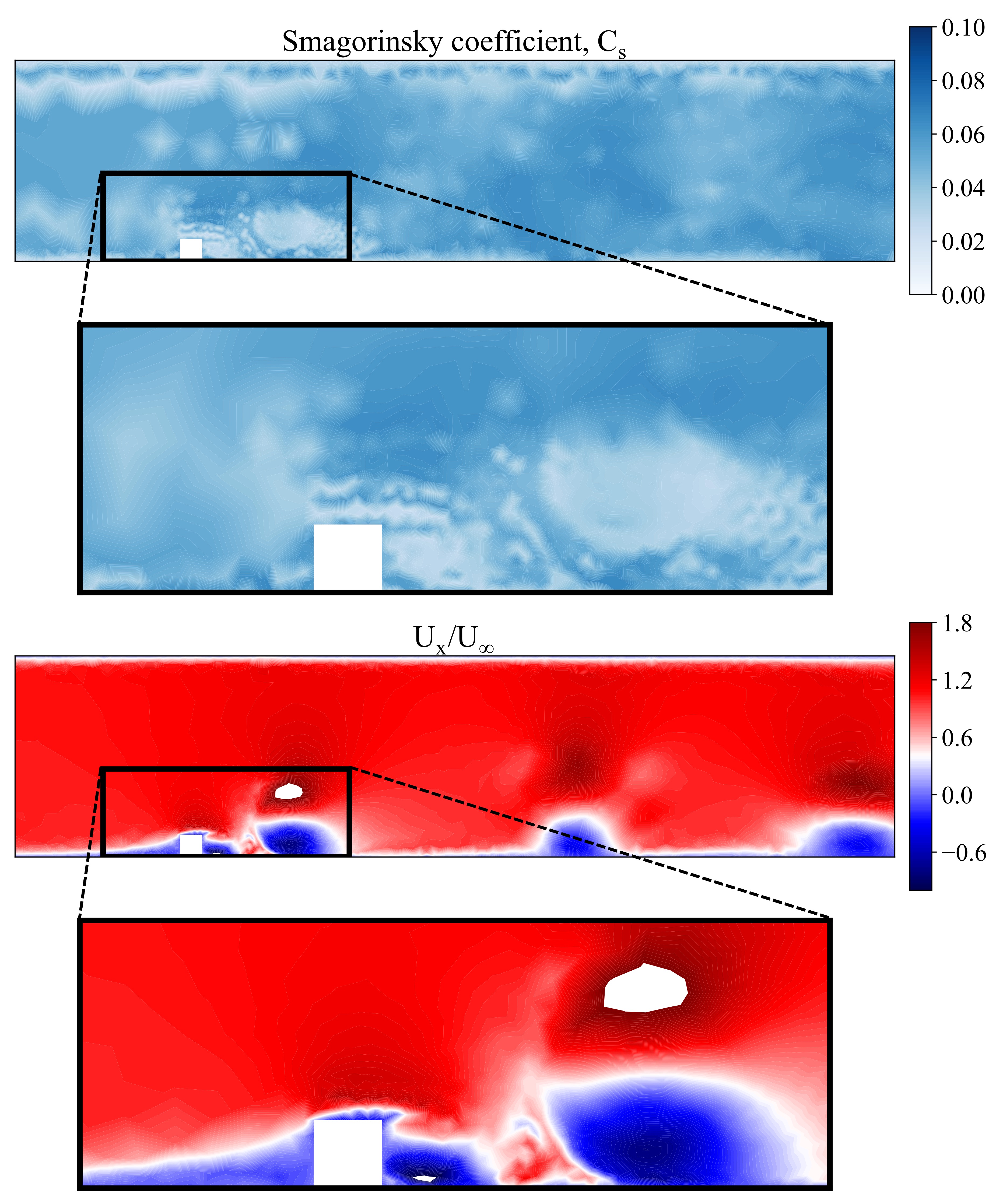}
  \end{center}
  \caption{\textbf{(Top row)} Smagorinsky coefficient and \textbf{(Bottom row)} Normalized $x$-direction velocity predicted by trained model across domain near the two-dimensional cube and their zoomed-in views at t=0.02}
  \label{fig:Smagorinsky_WMC}
\end{figure}

Figure \ref{fig:Flowfield_WMC} illustrates the normalized mean $x$-direction velocity field and the corresponding error distribution around the cube, along with the integrated error values in comparison with the LES-80k model. Across all models, the recirculation zones upstream and downstream of the cube are effectively captured. However, the LES-5k model exhibits the largest errors near the vortex trajectories and wall boundaries. In contrast, the LES-20k model shows significantly smaller errors in these regions. The ML-5k model achieves a lower integrated error compared to the LES-5k model by reducing errors near these areas. Figure \ref{fig:Smagorinsky_WMC} illustrates adaptive adjustment of the Smagorinsky coefficient and the normalized $x$-direction velocity across domain. This figures shows that the ML-5k model decreases $C_s$ value near the vortex-forming region and the wall boundary layer.

These results suggest that a GNN-based SGS closure model trained in an \textit{a-posteriori} manner can generalize to unseen geometries. Even on a geometry with significantly different separation physics, the model remains stable, recovers the baseline low-fidelity model, and improves predictability in the mean sense. This generalization capability can be attributed to the fact that the spatial mesh is perfectly represented by the graph connectivity in GNN, which inherently learns the relationships between nodes and edges through their structure. The \textit{a-posteriori} learning process enhances the stability and accuracy of the trained model by ensuring model-data consistency.

\subsubsection{Model training with constrained data availability}
\label{subsubsec:Model training with constrained data availability}

In this section, we perform experiments for training a GNN-based SGS model under a condition where only a portion of flowfield data is available. Sparse or incomplete experimental data is common in many scientific and engineering domains. This is often due to the limited number of sensors, as it is not feasible to install sensors throughout the entire field. For instance, in wind tunnel experiments, only a small number of sensors are located strategically, and weather prediction models are initialized or validated from sounding or satellite observations at specific locations. Furthermore, placing too many sensors in fluid dynamics experiments can affect the accuracy of experiments due to interference with the flow. 

\begin{figure*}
  \begin{center}
      \includegraphics[width=0.8\linewidth]{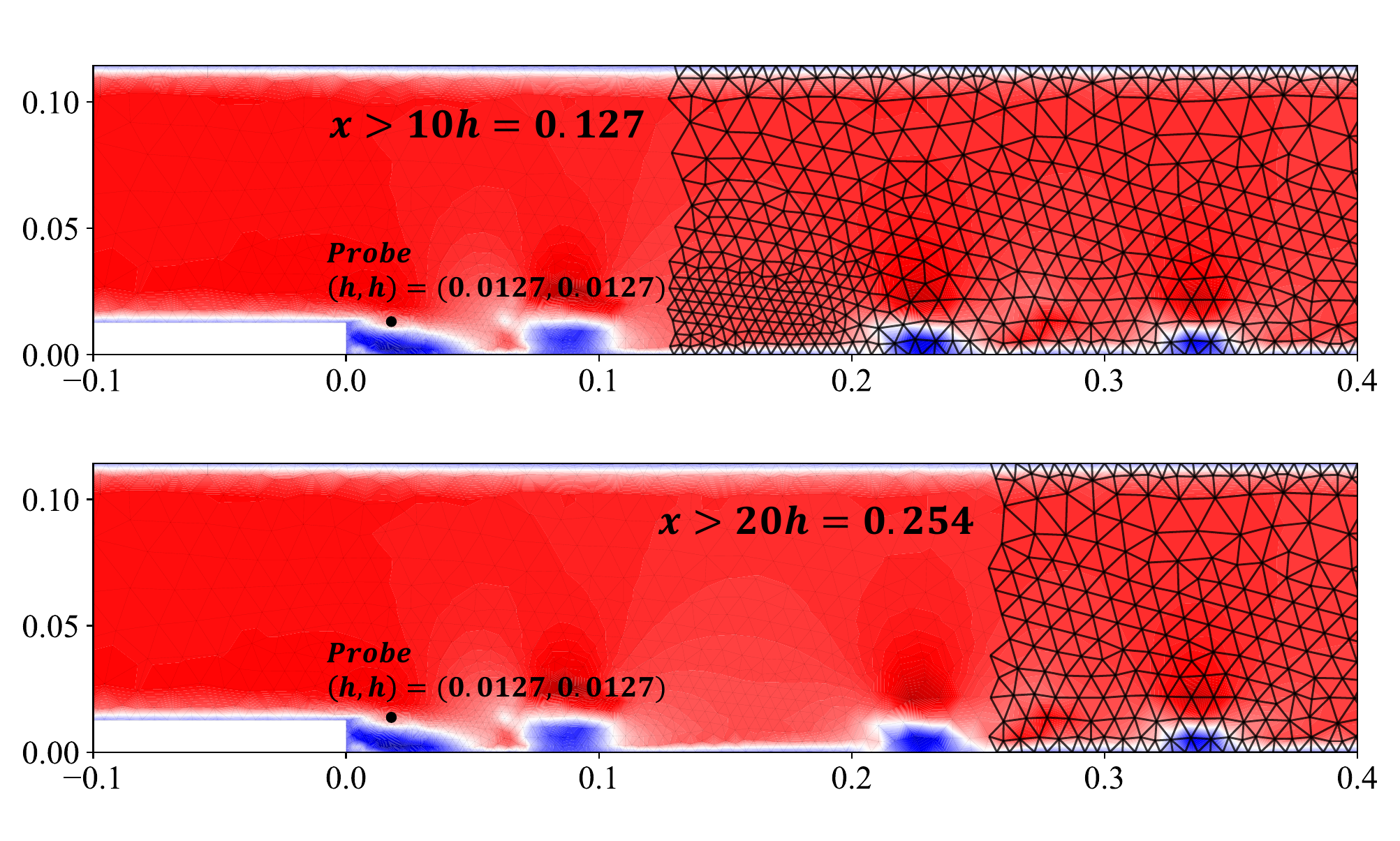}
  \end{center}
  \caption{Two examples of constrained data availability across the domain, where the mesh visualization indicates data being sampled for error computation; \textbf{(Top row)} when $x>10h=0.127m$ \textbf{(Bottom row)} when $x>20h=0.254m$}
  \label{fig:Sparse_domain}
\end{figure*}

We utilize the two-dimensional BFS case to train new models under several scenarios where only the flowfield data downstream of the step is available. Each scenario is defined by a different threshold that specifies the accessible portion of the domain for training. Figure \ref{fig:Sparse_domain} illustrates two example scenarios where the models are trained using only the flowfield data at $x>10h=0.127m$ and $x>20h=0.254m$, respectively. Each finite element cell is included in the training dataset when x-coordinates of all nodes in the element are included in the threshold range. In total, our scenarios include four different thresholds: $x>0, 5h, 10h, 20h$. We train four models under each of these scenarios and test them on different geometries used in the previous sections. For the training dataset, the initial 256 snapshots (i.e., until $t=0.0256s$) are utilized. Each trajectory in the training dataset is composed of 32 rollout snapshots, with initial conditions separated by 16 time steps. This results in a total of 15 trajectories. Models are trained under the same training setting as in Sec. \ref{subsubsec:Dataset generation for 2D flows}. Training history is shown in the Fig.~\ref{fig:Training_history} in Appendix. \ref{app:Training history}.

\begin{figure}
  \begin{center}
      \includegraphics[width=0.7\linewidth]{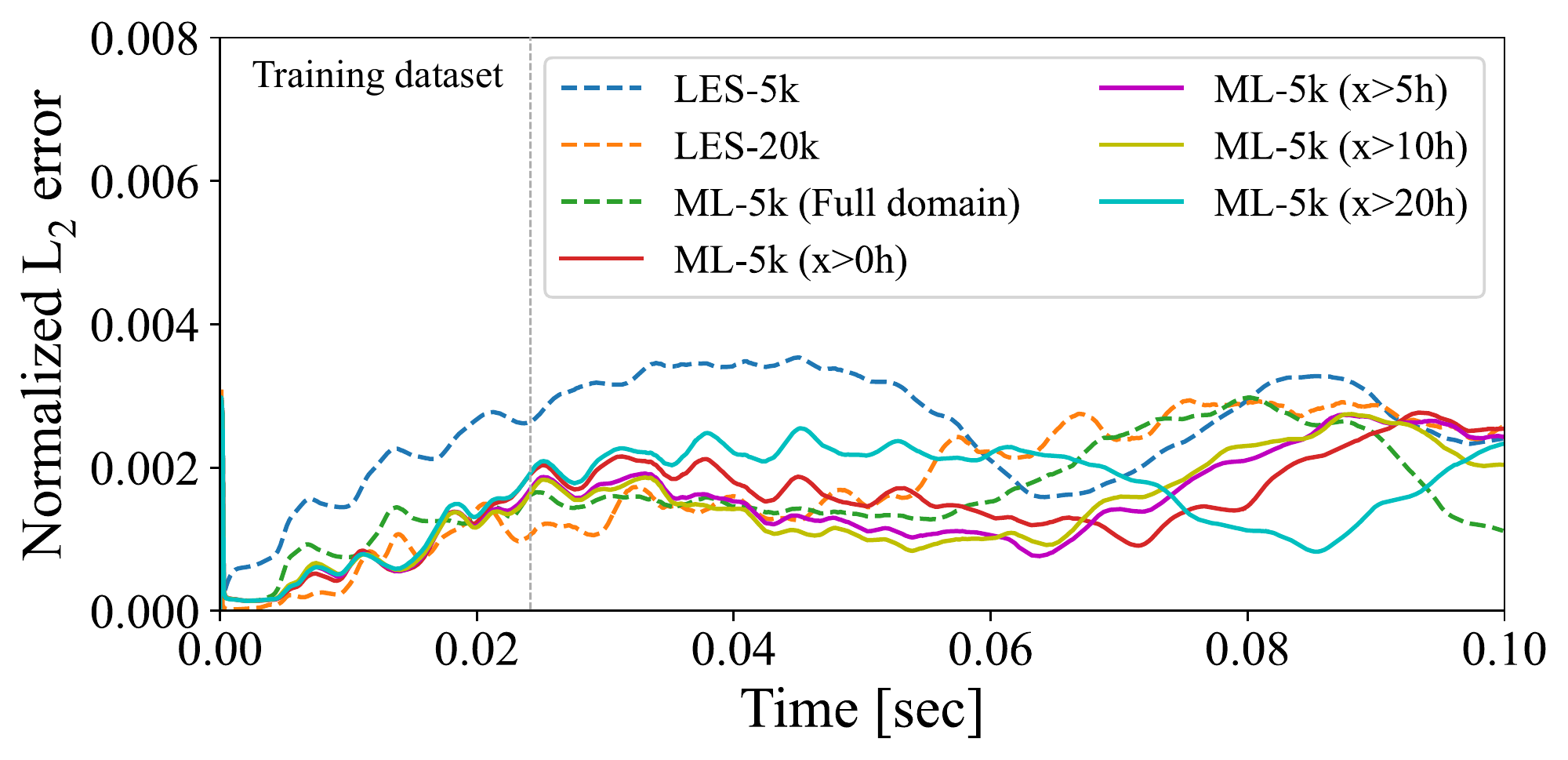}
  \end{center}
  \caption{Normalized $L_2$ error on the two-dimensional BFS case under different data availability conditions. The gray dashed line indicates the extent of the training data observed.}
  \label{fig:BFS_sparse_error}
\end{figure}

\begin{figure}
  \begin{center}
      \includegraphics[width=\linewidth]{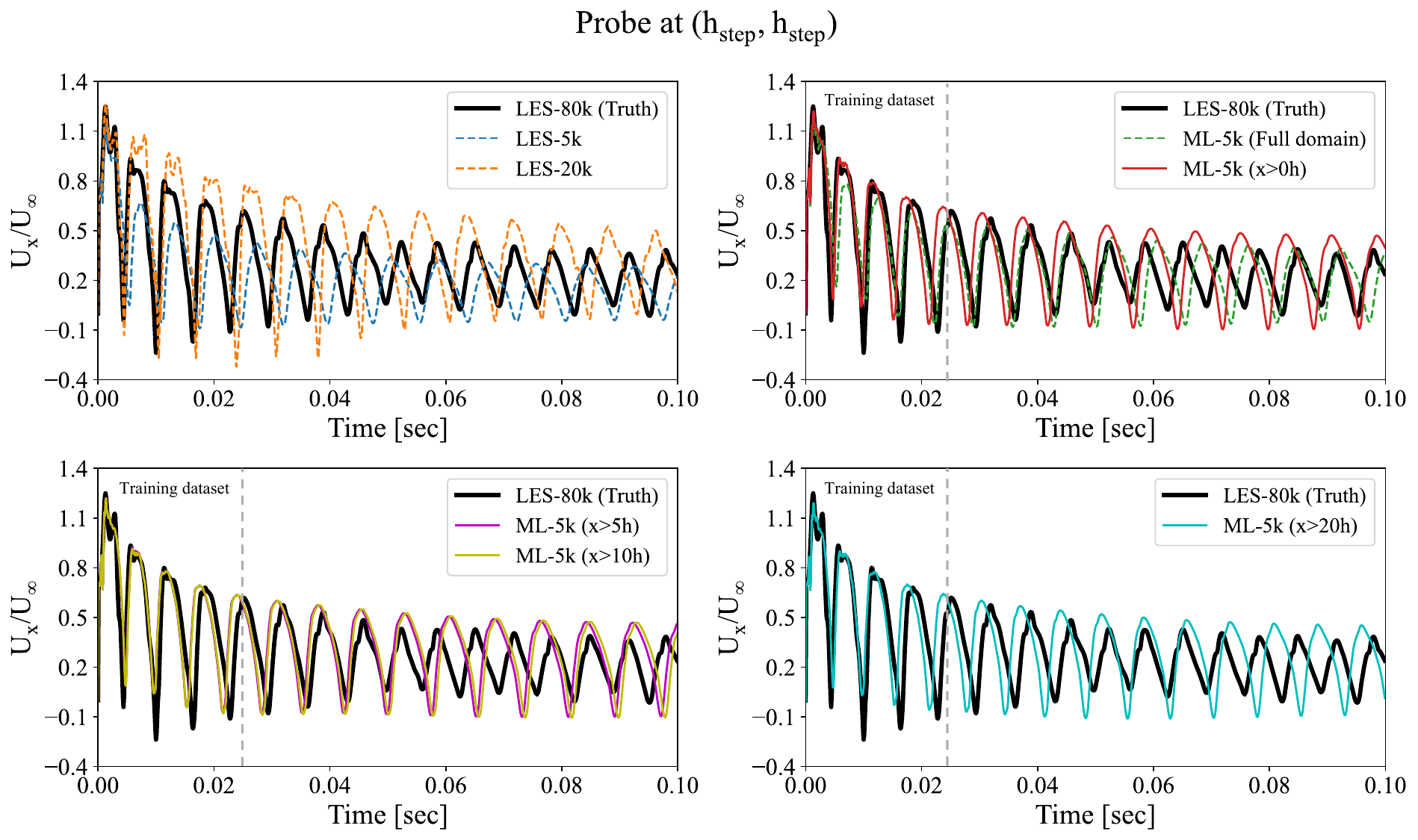}
  \end{center}
  \caption{$x$-direction velocity measurement from a probe in front of the step at $(x, y)=(h, h)$ under different data availability conditions.} 
  \label{fig:BFS_sparse_probe}
\end{figure}

Figure \ref{fig:BFS_sparse_error} shows the normalized $L^2$ error over time for the two-dimensional BFS case, comparing the performance of models trained under different data availability conditions. We observe that all models trained with incomplete domain data remain stable and achieve lower or comparable level of errors than the LES-20k model. Notably, even the model trained using flowfield data at $x>20h$, which constitutes a small portion of the entire domain, can improve upon the baseline LES-5k model. Furthermore, the ML-5k models trained with data from $x>5$ and $x>10h$ attain lower errors than the ML-5k model trained on the full domain, achieving lower errors than the LES-20k model. Figure \ref{fig:BFS_sparse_probe} compares the predictive accuracy using the normalized $x$-direction velocity data from the probe located at $(x,y)=(h,h)$, and similar trends with the $L^2$ error plot are observed. All models trained with the sparse training data are more accurate in capturing peaks of periodic velocity fluctuations than the LES-5K and LES-20K models. This indicates that far-field data can help calibrate models to perform well near the step.

\begin{figure}
  \begin{center}
      \includegraphics[width=0.7\linewidth]{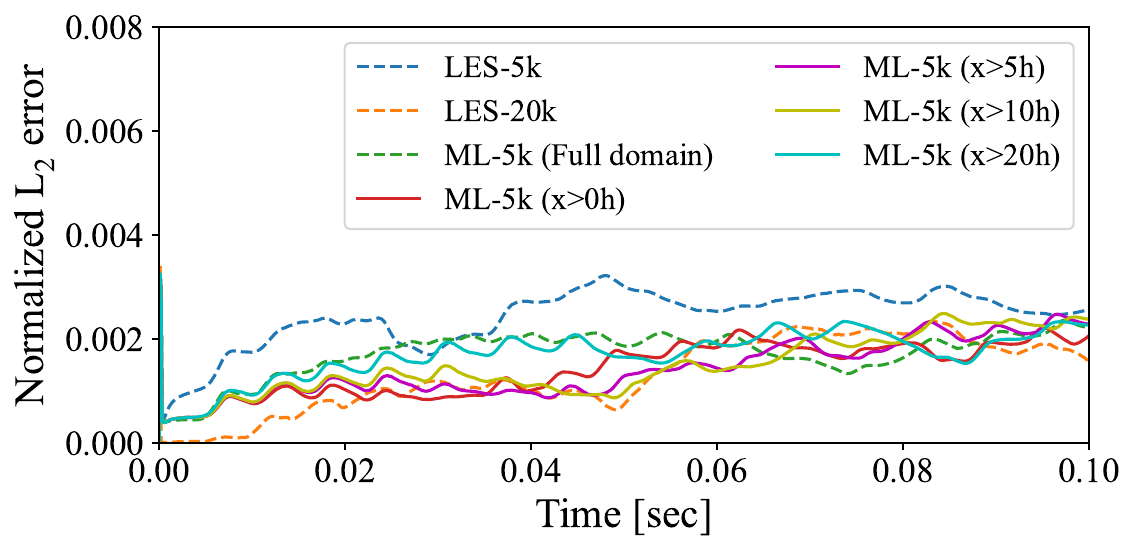}
  \end{center}
  \caption{Normalized $L_2$ error on the two-dimensional ramp case under different data availability conditions. This geometry was unseen during training.}
  \label{fig:BFR_sparse_error}
\end{figure}

\begin{figure}
  \begin{center}
      \includegraphics[width=\linewidth]{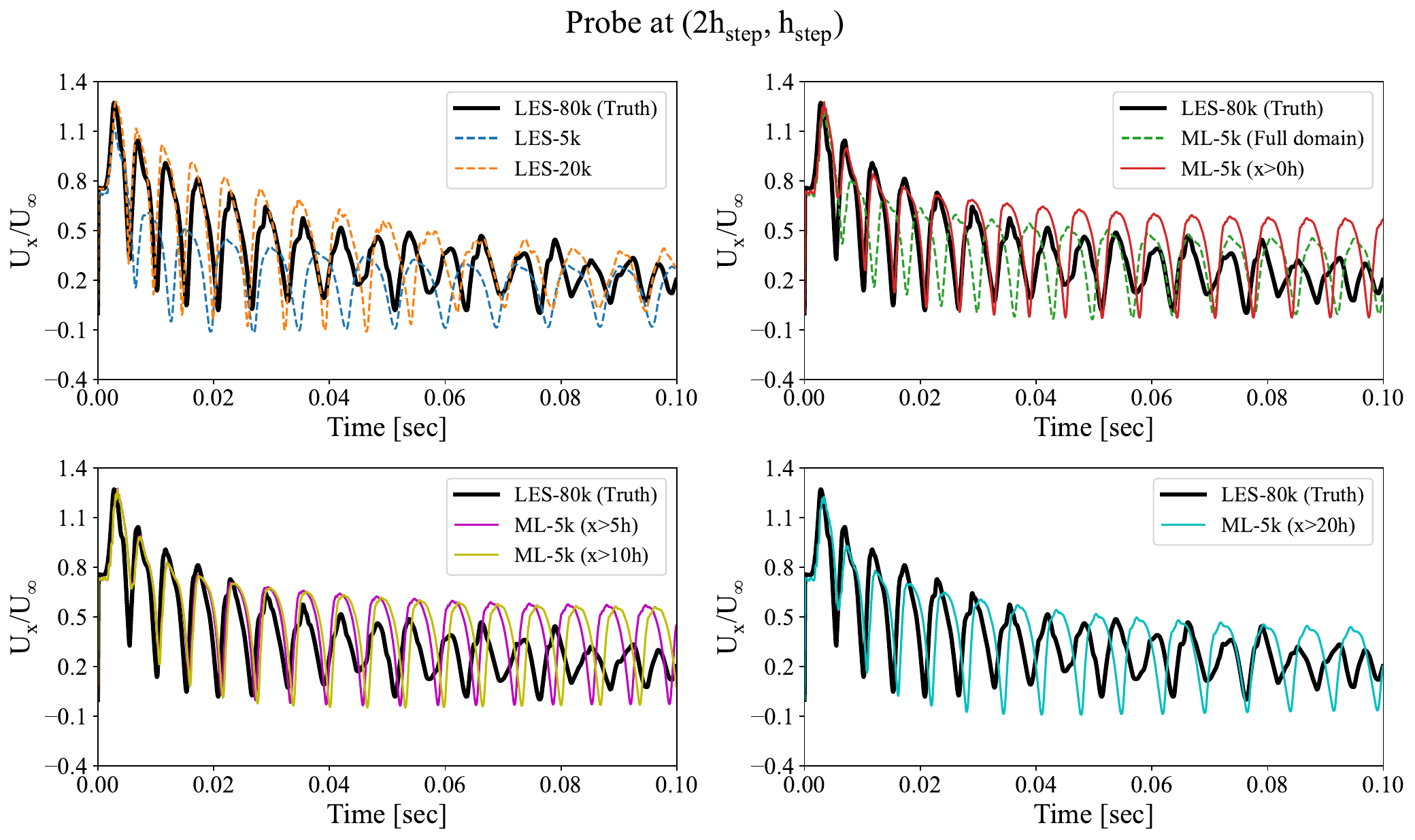}
  \end{center}
  \caption{$x$-direction velocity measurement from a probe in front of the two-dimensional ramp at $(x, y)=(2h, h)$ under different data availability conditions} 
  \label{fig:BFR_sparse_probe}
\end{figure}

We test our models, trained with incomplete data from the two-dimensional BFS case, on the ramp and WMC cases to assess generalization capability across different geometries, as discussed in Sec. \ref{subsubsec:Generalization to new geometries}. Figure \ref{fig:BFR_sparse_error} presents the normalized $L^2$ error over time for the two-dimensional ramp case, comparing the performance of models trained under varying data availability with other models. All models trained with incomplete flowfield data show lower errors than the LES-5k model and remain stable for longer time integrations than during training. The ML-5k models trained with data from $x>5h$ and $x>10h$ achieve lower errors than the ML-5K model trained on the full domain data, exhibiting a similar level or errors with the LES-20k model. Figure \ref{fig:BFR_sparse_probe} presents the pointwise accuracy of the models based on normalized $x$-direction velocity data obtained from a probe located at $(x,y)=(2h,h)$. The ML-5k models trained with partial flowfield data show improvement in capturing velocity fluctuations just downstream of the ramp until approximately the midpoint of the simulation, while the ML-5k model trained with full domain data shows better performance in capturing the decaying magnitude of the peaks.

\begin{figure}
  \begin{center}
      \includegraphics[width=0.7\linewidth]{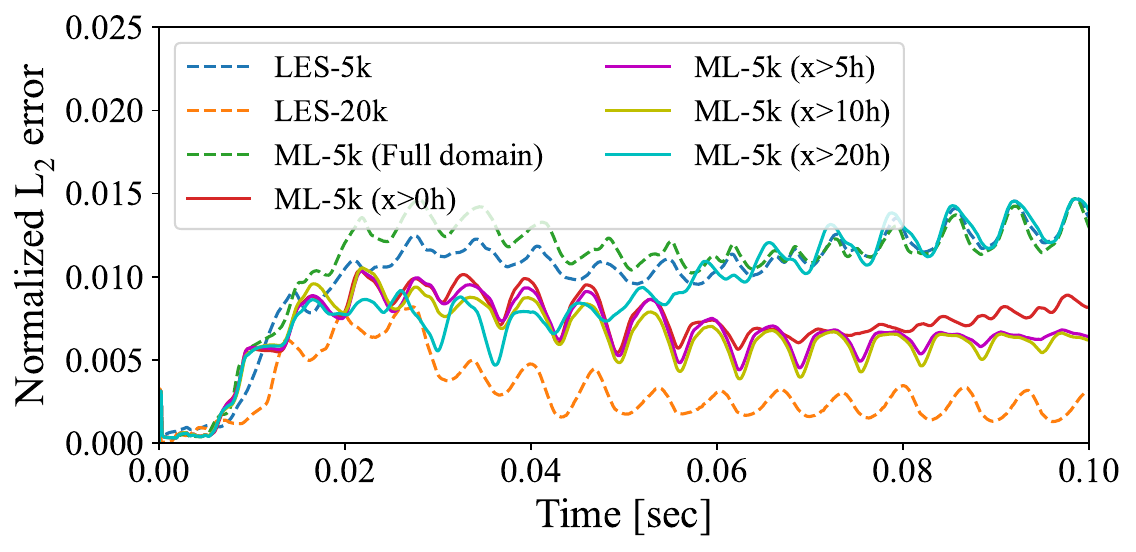}
  \end{center}
  \caption{Normalized $L_2$ error on the two-dimensional WMC case under different data availability conditions. This geometry was unseen during training.}
  \label{fig:WMC_sparse_error}
\end{figure}

\begin{figure}
  \begin{center}
      \includegraphics[width=\linewidth]{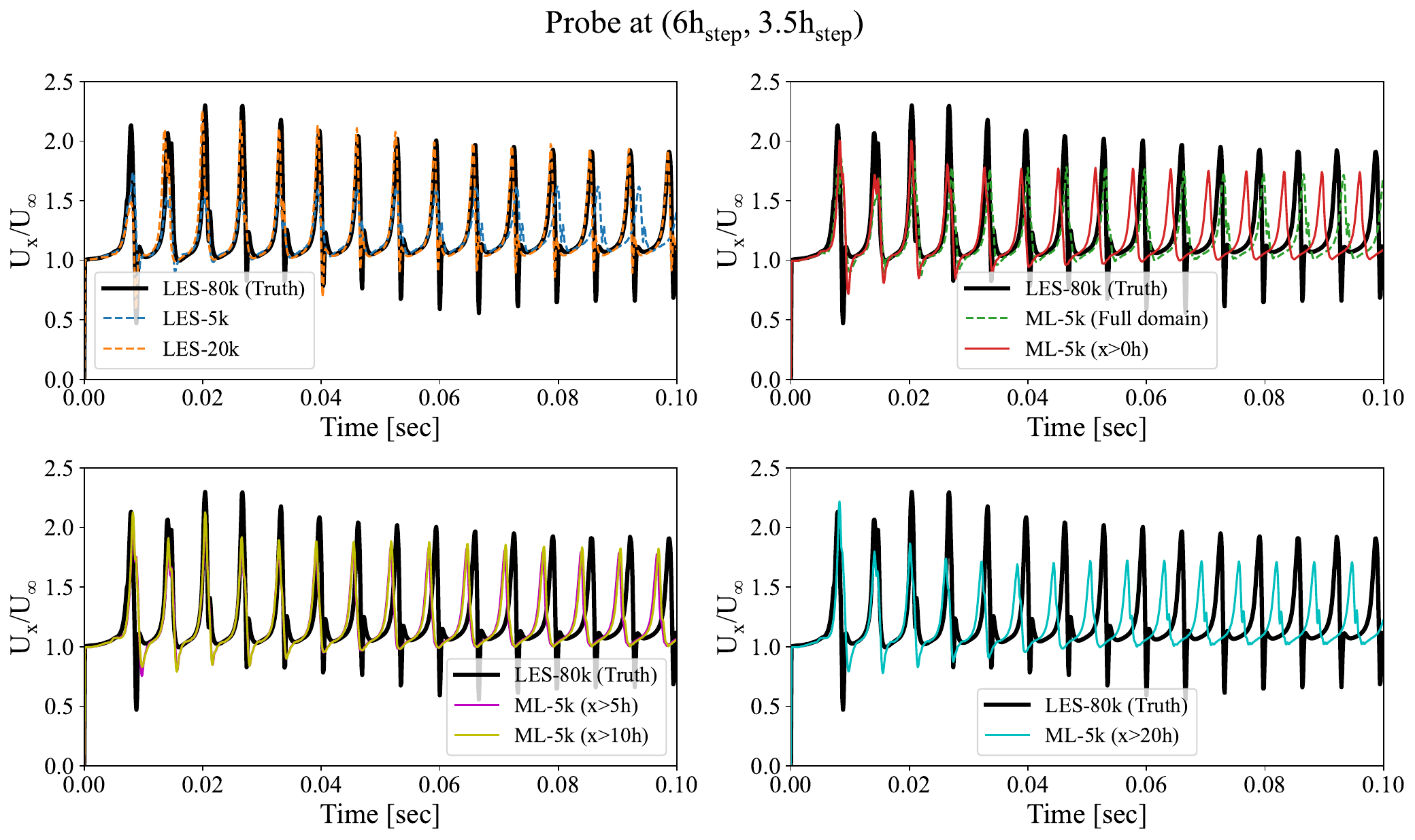}
  \end{center}
  \caption{$x$-direction velocity measurement from a probe in front of the two-dimensional cube at $(x, y)=(6h, 3.5h)$ under different data availability conditions} 
  \label{fig:WMC_sparse_probe}
\end{figure}

Figure \ref{fig:WMC_sparse_error} compares the accuracy of models trained under data availability constraints with other models using the normalized $L^2$ errors over time for the two-dimensional WMC case. All models trained with data constraints remain stable during extended time integration. Furthermore, while the ML-5k model trained with full-domain data does not show much improvement over the LES-5k model, all ML models trained with incomplete data achieve significantly lower $L^2$ errors. Specifically, the ML-5k models trained with data from $x>5h$ and $x>10h$ maintain errors at approximately half the magnitude of those from the full-domain-trained model throughout the simulation. Figure \ref{fig:WMC_sparse_probe} compares the pointwise prediction accuracy of each model using the normalized $x$-direction velocity data from a probe located at $(x,y)=(6.5h,3h)$. We observe that the ML-5k models trained with data from $x>5h$ and $x>10h$ demonstrate enhanced accuracy in capturing the peaks of velocity fluctuations caused by separation at the cube and the resulting vortices.

These results suggest that a subgrid closure model using GNN can learn general flow physics and remain stable even when trained only with flowfield data near the main vortex trajectory and wall boundaries. This is in accordance with the flow physics that is affected by closure design. Therefore, our formulation allows for the utilization of localized quantities of interest for closure calibration.

The model trained using only the downstream portion of the flow also shows improved generalization to unseen geometries. This trend suggests that limiting the training data to regions governed by universal turbulent dynamics, rather than geometry-specific boundary layer features, helps prevent overfitting to a particular configuration. Such behavior is observed within the present GNN-based PDE-constrained framework, although further investigation would be required to determine the extent to which the architecture itself contributes to this generalization.


A comprehensive examination of the metrics presented for the cases mentioned in the previous sections allows us to conclude that the ML-5k model, using our learned GNN-based closure, is capable of outperforming the more expensive LES-20k model in terms of accuracy, or improving the accuracy of simulations on unseen geometries with a coarse mesh. We compute an effective speedup by measuring the evaluation time of models, which is tabulated in Tab. \ref{tab:time}. Time costs of the ML-5k model for each case are calculated by averaging simulation times of the ML-5k models trained under different constrained data. For all geometries, the ML-5k model is able to achieve a reduction in cost of approximately a factor of 7 compared to the LES-20k model. Especially, for the ramp case, the ML-5k models trained with data downstream of the step can achieve a similar level of accuracy with the LES-20k model while significantly reducing the time cost. This suggests that in the engineering design process where detailed design parameter optimization follows broader design exploration tasks, a GNN-based closure model can accelerate forward model evaluations while preserving high accuracy. Furthermore, the generalizable results indicate greater reliability during extrapolation.

\def\arraystretch{1.3}
\begin{table}[ht!]
\centering
\caption{Core/GPU-s per $10^{-4}$ s of physical time in simulation}
\begin{tabular}{l|l|l|l}
\textbf{Model}    & \textbf{BFS} & \textbf{Ramp} & \textbf{WMC} \\ \hline
LES-80k & 48.57s & 50.74s & 65.16s \\
LES-20k & 3.16s & 2.69s & 4.17s \\
LES-5k & 0.32s & 0.34s & 0.52s \\
ML-5k & 0.4s & 0.43s & 0.63s\\
\end{tabular}
\label{tab:time}
\end{table}

\subsection{3D turbulence results: main contribution}
\label{subsec:3D turbulence results: main contribution}
The purpose of this section is to demonstrate the main contribution of this study, which involves training and evaluating the proposed framework on three-dimensional turbulent flows.

\subsubsection{Dataset generation for 3D turbulence}
\label{subsubsec:Dataset generation for 3D turbulence}
\begin{figure}
  \begin{center}
      \includegraphics[width=1.0\linewidth]{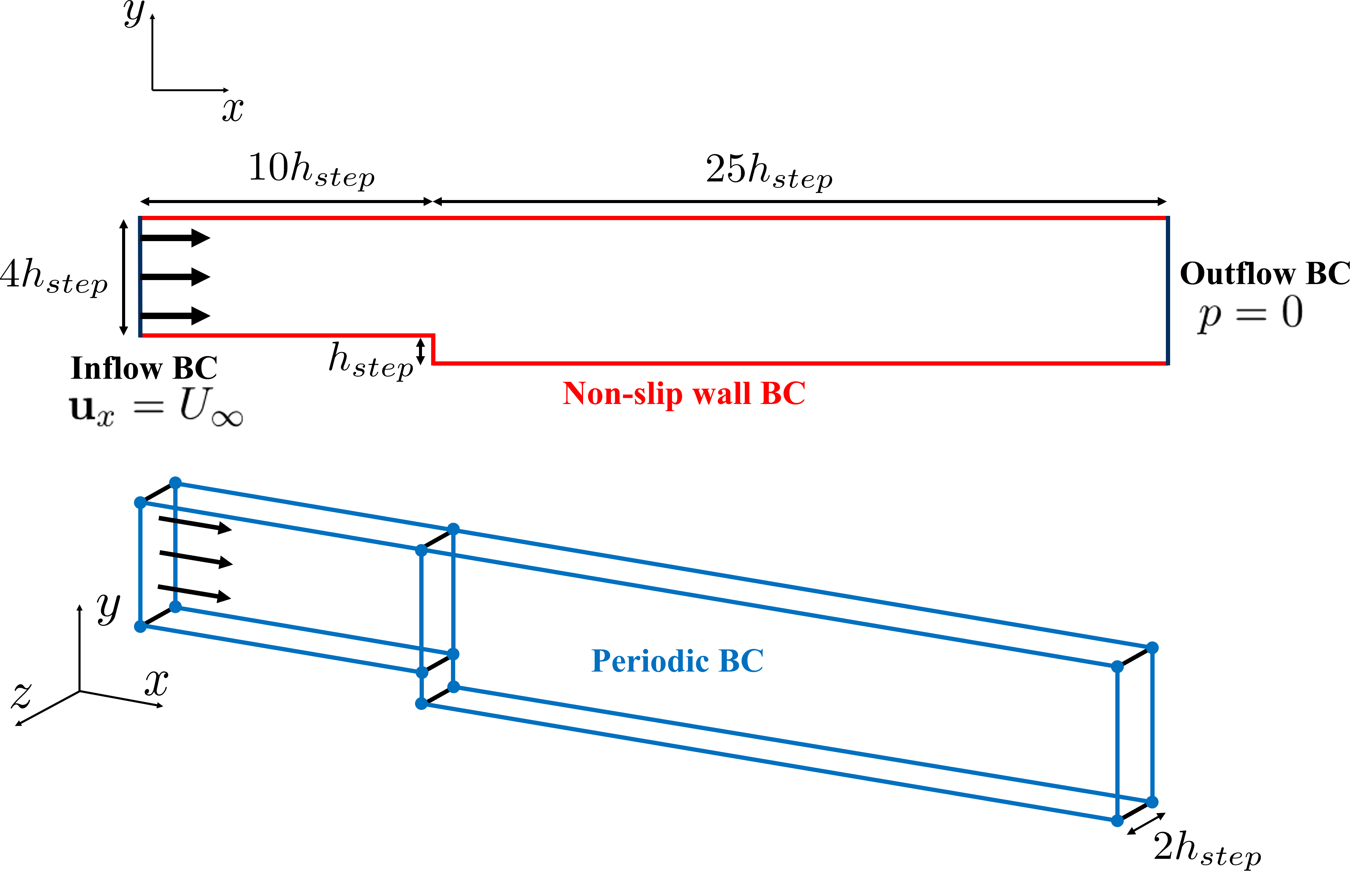}
  \end{center}
  \caption{Schematic of the three-dimensional backward-facing-step (BFS) domain showing the boundary conditions and labeled dimensions.}
  \label{fig:3d_dom}
\end{figure}

For the training dataset of three-dimensional turbulent flows, a three-dimensional BFS configuration is employed, following the setup used in previous study \cite{barwey2025mesh}. As shown in Fig.~\ref{fig:3d_dom}, all dimensions are normalized by the step height, $h_{step}$. The total channel length is $35h_{step}$, with a fixed streamwise inflow velocity prescribed at the inlet and a zero-pressure boundary condition imposed at the outlet. The step is located $10h_{step}$ downstream of the inlet, and the upstream channel height is $4h_{step}$. The spanwise width of the channel is $2h_{step}$, where periodic boundary conditions are applied in the spanwise direction. No-slip wall boundary conditions are enforced on the top and bottom surfaces of the channel.

Based on this configuration, a high-resolution LES under the Reynolds number, $Re=U_{\infty}h_{step}/\nu=3,600$, is performed using the finite element method (FEM) without any explicit SGS model to generate the ground-truth dataset. For the FEM solver, FeniCSx \cite{baratta2023dolfinx} is used. The fine grid for the three-dimensional BFS consists of approximately 432,000 hexahedral Taylor-hood elements with local refinement near the step and the walls. The near-wall grid resolution corresponds to $\Delta y^+\approx3.5$, computed by considering the mid-side node locations of the quadratic hexahedral elements. Although this resolution does not strictly satisfy the requirements of a DNS at $Re=3,600$, it provides sufficient fidelity to resolve the essential turbulent flow features. The timestep for the ground truth is set to $\Delta t = 5 \times 10^{-5}s$, which is sufficient to keep the CFL number less than 1. Snapshots of the velocity and pressure fields are saved every  $10^{-4}$ seconds, and a total of 1200 snapshots are generated.

To establish baseline comparisons, the Smagorinsky \cite{SMAGORINSKY1963} and wall-adapting local eddy viscosity (WALE) \cite{nicoud1999subgrid} models are applied on a coarser grid under identical flow and boundary conditions. The coarse grid comprises approximately 46,500 hexahedral Taylor-hood elements, yielding a near-wall resolution of $\Delta y^+\approx10$. This same coarse mesh is subsequently used to evaluate the machine-learning-based models, enabling a direct and consistent comparison among the ground truth, traditional SGS closures, and the proposed ML-based approaches. The timestep size for the coarse grid is set to $\Delta t = 10^{-4}s$. The details for both grids are described in Table. \ref{table:Computational Grid Details}, and visualizations of fine and coarse grids are provided in Appendix. \ref{app:Grids used for data generation}.

\begin{table}[h!]
\centering
\caption{Computational Grid Details. The streamwise spacing $\Delta x^+$ is evaluated at $x=3h_{step}$ downstream of the step, and the spanwise spacing (z-direction) is uniform throughout the domain, resulting in a constant $\Delta z^+$ for each case.}
\setlength{\tabcolsep}{8pt}
\begin{tabular}{lrrrrrrrrrr}
\toprule
\multirow{3}{*}{Case} & \multicolumn{3}{c}{Block 1} & \multicolumn{3}{c}{Block 2} & \multicolumn{3}{c}{\multirow{2}{*}{Near-wall}} & \multirow{3}{*}{Total Cells} \\
& \multicolumn{3}{c}{Upstream} & \multicolumn{3}{c}{Downstream} & & \\
\cmidrule(lr){2-4} \cmidrule(lr){5-7} \cmidrule(lr){8-10}
&  $i_{max}$ & $j_{max}$ & $k_{max}$ & $i_{max}$ & $j_{max}$ & $k_{max}$ & $\Delta x^+$ & $\Delta y^+$ & $\Delta z^+$ & \\
\midrule
Fine & 51 & 41 & 41 & 111 & 81 & 41 & 17.5 & 3.5 & 7.5 & 432,000 \\
Coarse & 21 & 21 & 21 & 56 & 36 & 21 & 25 & 10 & 15 & 46,500 \\
\bottomrule
\end{tabular}
\label{table:Computational Grid Details}
\end{table}

\subsubsection{Training procedure and loss formulation}

We optimize the ML model by minimizing the $L^2$ norm error between the predicted \textit{a-posteriori} velocity field and the ground truth velocity field. During training, a predicted solution trajectory $\hat{\mathbf{u}}$ is obtained by solving the LES equation using the GNN-based SGS closure model. The training dataset consists of 128 snapshots from the fine grid LES after $t=0.15s$, when the initial vortices have exited the domain. Multiple trajectories are constructed for training, where each of the trajectories consists of four consecutive snapshots with different initial conditions separated by three timesteps. This results in a total of 42 trajectories used for training. For each trajectory, the velocity field {$\{ \mathbf{u}_k \}_{k=0}^{n}$} is projected using the Galerkin projection method(see Sec. \ref{subsubsec:Training procedure and loss formulation}) onto the coarse mesh to obtain the target velocity field {$\{ \overline{\mathbf{u}}_k \}_{k=0}^{n}$}. The pressure field is similarly projected for the initial condition, which is then advanced for four timesteps using the GNN-based turbulence model. The overall loss for each predicted trajectory is computed as
\begin{equation}
    L = \frac{1}{T}\sum_{t=0}^T \int_\Omega ||\overline{\mathbf{u}}_t - \hat{\mathbf{u}}_t||_2^2 \; d\Omega,
\end{equation}

Since the integral in the above equation is computed in FEniCSx, an adjoint can be obtained using the same method as described Sec. \ref{subsec:Implementation of the adjoint method}. The loss is then backpropagated in an end-to-end fashion, through all PDE evaluations and timesteps, to update the parameters of the GNN. The model is trained for 5 epochs using the Adamw optimizer \cite{loshchilov2017decoupled} with a learning rate of $10^{-4}$. Training takes about 2.5 hours on 28 NVIDIA A100 GPUs and 175 AMD EPYC 7543P CPU-cores of the Polaris high-performance computer at Argonne National Laboratory. Training history is shown in the Fig.~\ref{fig:Training_history_3d} in Appendix. \ref{app:Training history}.

\subsubsection{Training results for 3D turbulence}
\label{subsub:Training results for 3D turbulence}
\begin{figure}
  \begin{center}
      \includegraphics[width=0.7\linewidth]{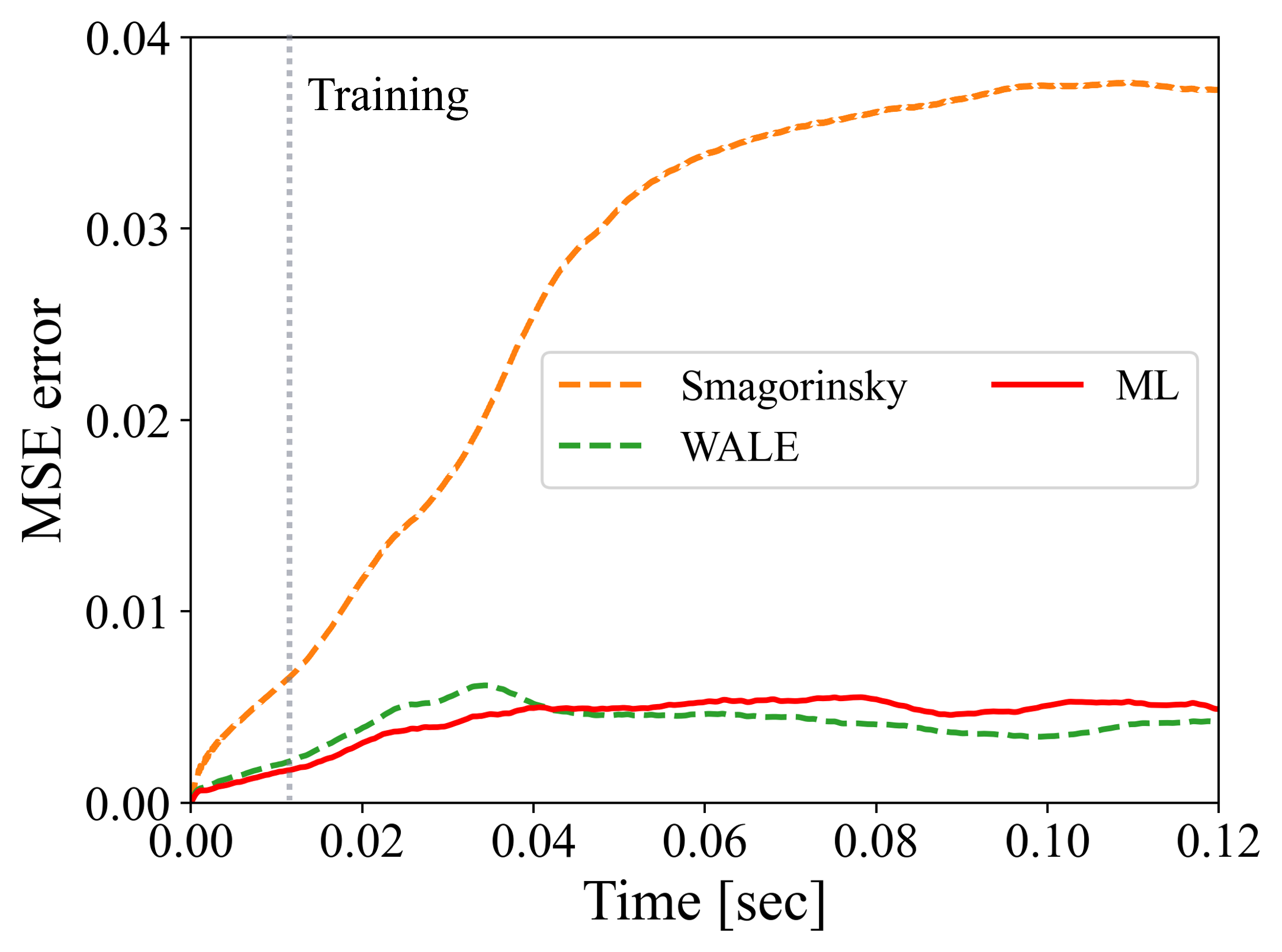}
  \end{center}
  \caption{Mean-squared error of velocity predictions for the three-dimensional BFS case obtained by different models. The gray dashed line represents the extent of training data.}
  \label{fig:BFS_error_3d}
\end{figure}

Figure~\ref{fig:BFS_error_3d} compares the mean-squared error (MSE) of velocity predictions between traditional SGS models and the proposed ML-based model. Each model is initialized with the ground truth velocity and pressure fields at $t=0.15s$ ,which is projected onto the coarse grid and subsequently advanced for 1200 timesteps. As shown in the figure, the Smagorinsky model exhibits the largest error with the steepest growth rate, continuing to increase until approximately $t=0.08s$. In contrast, the WALE model shows lower error than the Smagorinsky model. This is because, unlike the Smagorinsky model that relies solely on the local rate-of-strain tensor in its formulation, the WALE model accounts for both the strain-rate and the rotation-rate tensors, enabling it to capture turbulence structures relevant for the kinetic energy dissipation. Moreover, due to its formulation, the model yields zero eddy viscosity in the vicinity of walls and in pure shear flows, thereby preventing the overdissipation typically observed in the Smagorinsky model \cite{nicoud1999subgrid}. The trained ML model achieves a comparable level of low error to that of the WALE model. Furthermore, it remains numerically stable for timesteps well beyond those used during training, as indicated by the gray dashed line.

\begin{figure}
  \begin{center}
      \includegraphics[width=1.0\linewidth]{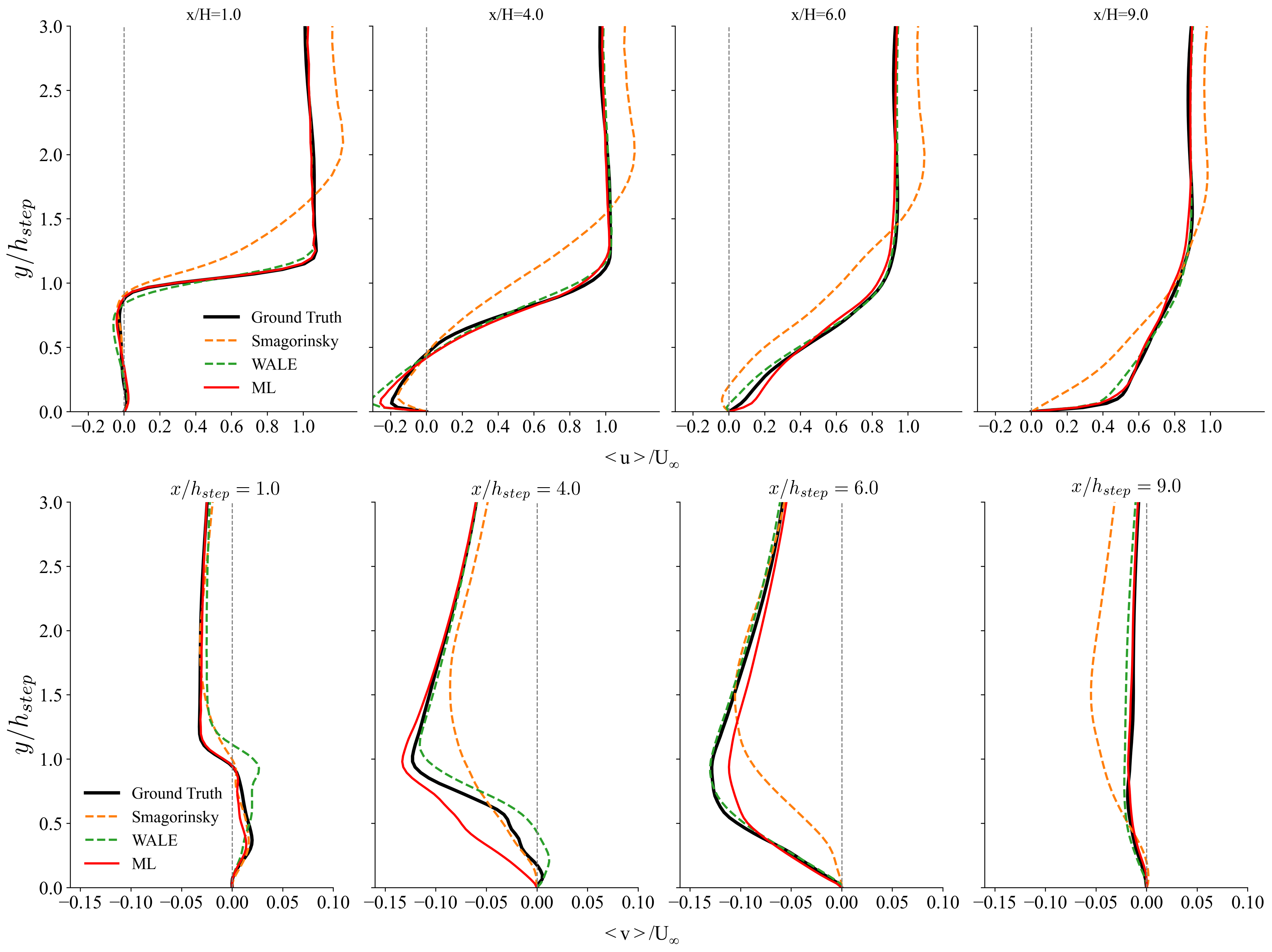}
  \end{center}
  \caption{Predictions of the mean velocity profiles at different downstream locations of the three-dimensional BFS case obtained by different models. \textbf{(Top)} Streamwise velocity, $<u> / U_{\infty}$; \textbf{(Bottom)} Wall-normal velocity, $<v>/ U_{\infty}$}
  \label{fig:1st_stat_bfs3d}
\end{figure}

While the MSE provides a basic measure of model accuracy, turbulence is inherently chaotic, making pointwise comparisons less informative over long integration times. Therefore, to more rigorously assess the model’s predictive capability for turbulent flows, we additionally compare turbulence statistics such as mean velocity profiles, Reynolds stresses, and triple product velocity correlations. These quantities provide a quantitative measure of the models' ability to reproduce both the mean flow behavior and higher-order turbulent interactions, which are essential for assessing the fidelity of LES closures. The statistical quantities are obtained by performing temporal averaging over the 1000 snapshots and spatial averaging in the spanwise direction. Figure~\ref{fig:1st_stat_bfs3d} presents the comparison of \textbf{(top)} streamwise and \textbf{(bottom)} wall-normal velocity profiles predicted by each model at several downstream locations. The ground truth profiles show the typical behavior of the backward-facing step flow: flow separation near downstream of the step, followed by gradual reattachment and recovery toward a fully developed turbulent channel profile. The Smagorinsky model exhibits substantial discrepancies in predicting both velocity components. For the streamwise velocity, it underpredicts the mean velocity in the near-wall region ($y/h<1.5)$, while overpredicting it in the outer layer ($1.5<y/h<3.0)$. At $x/h_{step} = 6.0$, although the ground truth indicates that the flow has already reattached, the Smagorinsky model still predicts negative streamwise velocity near the wall, implying a delayed reattachment and persistent separation. For the wall-normal velocity, the Smagorinsky model generally underpredicts the magnitude across most downstream locations, failing to accurately reproduce the upward motion associated with recirculating flow. In contrast, both the WALE and ML-based models show markedly improved performance in predicting the mean velocity profiles. For the streamwise mean velocity, the ML-based model exhibits better agreement with the ground truth at $x/h_{step} =1.0, 4.0$ and $9.0$ compared to the WALE model. For the wall-normal velocity, the ML-based model performs better at $x/h_{step} = 1.0$ than the WALE model, but slightly overpredicts the magnitude at $x/h_{step}=4.0$ and underpredicts at $x/h_{step}=6.0$, where the WALE model shows better agreement with the ground truth. Overall, while the WALE model performs marginally better in certain wall-normal components, the ML-based closure achieves comparable or better accuracy across most regions than traditional SGS models.

\begin{figure}
  \begin{center}
      \includegraphics[width=1.0\linewidth]{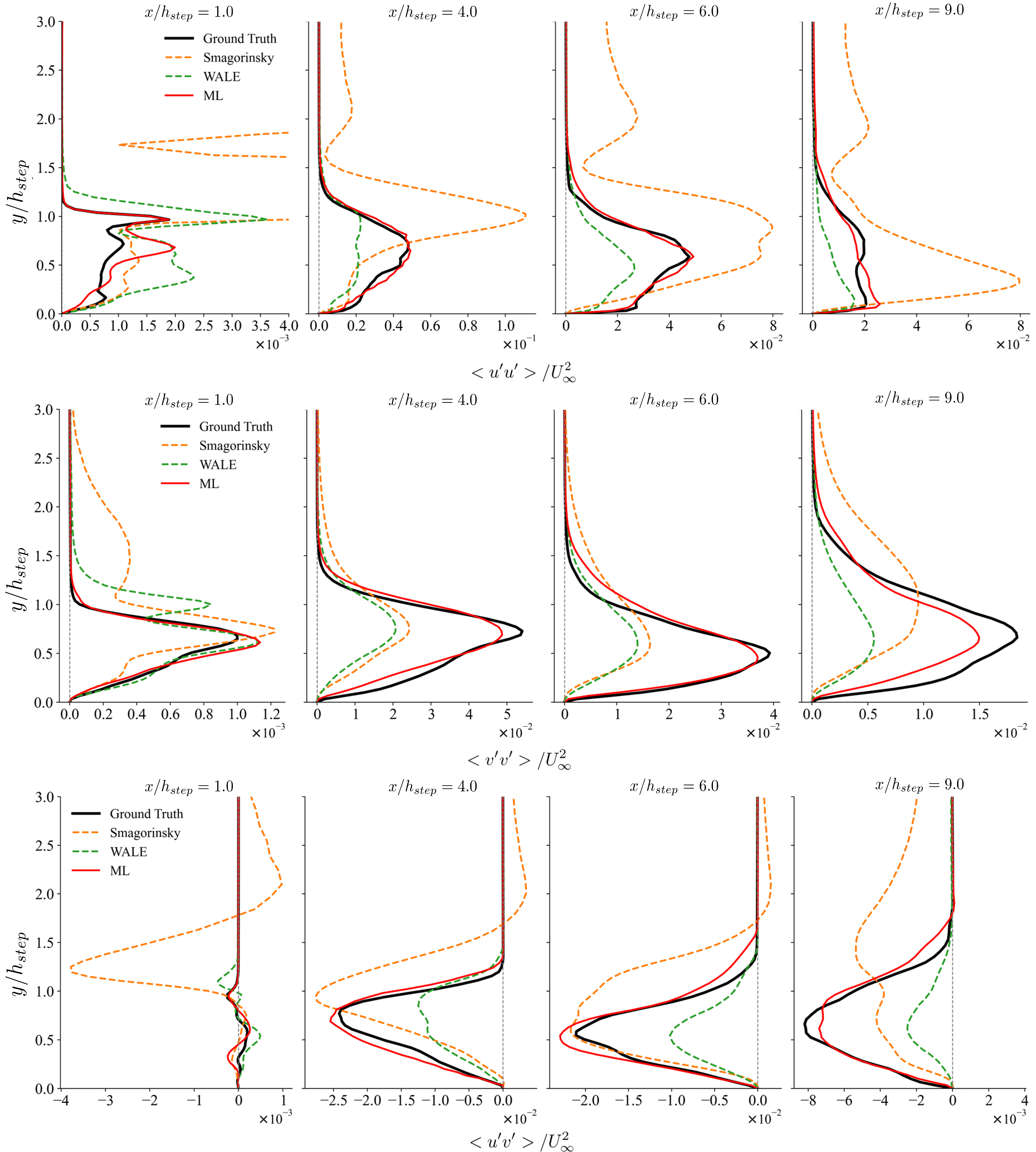}
  \end{center}
  \caption{Predictions of the Reynolds stress profiles at different downstream locations of the three-dimensional BFS case obtained by different models. \textbf{(Top)} Streamwise normal stress, $<u'u'>/ U_{\infty}^2$; \textbf{(Middle)} Vertical normal stress, $<v'v'>/ U_{\infty}^2$; \textbf{(Bottom)} Shear stress, $<u'v'>/ U_{\infty}^2$}
  \label{fig:2nd_stat_bfs3d}
\end{figure}

Fig.~\ref{fig:2nd_stat_bfs3d} compares the profiles of the Reynolds stresses at different downstream locations: \textbf{(Top)} streamwise normal stress, $\langle u'u' \rangle \ / U_{\infty}^2$; \textbf{(Middle)} vertical normal stress, $\langle v'v' \rangle / U_{\infty}^2$; \textbf{(Bottom)} shear stress, $\langle u'v' \rangle / U_{\infty}^2$. For the streamwise component, the Smagorinsky model exhibits excessively high predictions of turbulent kinetic energy, a common characteristic of under-resolved LES where subgrid dissipation is insufficient to remove small-scale energy \cite{debonis2022large}. The WALE model provides improved agreement with the ground truth but still shows noticeable discrepancies, particularly in the near-wall and reattachment regions. In contrast, the ML-based model demonstrates closer agreement with the ground truth in capturing the turbulent momentum transfer along the streamwise direction. Similarly, for the wall-normal and shear stress components, the ML-based closure consistently outperforms the traditional SGS models, showing improved accuracy across all downstream locations.

\begin{figure}
  \begin{center}
      \includegraphics[width=1.0\linewidth]{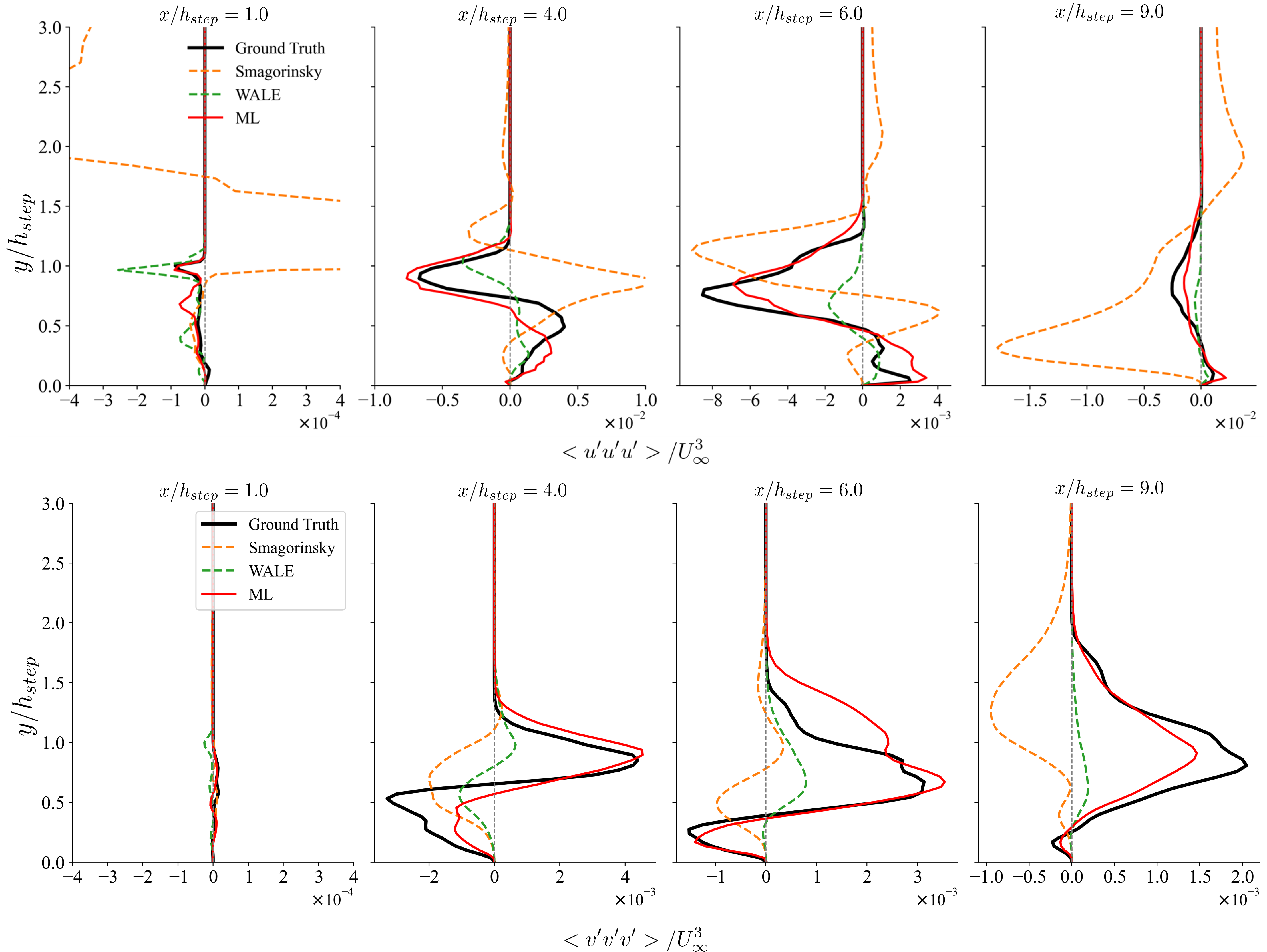}
  \end{center}
  \caption{Predictions of the triple product velocity correlations at different downstream locations of the three-dimensional BFS case obtained by different models. \textbf{(Top)} Streamwise triple correlation, $<u'u'u'>$; \textbf{(Bottom)} Wall-normal triple correlation, $<v'v'v'>$}
  \label{fig:3rd_stat_bfs3d}
\end{figure}

Figure~\ref{fig:3rd_stat_bfs3d} compares the predicted triple product velocity correlation profiles at different downstream positions:
\textbf{(Top)} $\langle u'u'u' \rangle / U_{\infty}^3$;
\textbf{(Bottom)} $\langle v'v'v' \rangle / U_{\infty}^3$. These higher-order statistics provide a more sensitive measure of the models’ ability to capture turbulent momentum transfer. While both the Smagorinsky and WALE models exhibit significant deviations from the ground truth, failing to reproduce the characteristic "S-shaped" profiles, the ML-based model achieves substantially better agreement across all components and downstream locations.

\begin{figure}
  \begin{center}
      \includegraphics[width=1.0\linewidth]{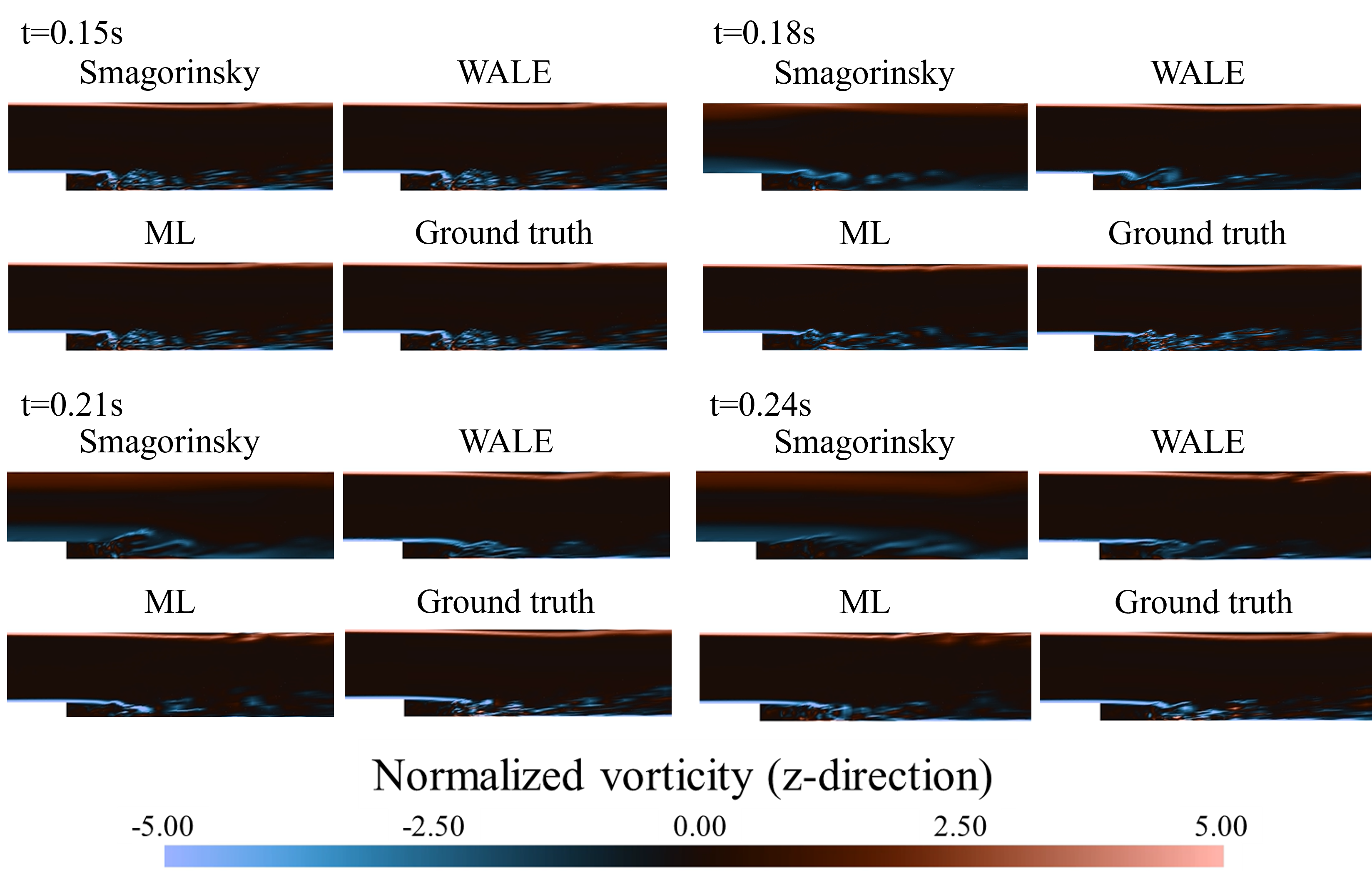}
  \end{center}
  \caption{Predictions of the normalized vorticity field (z-direction) for the three-dimensional BFS case on the $xy$-plane at $z = 0$ at different time instants obtained by different models. The ground-truth data at $t=0.15s$ is used as the common initial condition.}
  \label{fig:vorticity_bfs3d}
\end{figure}

In addition to the turbulence statistics, we evaluate each model's predictive accuracy in capturing the unsteady and complex turbulent structures. Figure~\ref{fig:vorticity_bfs3d} compares the normalized z-direction vorticity fields predicted by different models at various time instants on the $xy$-plane at $z=0$, corresponding to the spanwise centerline of the channel. The vorticity is normalized as $\omega^* = \omega/(U_{\infty}/h_{step})$. All models are initialized with the same velocity and pressure fields projected from the ground-truth data at $t=0.15s$. The ground truth results exhibit flow separation due to the step anchor, followed by reattachment further downstream, where complex multiscale turbulent structures develop. The Smagorinsky model rapidly dissipates vorticity and fails to reproduce the downstream turbulent structures, primarily due to excessive eddy-viscosity production in shear-dominated regions such as near walls and vortex cores. In contrast, the WALE model performs better by adaptively reducing the eddy viscosity near shear regions, allowing the larger-scale vortical structures to persist for a longer time. However, it still cannot capture the complex multiscale interactions present in the ground truth data. Meanwhile, the trained ML-based model successfully reproduces both the behavior of large-scale structures and their complex interactions with smaller scales, exhibiting significantly improved fidelity in representing turbulent structures. Notably, although the training data covers only from $t=0.15s$ to $t=0.1628s$, the ML model remains stable and continues to generate physically realistic turbulent flow structures for much longer timesteps.

To further assess the models’ ability to capture three-dimensional turbulent structures, we visualize the iso-surfaces of the Q-criterion colored by the vorticity magnitude at different time instants, obtained from different models. As in Fig.~\ref{fig:vorticity_bfs3d}, all models are initialized with the same velocity and pressure fields from the ground truth at $t=0.15s$. The results from the ground truth exhibit complex multiscale vortical structures that develop downstream of the step, where the flow separates and reattaches to the wall. In contrast, both the Smagorinsky and WALE models fail to reproduce such rich turbulent structures in the downstream region. The Smagorinsky model excessively damps small-scale motions due to its strong eddy-viscosity near high-shear regions, while the WALE model preserves large-scale vortices slightly better but still lacks the intricate multiscale interactions observed in the ground truth. Meanwhile, the trained ML-based model reconstructs substantially richer and more coherent vortical structures compared to the traditional models. Although the reconstructed flowfield does not entirely match the complexity of the ground truth, the ML-based model demonstrates a clear improvement in reproducing the essential features of three-dimensional turbulence.

\begin{figure}
  \begin{center}
      \includegraphics[width=1.0\linewidth]{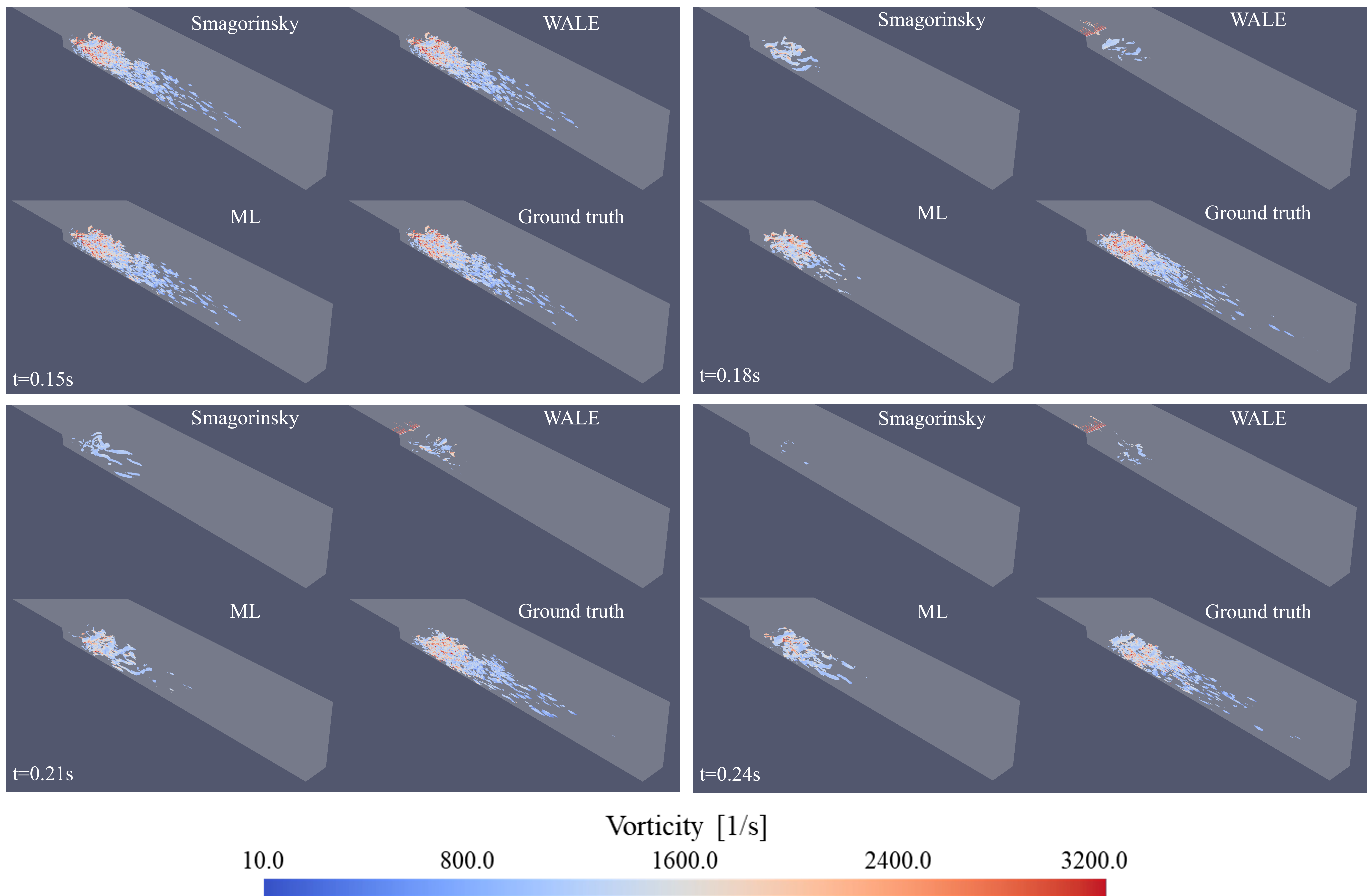}
  \end{center}
  \caption{Visualizations of the predicted iso-surfaces of the Q-criterion, colored by vorticity magnitude, for the three-dimensional BFS case at different time instants obtained by different models. The ground-truth data at $t=0.15s$ is used as the common initial condition.}
  \label{fig:iso_bfs3d}
\end{figure}

\begin{figure}
  \begin{center}
      \includegraphics[width=0.85\linewidth]{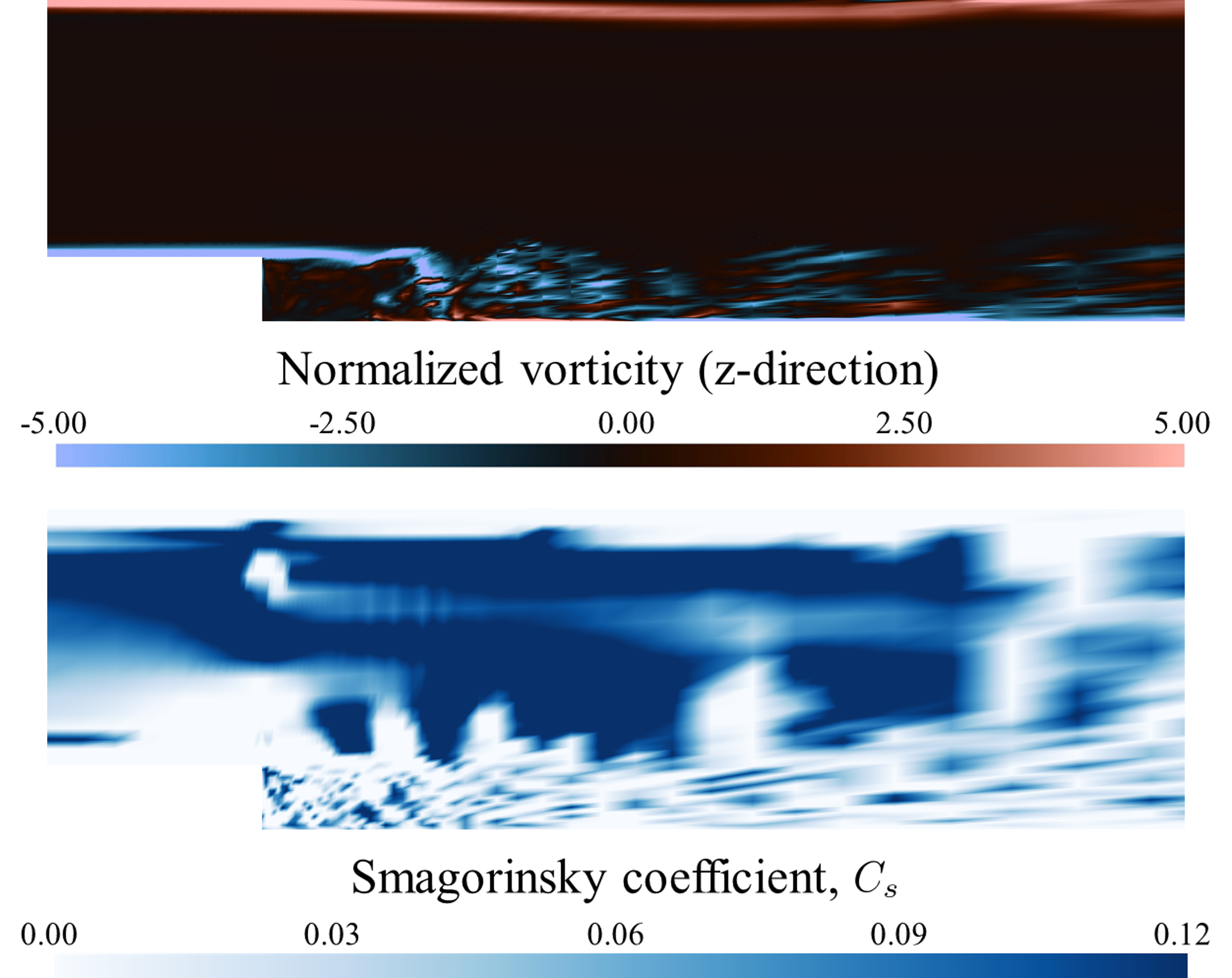}
  \end{center}
  \caption{\textbf{(Top row)} Normalized vorticity field (z-direction) and \textbf{(Bottom row)} Smagorinsky coefficient, $C_s$, predicted by the ML-based model for the three-dimensional BFS case on the $xy$-plane at $z = 0$ at $t=0.15s$}
  \label{fig:bfs_3d_Cs}
\end{figure}

Figure \ref{fig:bfs_3d_Cs} shows visualizations of the normalized z-direction vorticity field and Smagorinsky coefficient, $C_s$, predicted by the trained ML model on the $xy$-plane at $z=0$ at $t=0.15s$. In the LES simulation with the conventional Smagorinsky model, $C_s$ is predetermined and spatially uniform across the entire computational domain, remaining fixed throughout the simulation. This often causes excessive subgrid-scale dissipation near walls and within high-shear regions, leading to the damping of physically relevant turbulent motions. In contrast, the trained ML-based model adaptively adjusts $C_s$ downstream of the step, where complex three-dimensional multiscale turbulent structures develop. In particular, the model predicts reduced $C_s$ values near the step anchor and vortex cores, where strong local shear and recirculation occur. This adaptive modulation of subgrid dissipation enables the model to better preserve energetic vortical structures and capture the intricate interactions of turbulence in the downstream region as in Figs. \ref{fig:vorticity_bfs3d} and \ref{fig:iso_bfs3d}.

\subsubsection{Generalization to new geometries}
We apply the GNN-based closure model, trained on the original three-dimensional BFS case, to a configuration with an increased step height to evaluate its generalization capability. The step height is increased by 20$\%$, and fine and coarse computational meshes are generated while maintaining the same near-wall resolution as in Table~\ref{table:Computational Grid Details} of Sec. \ref{subsubsec:Dataset generation for 3D turbulence}. The visualizations of the computational meshes for the new configuration are shown in Appendix \ref{app:Grids used for data generation}. The total numbers of cells for the fine and coarse meshes are 476,000 and 50,900, respectively. Since GNNs naturally support inputs with different graph topologies-i.e., varying numbers of nodes and edges—the trained model can be directly applied to the new geometry without any retraining, despite the increase in the total number of cells in the coarse grid.

\begin{figure}
  \begin{center}
      \includegraphics[width=0.7\linewidth]{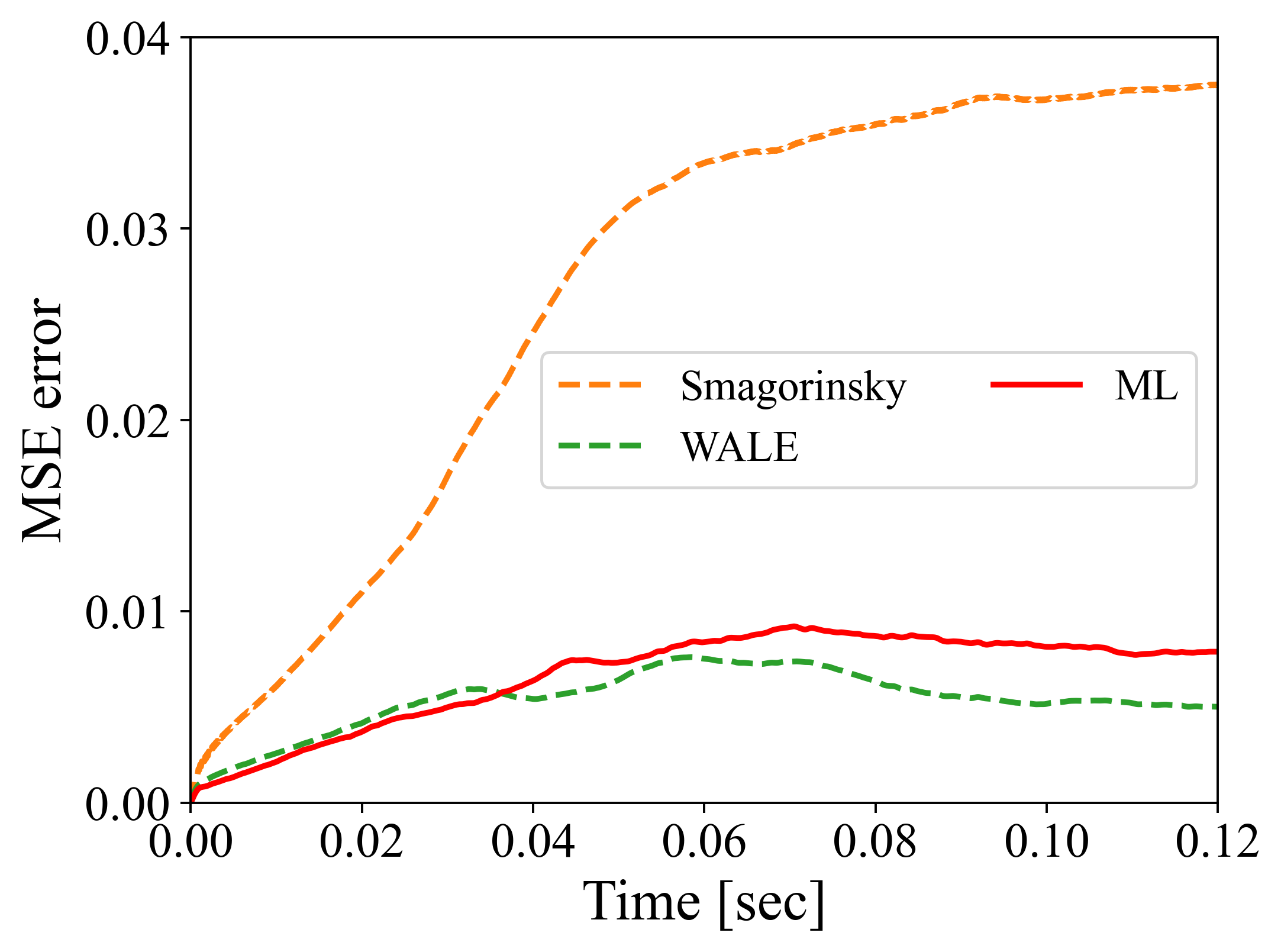}
  \end{center}
  \caption{Mean-squared error of velocity predictions for the three-dimensional BFS case with the increased step height, obtained by different models.}
  \label{fig:BFS_error_3d_1.2h}
\end{figure}

Figure~\ref{fig:BFS_error_3d_1.2h} compares MSE of the predicted velocity fields for the new configuration obtained from different models. The ground truth velocity and pressure fields at $t=0.15s$ are used as the common initial condition for all models, and each simulation is advanced for 1200 time steps. While the WALE model maintains relatively low errors, the Smagorinsky model exhibits the largest error with a steep growth rate, consistent with the trend observed in the original BFS case. The ML-based model, trained on the original BFS geometry, also maintains a low level of error and remains stable when applied to the new configuration with the increased step height, demonstrating its generalization capability.

\begin{figure}
  \begin{center}
      \includegraphics[width=1.0\linewidth]{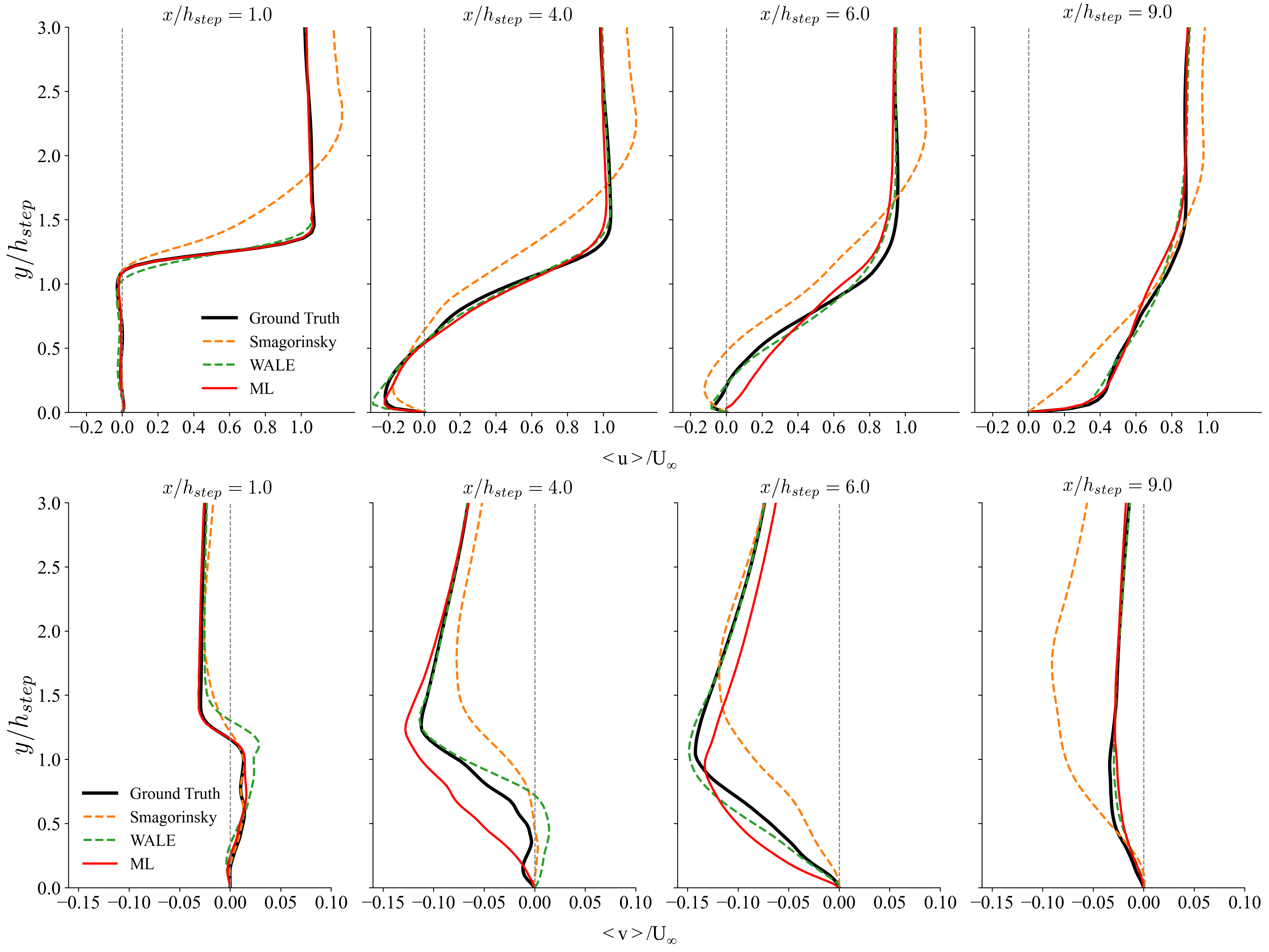}
  \end{center}
  \caption{Predictions of the first-order turbulence statistics at different downstream locations of the three-dimensional BFS case with the increased step height, obtained by different models. \textbf{(Top)} Streamwise velocity, $<u>$; \textbf{(Bottom)} Wall-normal velocity, $<v>$}
  \label{fig:1st_stat_bfs3d_1.2h}
\end{figure}

Figure~\ref{fig:1st_stat_bfs3d_1.2h} compares the predicted mean velocity profiles for the new configuration, obtained by different models.
The ML-based model, trained on the original geometry, captures both the streamwise and wall-normal velocity profiles with good accuracy across all downstream locations. At $x/h_{step}=6.0$, the ML-based model exhibits slightly earlier reattachment compared to the ground truth but overall reproduces the main flow features, including separation and recovery, without any additional training.

\begin{figure}
  \begin{center}
      \includegraphics[width=1.0\linewidth]{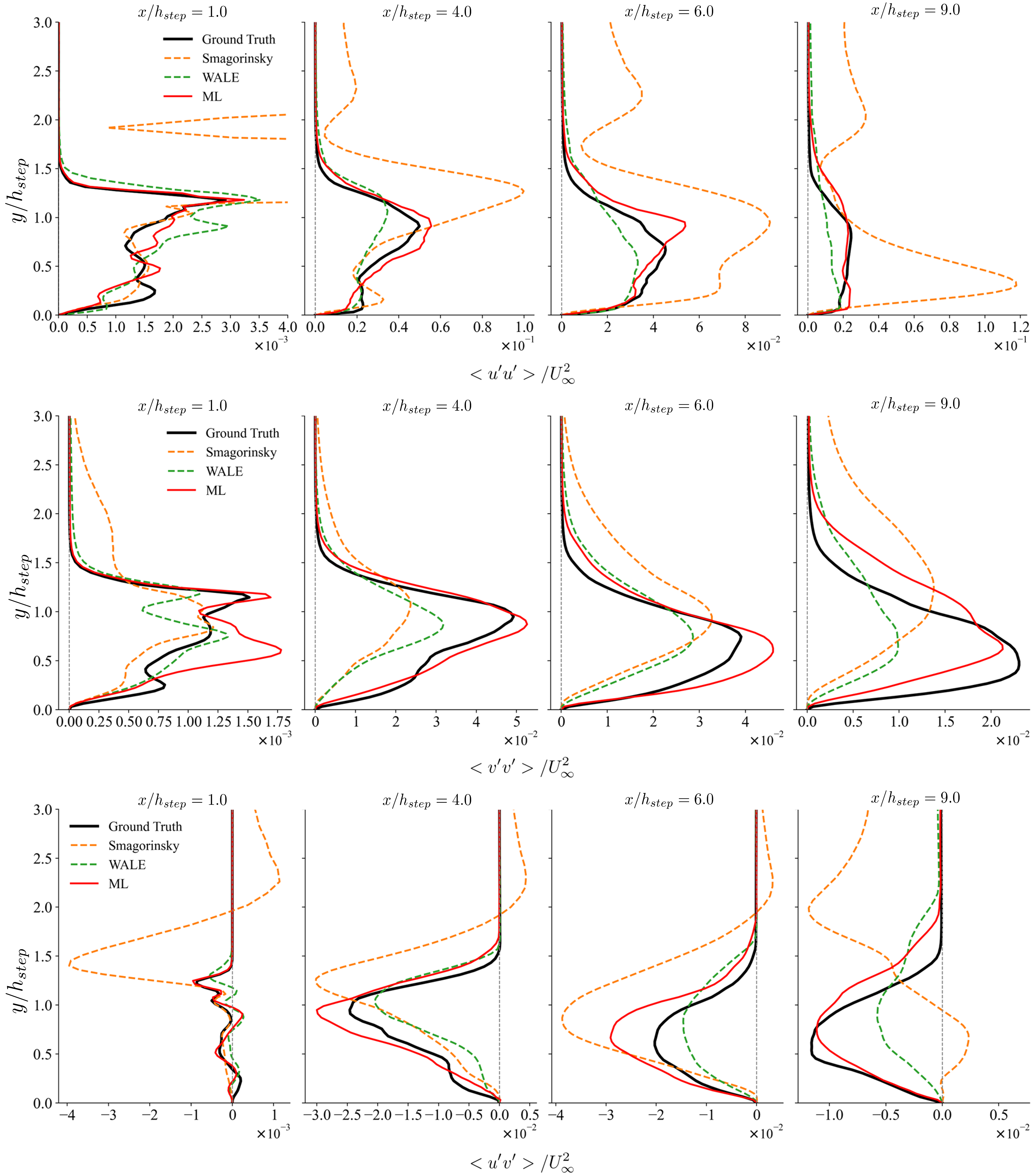}
  \end{center}
  \caption{Predictions of the Reynolds stress profiles at different downstream locations of the three-dimensional BFS case with the increased step height, obtained by different models. \textbf{(Top)} Streamwise normal stress, $<u'u'>$; \textbf{(Middle)} Wall-normal normal stress, $<v'v'>$; \textbf{(Bottom)} Shear stress, $<u'v'>$}
  \label{fig:2nd_stat_bfs3d_1.2h}
\end{figure}

The comparison of Reynolds stress profiles for the new geometry is shown in Fig.~\ref{fig:2nd_stat_bfs3d_1.2h}. The ML-based closure, trained on the original geometry, shows better agreement with the ground truth than the traditional SGS models across all three components-streamwise normal, vertical normal, and shear stresses. While the Smagorinsky and WALE models continue to show noticeable discrepancies in both shape and magnitude, the ML-based model accurately captures the overall shape and magnitude of the stress profiles without any additional training.

\begin{figure}
  \begin{center}
      \includegraphics[width=1.0\linewidth]{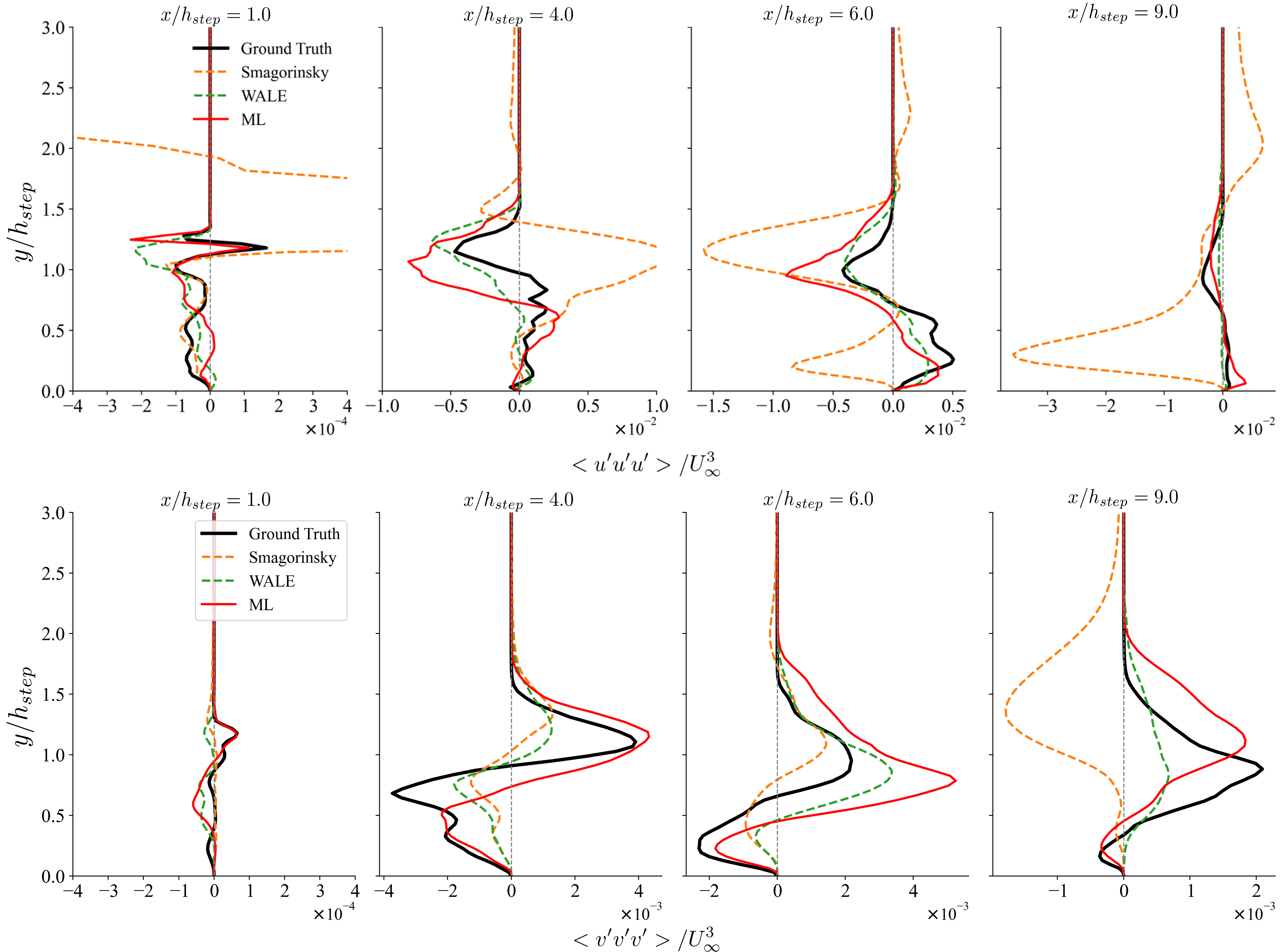}
  \end{center}
  \caption{Predictions of the third-order turbulence statistics at different downstream locations of the three-dimensional BFS case with the increased step height, obtained by different models. \textbf{(Top)} Streamwise triple correlation, $<u'u'u'>$; \textbf{(Bottom)} Wall-normal triple correlation, $<v'v'v'>$}
  \label{fig:3rd_stat_bfs3d_1.2h}
\end{figure}

Figure~\ref{fig:3rd_stat_bfs3d_1.2h} compares the triple product velocity correlation profiles for the new geometry obtained by different models. The ML-based closure, trained on the original geometry, shows better agreement with the ground truth and its characteristic “S-shaped” behavior of the triple correlations, compared to the traditional SGS models. 

\begin{figure}
  \begin{center}
      \includegraphics[width=1.0\linewidth]{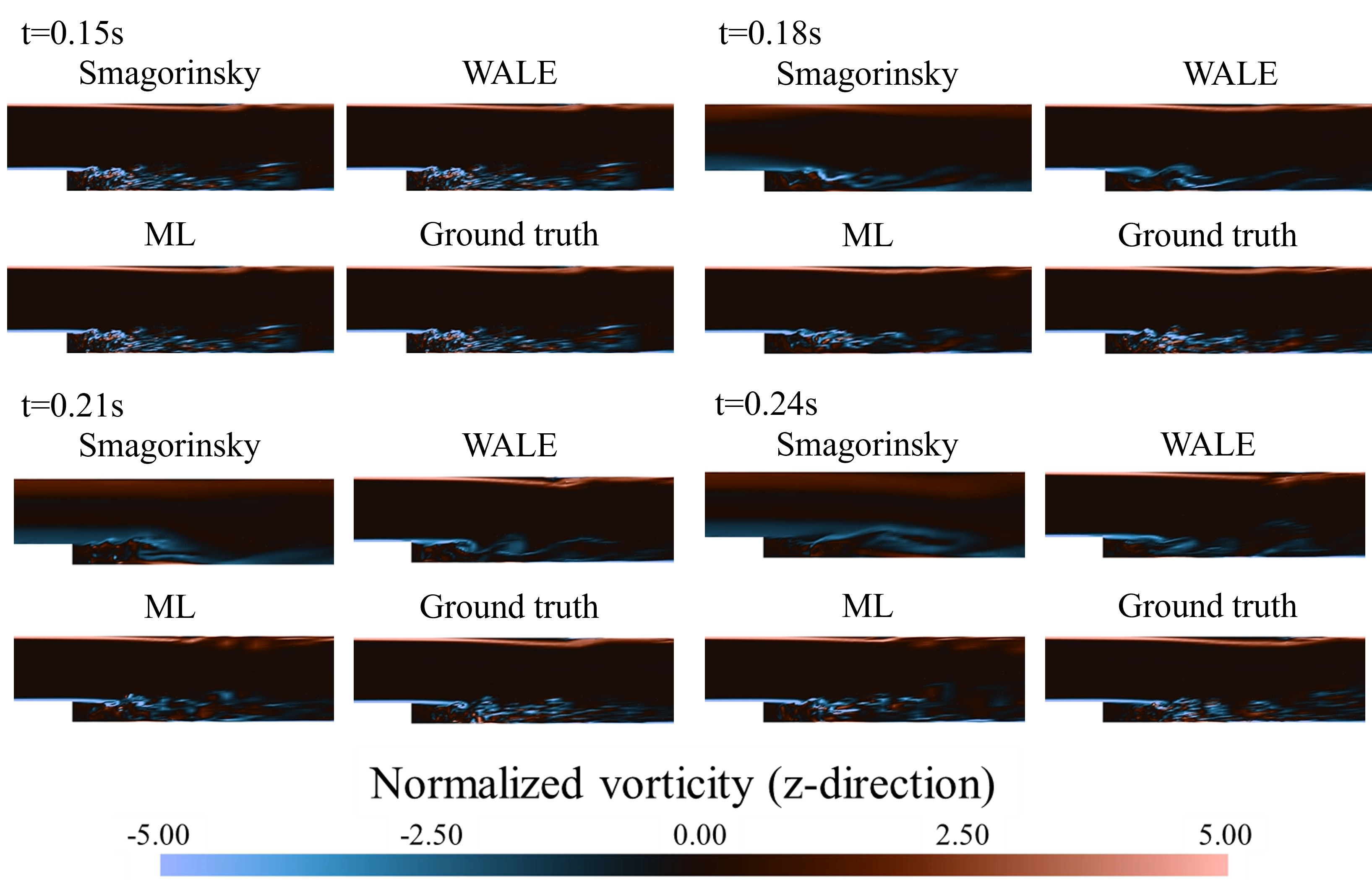}
  \end{center}
  \caption{Predictions of the normalized vorticity field (z-direction) for the three-dimensional BFS case with the increased step height on the $xy$-plane at $z = 0$ at different time instants obtained by different models. The ground truth data at $t=0.15s$ is used as the common initial condition.}
  \label{fig:vorticity_bfs3d_1.2h}
\end{figure}

Figure. \ref{fig:vorticity_bfs3d_1.2h} shows the visualizations of the normalized z-direction vorticity on the $xy$-plane at $z=0$ at different timesteps for the new configuration, obtained by different models. Similar to the original BFS geometry, the results from the ground truth show that multiscale complex turbulent structures are developed downstream of the step after flows from the inlet are separated by the step anchor. The Smagorinsky model again exhibits strong over-dissipation, leading to a rapid loss of vortical structures and failure to capture downstream turbulence. The WALE model performs slightly better by preserving some large-scale vortices but still lacks sufficient resolution of small-scale features. In contrast, the ML-based model, trained solely on the original geometry, successfully captures the multiscale turbulent structures downstream of the step and maintains physically consistent flow evolution over time.

\begin{figure}
  \begin{center}
      \includegraphics[width=1.0\linewidth]{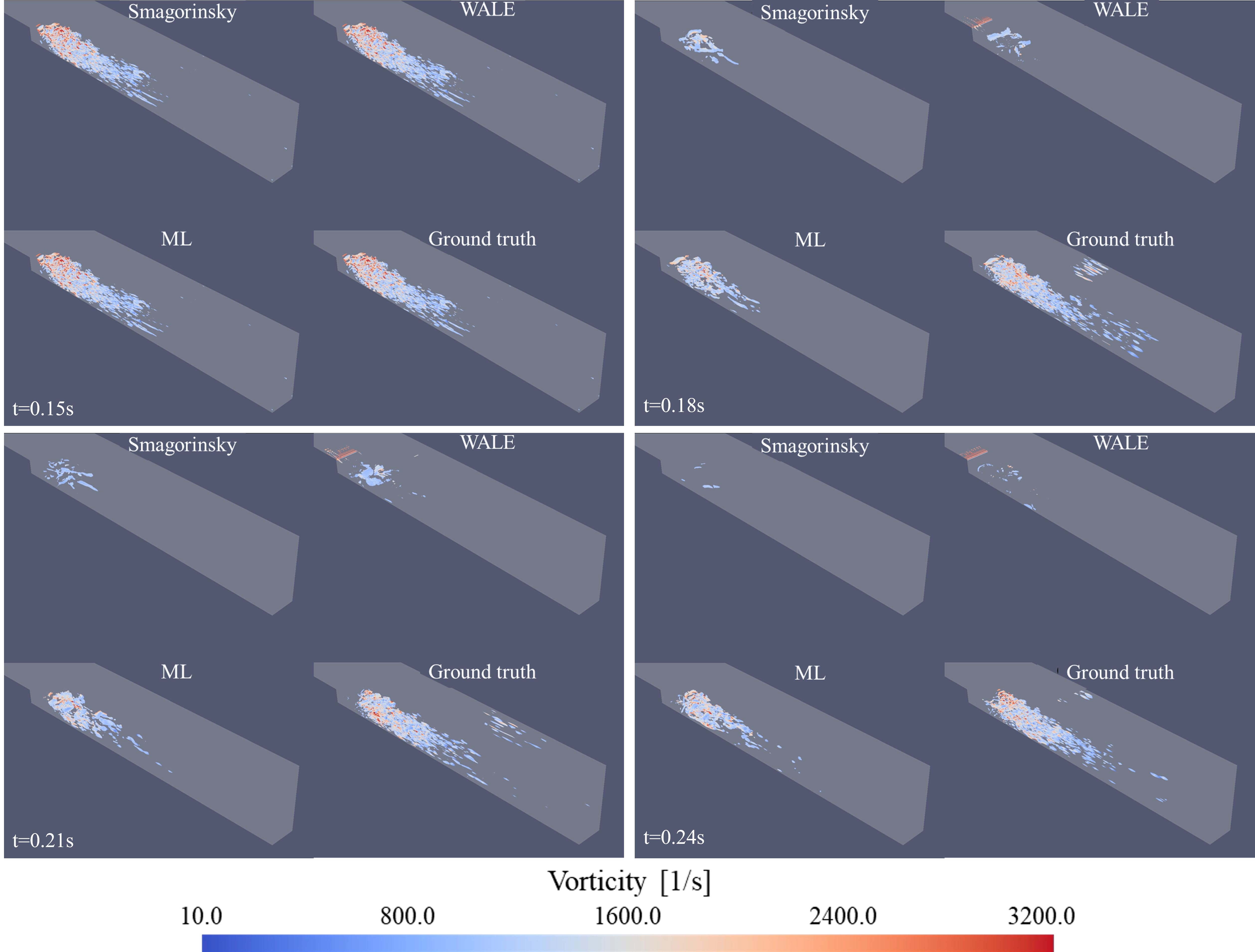}
  \end{center}
  \caption{Visualizations of the predicted iso-surfaces of the Q-criterion, colored by vorticity magnitude, for the three-dimensional BFS case with the increased step height at different time instants obtained by different models. The ground truth data at $t=0.15s$ is used as the common initial condition.}
  \label{fig:iso_bfs3d_1.2h}
\end{figure}

Figure \ref{fig:iso_bfs3d_1.2h} visualizes the iso-surfaces of the Q-criterion colored by vorticity magnitude at different time instants, obtained from different models. The Smagorinsky model again exhibits strong over-dissipation, rapidly damping small-scale vortices and failing to reproduce the downstream turbulence. The WALE model performs better by retaining some large-scale coherent structures but still misses the intricate multiscale interactions observed in the ground truth.
In contrast, the ML-based model reconstructs richer and more coherent vortical structures, demonstrating an improved ability to capture three-dimensional turbulent dynamics.

\begin{figure}
  \begin{center}
      \includegraphics[width=0.85\linewidth]{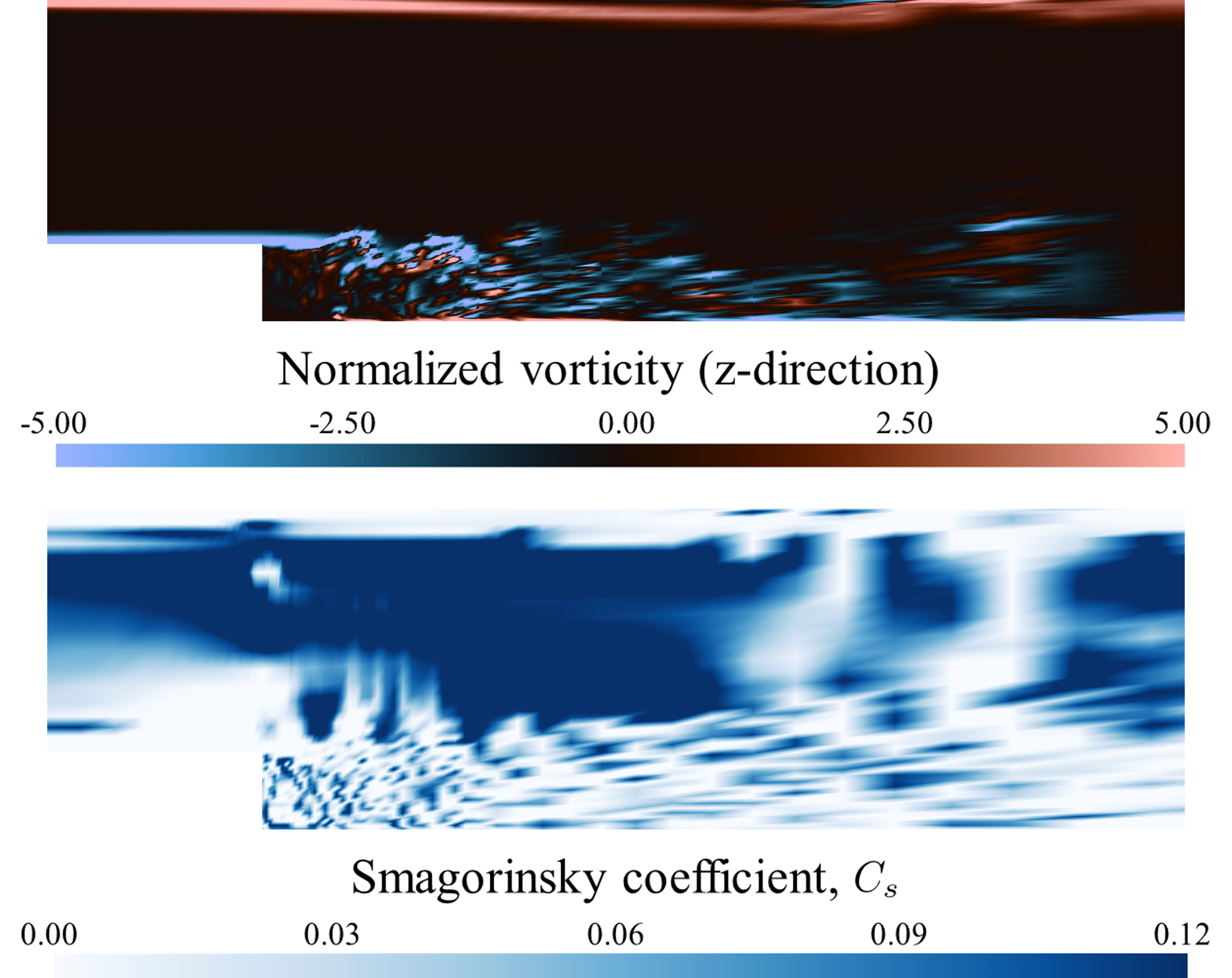}
  \end{center}
  \caption{\textbf{(Top row)} Normalized vorticity field (z-direction) and \textbf{(Bottom row)} Smagorinsky coefficient, $C_s$, predicted by the ML-based model for the three-dimensional BFS with the increased step height on the $xy$-plane at $z = 0$ at $t=0.15s$}
  \label{fig:bfs_3d_1.2h_Cs}
\end{figure}

Figure~\ref{fig:bfs_3d_1.2h_Cs} presents the normalized z-direction vorticity field and the predicted Smagorinsky coefficient, $C_s$, by the trained ML-based model on the $xy$-plane at $z=0$ at $t=0.15s$ for the new configuration. As in Fig.~\ref{fig:bfs_3d_Cs} for the original BFS geometry, the ML-based model adaptively adjusts the subgrid-scale dissipation near the step anchor and vortex cores, thereby preserving the complex multiscale turbulent structures observed in Figs.~\ref{fig:vorticity_bfs3d_1.2h} and~\ref{fig:iso_bfs3d_1.2h}, even when applied to a geometry unseen during training.

Overall, the generalization experiment demonstrates that the proposed ML-based closure maintains strong predictive performance even when applied to a geometry different from that used for training. Without any additional retraining, the model accurately reproduces the mean flow characteristics, Reynolds stress distributions, and higher-order velocity correlations, outperforming traditional SGS models in both accuracy and physical consistency. Furthermore, the ML-based model successfully reconstructs complex multiscale vortical structures that develop downstream of the step, showing enhanced capability to capture three-dimensional turbulent dynamics. These results collectively confirm that the learned closure is not merely interpolative within the training regime but effectively generalizes to unseen geometries, preserving the key statistical and structural features of turbulence.

\subsubsection{Model training with constrained data availability}

\begin{figure*}
  \begin{center}
      \includegraphics[width=0.8\linewidth]{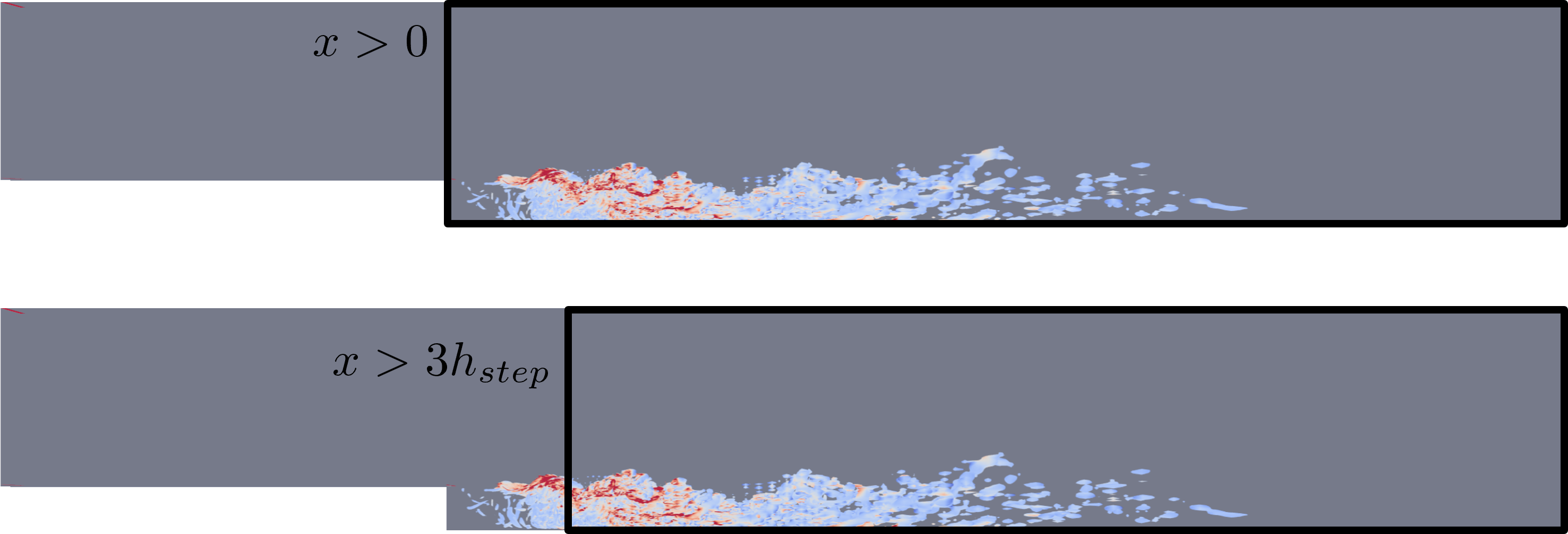}
  \end{center}
  \caption{Two examples of constrained data availability across the domain. The regions enclosed by the black boxes indicate where data are sampled for training: \textbf{(Top row)} for $x>0 $; \textbf{(Bottom row)} for $x>3h_{step}$}
  \label{fig:Sparse_domain_3d}
\end{figure*}

In this section, we train the GNN-based closure model for the three-dimensional BFS case under sparse data conditions, where only flowfield data downstream of the step are available, following the approach of the two-dimensional case (Sec.~\ref{subsubsec:Model training with constrained data availability}). Three training scenarios are examined, using the ground truth velocity data from regions $x>0$, $x>h_{step}$ and $x>3h_{step}$. Figure \ref{fig:Sparse_domain_3d} shows two examples of training scenarios.

\begin{figure}
  \begin{center}
      \includegraphics[width=0.7\linewidth]{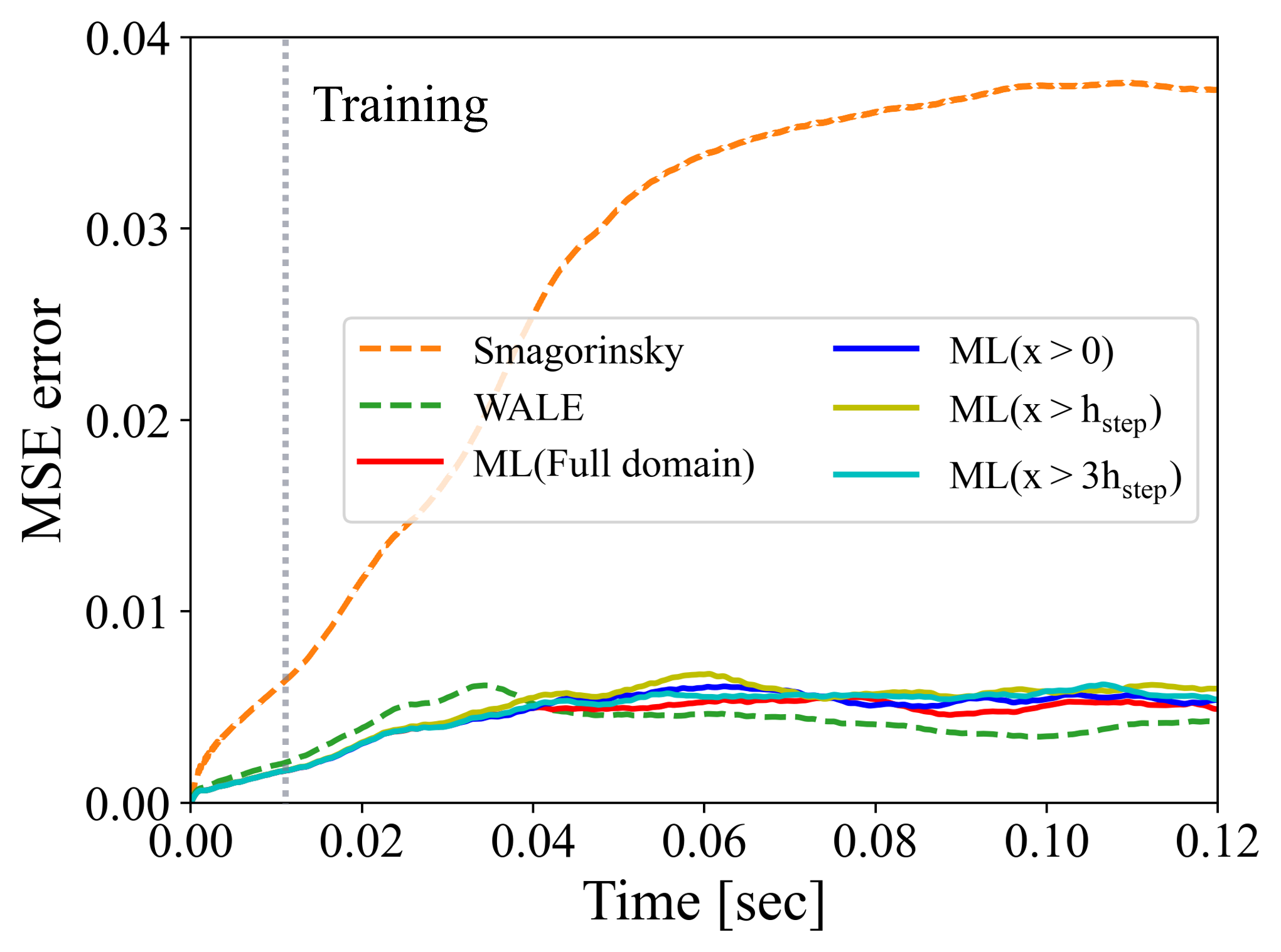}
  \end{center}
  \caption{Mean-squared error of velocity predictions for the three-dimensional BFS case obtained by both traditional and machine-learning-based models trained under varying data-availability conditions. The gray dashed line represents the extent of training data.}

  \label{fig:BFS_error_3d_sparse}
\end{figure}

Figure \ref{fig:BFS_error_3d_sparse} compares the MSE between the ground truth and the predicted velocity data, obtained from the traditional SGS models and ML models trained under different data availability conditions. All ML models trained only using the flowfield data downstream of the step show low levels of error, comparable to the ML model trained using the flow field data from the entire domain. In addition, these ML models remain stable beyond its training time horizon ($t=0.0128s$).

\begin{figure}
  \begin{center}
      \includegraphics[width=1.0\linewidth]{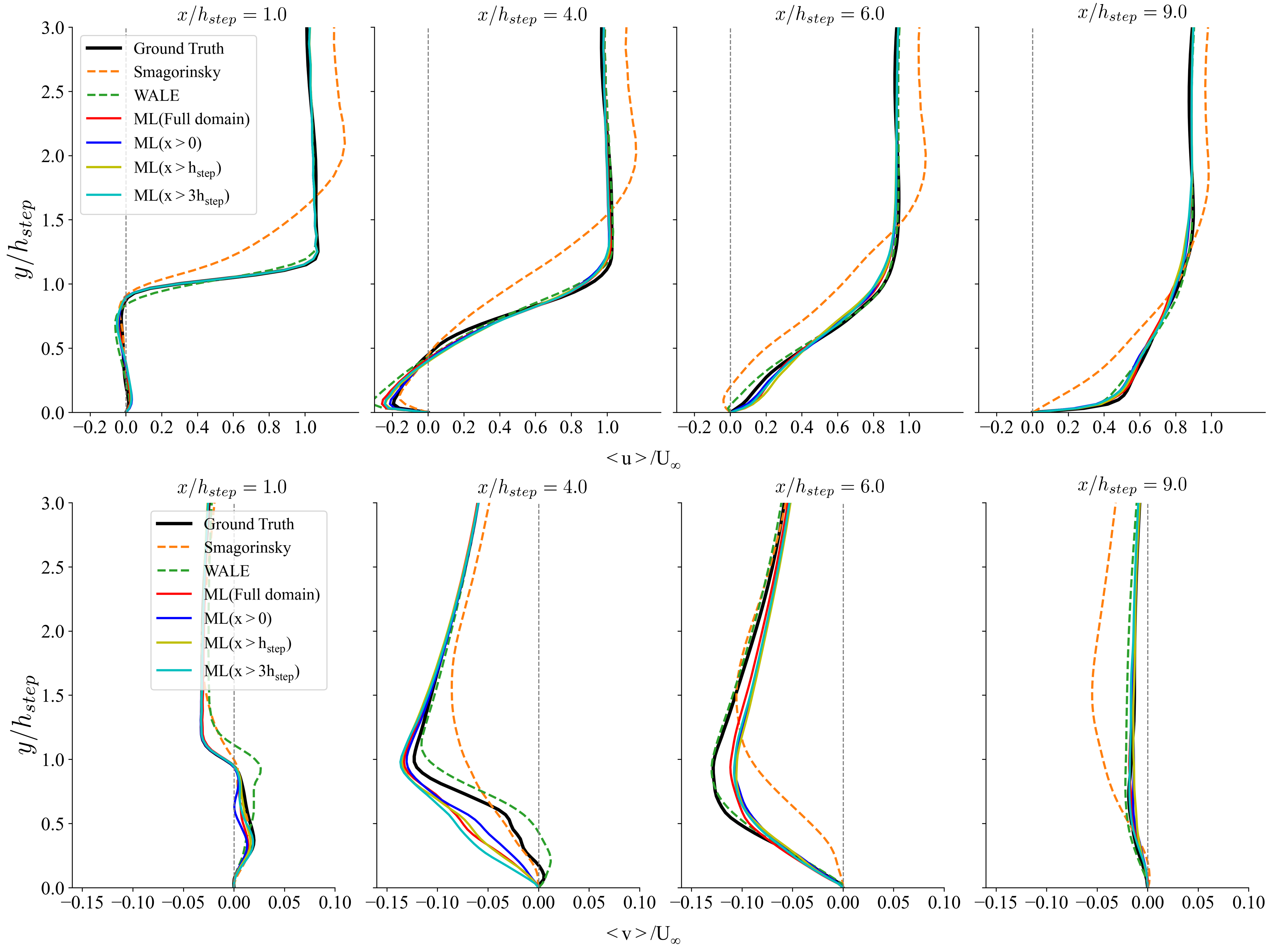}
  \end{center}
  \caption{Predictions of the first-order turbulence statistics at different downstream locations of the three-dimensional BFS case obtained by different models. \textbf{(Top)} Streamwise velocity, $<u>$; \textbf{(Bottom)} Wall-normal velocity, $<v>$}
  \label{fig:1st_stat_bfs3d_sparse}
\end{figure}

Figure~\ref{fig:1st_stat_bfs3d_sparse} compares the mean velocity profiles at different downstream locations. The ML-based models trained under sparse data conditions successfully recover the ground-truth velocity distributions, demonstrating robust predictive capability even when trained on limited flowfield information.

\begin{figure}
  \begin{center}
      \includegraphics[width=1.0\linewidth]{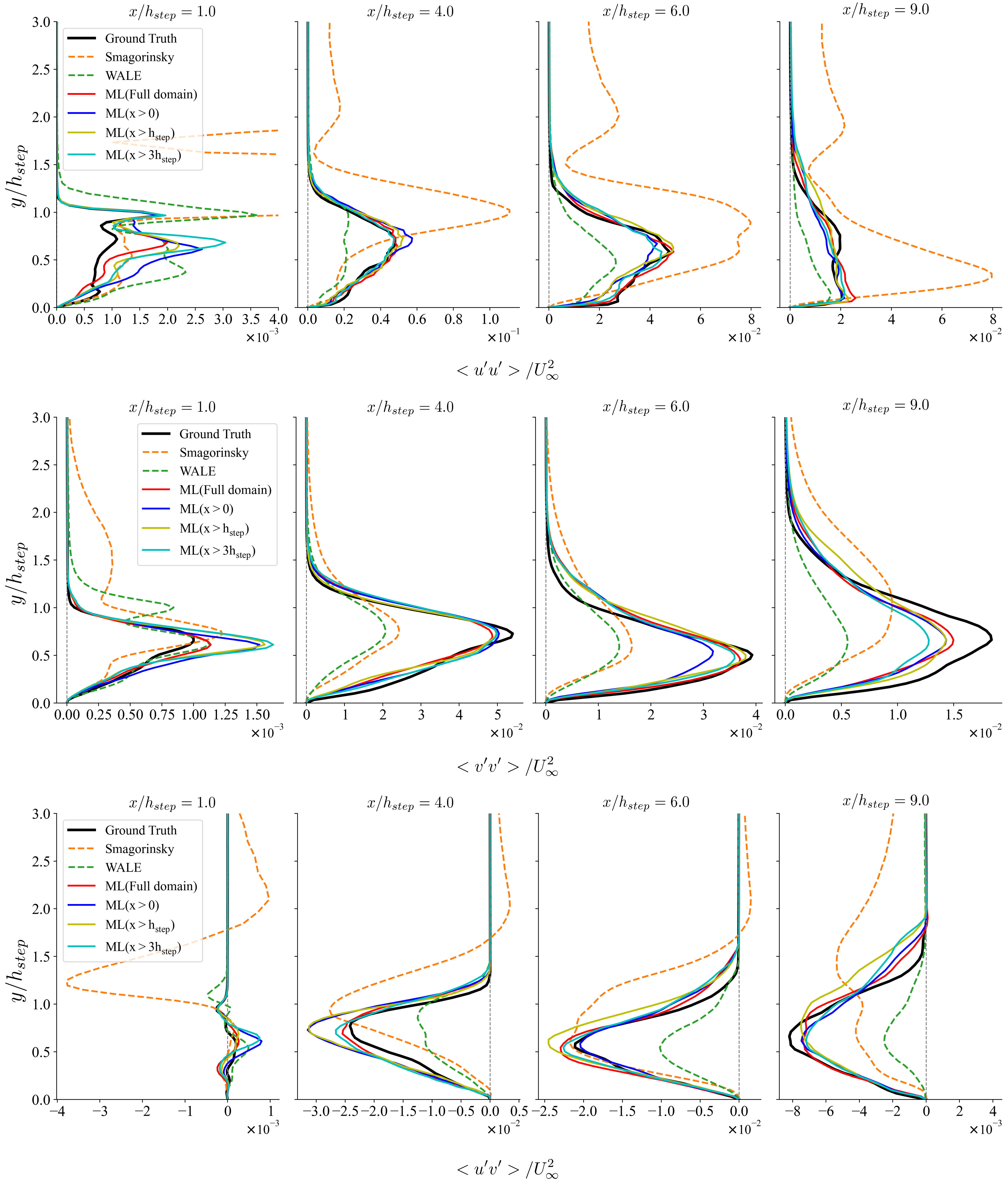}
  \end{center}
  \caption{Predictions of the Reynolds stress profiles at different downstream locations of the three-dimensional BFS case obtained by different models. \textbf{(Top)} Streamwise normal stress, $<u'u'>$; \textbf{(Middle)} Wall-normal normal stress, $<v'v'>$; \textbf{(Bottom)} Shear stress, $<u'v'>$}
  \label{fig:2nd_stat_bfs3d_sparse}
\end{figure}

Figure~\ref{fig:2nd_stat_bfs3d_sparse} presents the Reynolds stress profiles at the same downstream positions. While the ML-based models trained only with the downstream flowfield data slightly overpredict the streamwise and vertical components at $x/h_{step}=1.0$, they exhibit better agreement with the ground truth than the traditional SGS models at all other locations.
For the shear stress component, the ML-based models also provide predictions that are significantly closer to the ground-truth profiles compared to the Smagorinsky and WALE models.

\begin{figure}
  \begin{center}
      \includegraphics[width=1.0\linewidth]{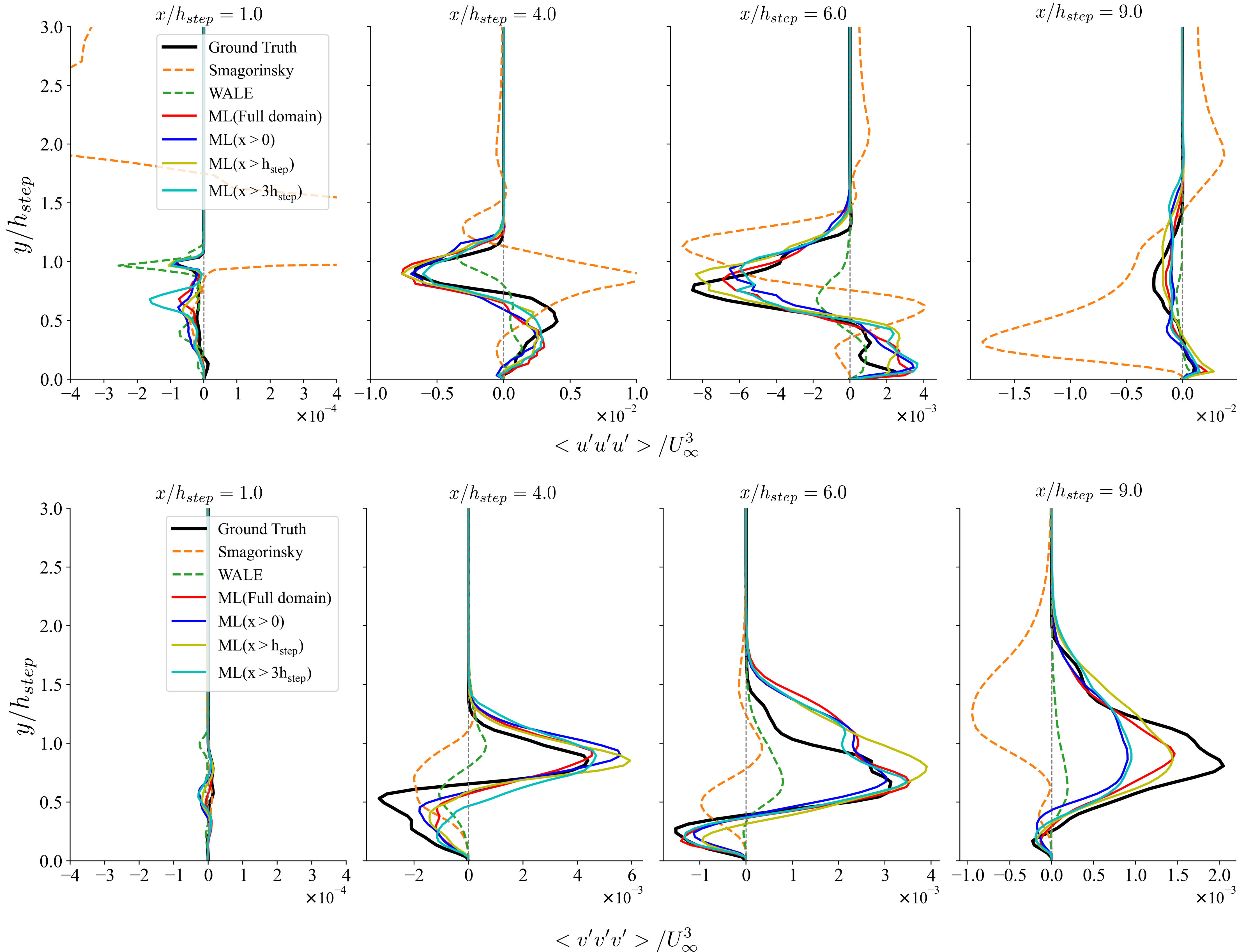}
  \end{center}
  \caption{Predictions of the third-order turbulence statistics at different downstream locations of the three-dimensional BFS case obtained by different models. \textbf{(Top)} Streamwise triple correlation, $<u'u'u'>$; \textbf{(Bottom)} Wall-normal triple correlation, $<v'v'v'>$}
  \label{fig:3rd_stat_bfs3d_sparse}
\end{figure}

Similarly, as shown in the triple-product velocity correlation profiles in Fig. \ref{fig:3rd_stat_bfs3d_sparse}, although minor differences exist among the ML-based models, their predictions overall align much more closely with the ground truth than those of the traditional SGS models, confirming the robustness of the proposed closure under constrained data availability.

The Table~\ref{tab:time_3d} compares the average computation time per iteration for each simulation using the coarse and fine grids in the three-dimensional BFS case. The results show that the Smagorinsky model on the coarse grid exhibits the fastest computation among the coarse-grid cases. However, as discussed in this section, the Smagorinsky model has limitations in capturing turbulence statistics and complex multiscale structures due to its excessive dissipation near walls and high-shear regions. In contrast, the ML model accurately predicts turbulence statistics and preserves intricate turbulent structures, while achieving approximately a fivefold speed-up compared to the fine-grid reference simulation. This demonstrates that the proposed ML-based SGS model can be effectively utilized in multi-query tasks such as uncertainty quantification and design optimization.

The WALE model, on the other hand, shows a significant slowdown compared to the Smagorinsky model because it involves more complex tensor operations. While the Smagorinsky model depends only on the strain-rate tensor, the WALE model requires both the symmetric and antisymmetric components of the velocity gradient tensor, as well as nonlinear operations such as tensor products, double contractions, and fractional powers. In FEM-based LES, these quantities must be evaluated at every quadrature point within each element, substantially increasing the computational cost. Furthermore, the viscosity in the WALE model is highly nonlinear, leading to additional Newton iterations and matrix reassembly during the solution process.

\begin{table}
\centering
\caption{Average computation time per iteration for each simulation.}
\begin{tabular}{c|c|c|c|c}
 & \multicolumn{3}{c|}{\textbf{Coarse grid}} & \textbf{Fine grid} \\
\hline
\textbf{Model} & Smagorinsky & WALE & ML & Ground truth \\
\hline
\textbf{Computation time} & 4.1\,s/it & 28.8\,s/it & 6.2\,s/it & 32.5\,s/it \\
\hline
\end{tabular}
\label{tab:time_3d}
\end{table}

\section{Conclusions}
\label{sec:Conclusions}

This study presents a framework for turbulence closure modeling within the paradigm of differentiable physics, where a data-driven machine learning closure is embedded within a differentiable solver for the incompressible Navier-Stokes equations. Specifically, we incorporate a graph neural network (GNN) into a large-eddy simulation to learn subgrid-scale (SGS) closures on coarse, unstructured meshes. The mesh-invariant nature of the GNN enables generalization to new geometries, while the differentiable solver, which is based on the finite element method (FEM) implemented in FEniCS/FEniCSx, allows exact coupling between the solver adjoint and the GNN gradients through custom vector-Jacobian products defined for each sub-step of the numerical scheme.

As a proof of concept, we first apply the framework to a two-dimensional backward-facing step (BFS) flows to demonstrate its feasibility and training stability. The trained model accurately captures both transient and mean flow behaviors on coarse grids, outperforming a conventional SGS model. These results confirm the validity of the proposed differentiable learning framework for turbulence closure discovery.

The main contribution of this work lies in extending the framework to three-dimensional turbulence. We train a GNN-based closure model for the three-dimensional BFS case and comprehensively evaluate its predictive performance, physical fidelity, and generalization capability. The trained model reproduces mean velocity profiles, Reynolds stress distributions, and higher-order velocity correlations with high accuracy, while also capturing rich multiscale vortical structures that persist over long integration times. Furthermore, without any retraining, the same model generalizes robustly to a new geometry with an increased step height, maintaining physical consistency and numerical stability. Notably, these results demonstrate that training the closure model using instantaneous flowfield quantities enables the recovery of more accurate statistical properties within the differentiable physics framework.

Finally, to assess data efficiency, we train the model using sparse downstream flowfield data and find that it can achieve comparable or even improved accuracy compared to models trained on the full domain. This demonstrates that the proposed differentiable GNN-LES framework can learn effective turbulence closures even under limited data availability, offering a pathway toward data-efficient turbulence modeling.

While this study shows the potential of closure modeling using GNN and differentiable physics solver for generalizability, further research is required to fully achieve robust generalization capability. Our trained models are only generalizable to geometries where computational meshes have same order of density as the meshes used for training as shown in Figs. \ref{fig:BFS_error_gridconvergence} and \ref{fig:BFS_probe_gridconvergence} in Appendix. \ref{app:Grid convergence test}. Further research can improve generalization capability by making trained models applicable to other meshes with different order of density. In addition, incorporating interpretability of graph structures into the training procedure can make a model's behavior more understandable especially when tested on unseen geometries. Lastly, the proposed framework also has the potential to be extended to high-speed flows and for wall-modeling applications, suggesting its applicability to a wider range of realistic engineering problems. Preliminary results presented in Appendix~\ref{app:Learning SGS boundary conditions} further demonstrate this potential, where a wall-localized GNN-based SGS closure model adaptively applies subgrid-scale corrections near wall boundaries while retaining the baseline LES model in the interior region, improving predictive accuracy and stability even for unseen geometries.

\color{black}

\section{Acknowledgments}
This material is based on work supported by the National Science Foundation Graduate Research Fellowship under Grant No. DGE 1745016 awarded to VS. VS and VV acknowledge the support from the Technologies for Safe and Efficient Transportation University Transportation Center, and Mobility21, A United States Department of Transportation National University Transportation Center. 
RM and HK acknowledge funding support from ASCR for DOE-FOA-2493 ``Data-intensive scientific machine learning'' and for SCIDAC RAPIDS2. RM and HK also acknowledge support from the Penn State Institute for Computational and Data Sciences and a Young Investigator Program award from the Army Research Office (PM - Dr. Rob Martin). 

\clearpage
\appendix

\section{Training history}
\label{app:Training history}
\begin{figure}[H]
  \centering
  \includegraphics[width=.6\linewidth]{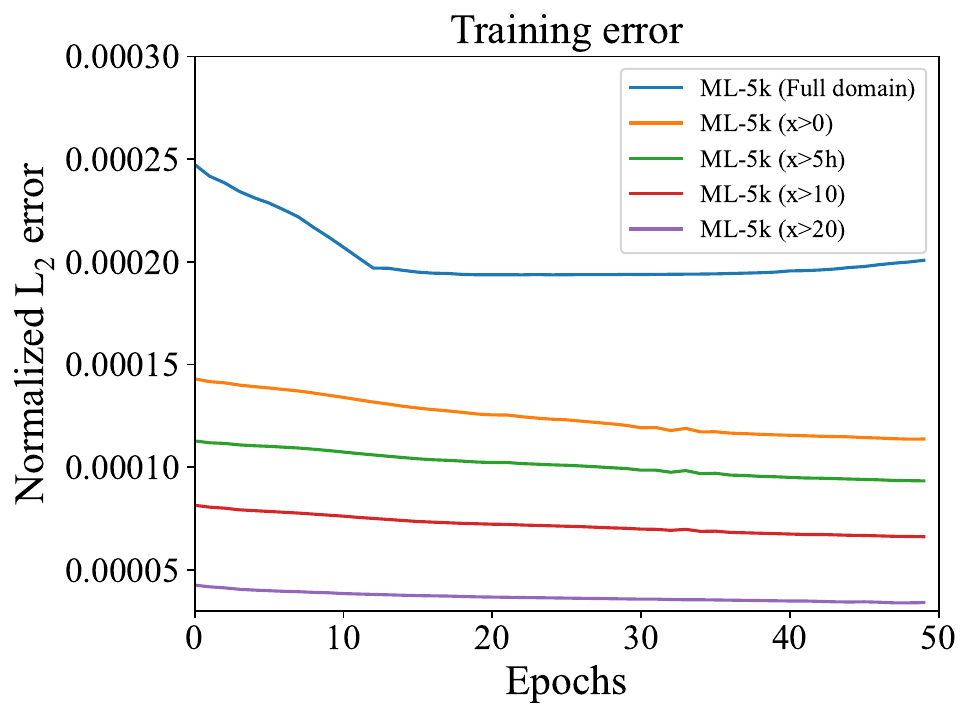}
  \caption{Training history of GNN-based subgrid closure models for the two-dimensional BFS case} 
  \label{fig:Training_history}
\end{figure}

Figure \ref{fig:Training_history} shows the training error history of each model under different incomplete data scenarios for the two-dimensional BFS case.

\begin{figure}[H]
  \centering
  \includegraphics[width=.6\linewidth]{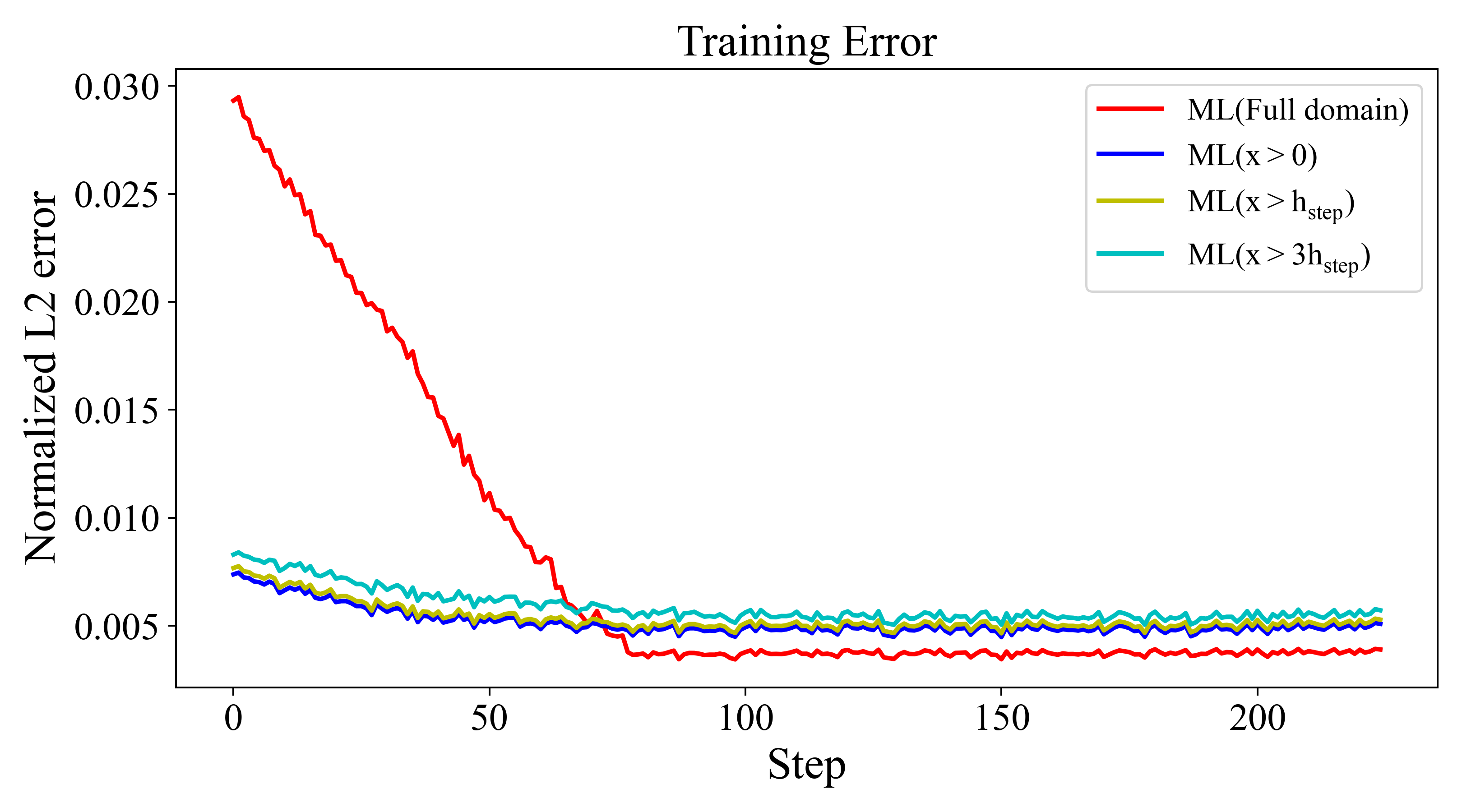}
  \caption{Training history of GNN-based subgrid closure models for the three-dimensional BFS case} 
  \label{fig:Training_history_3d}
\end{figure}

Figure \ref{fig:Training_history_3d} shows the training error history of each model under different incomplete data scenarios for the three-dimensional BFS case.

\section{Grids used for data generation}
\label{app:Grids used for data generation}
\begin{figure}
  \includegraphics[width=\linewidth]{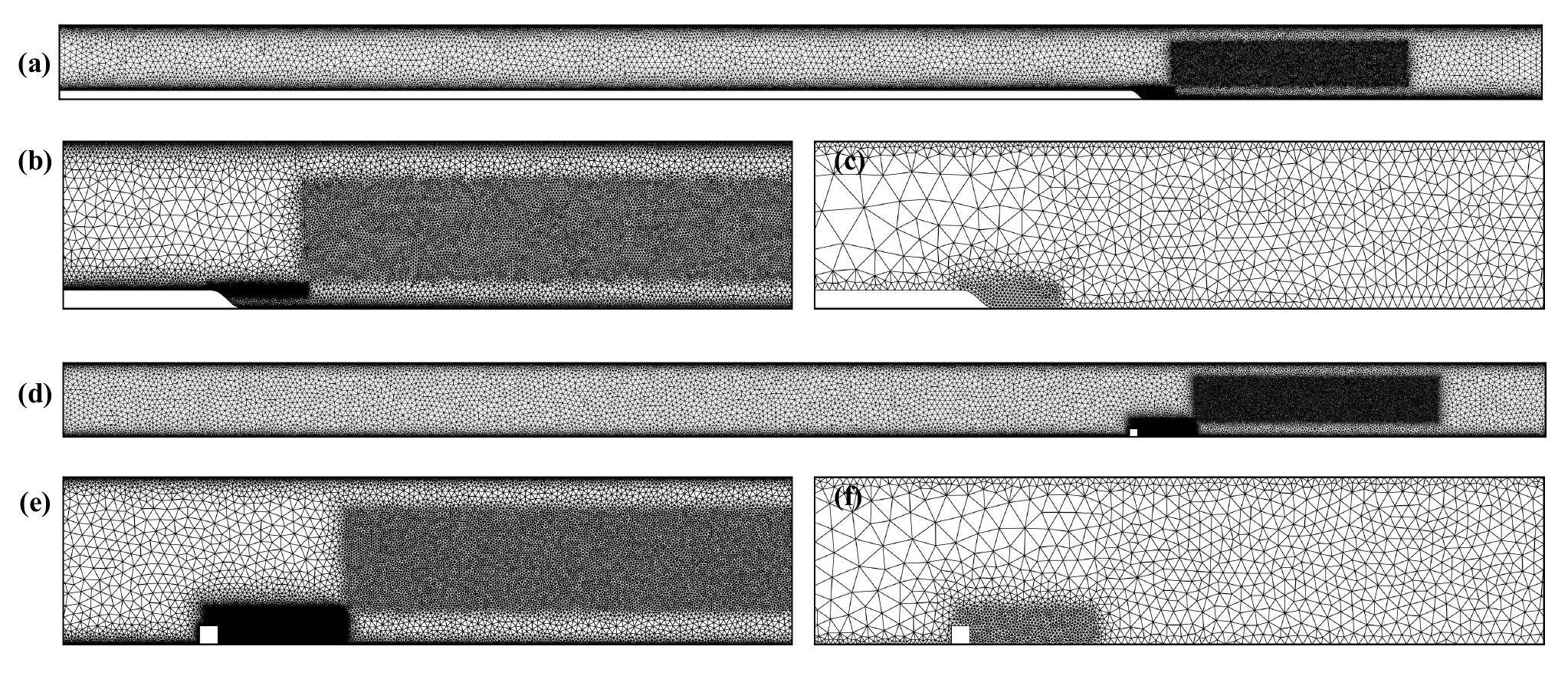}
  \caption{\textbf{(a)} Image of the ramp geometry \textbf{(b)} Zoomed-in image of the ramp used for LES-80k \textbf{(c)} Zoomed-in image of the ramp used for LES-5k \textbf{(d)} Image of the wall-mounted cube (WMC) geometry \textbf{(e)} Zoomed-in image of the WMC used for LES-80k \textbf{(f)} Zoomed-in image of the WMC used for LES-5k} 
  \label{fig:Grid_BFR_WMC}
\end{figure}

Figure \ref{fig:Grid_BFR_WMC} depicts the images of the grids around the ramp and the wall-mounted cube, which were utilized for data generation.


\begin{figure}
  \includegraphics[width=\linewidth]{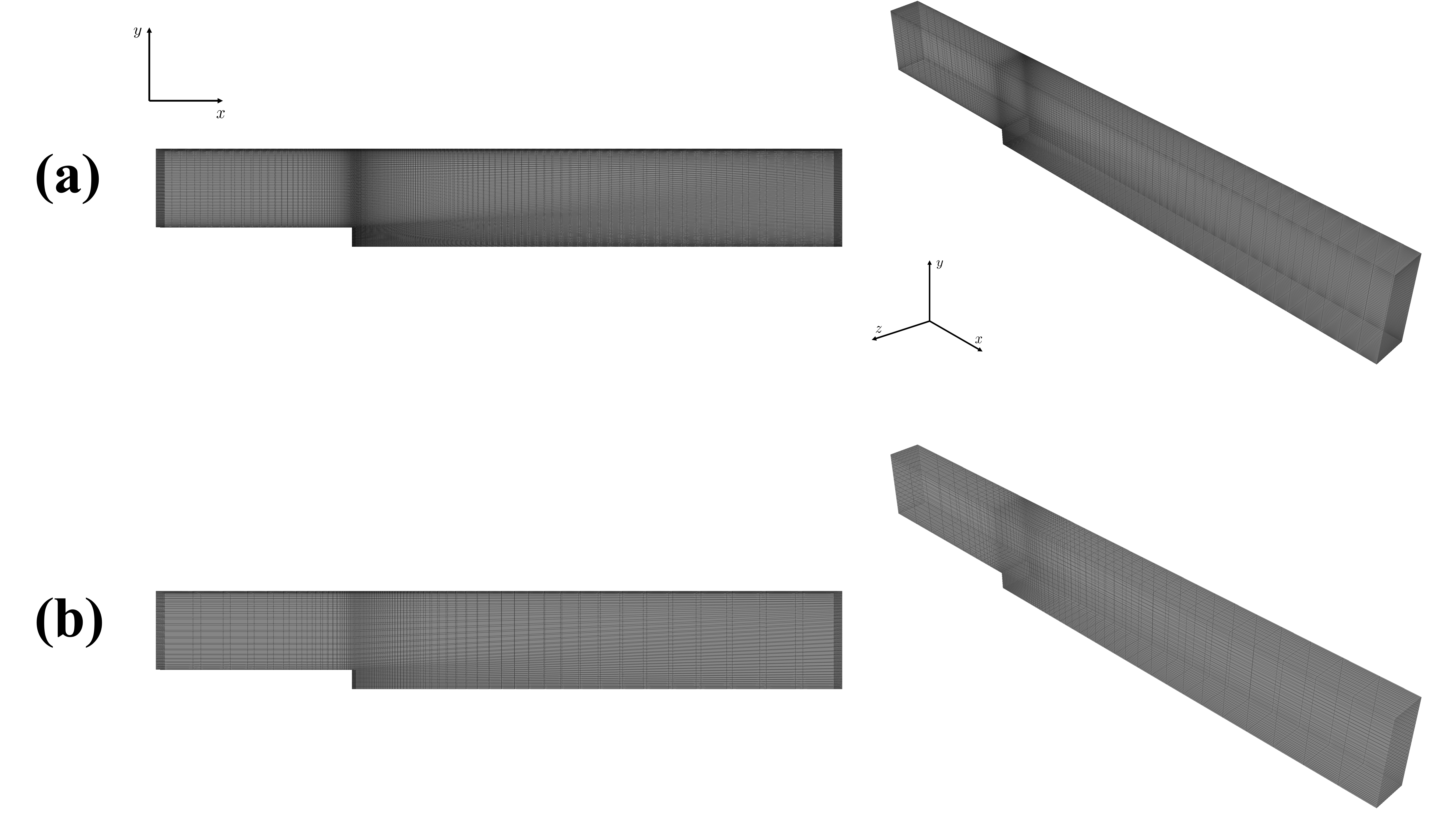}
  \caption{\textbf{(a)} Image of the fine grid for the three-dimensional BFS \textbf{(b)} Image of the coarse grid for the three-dimensional BFS} 
  \label{fig:Grid_BFS_3d}
\end{figure}

\begin{figure}
  \includegraphics[width=\linewidth]{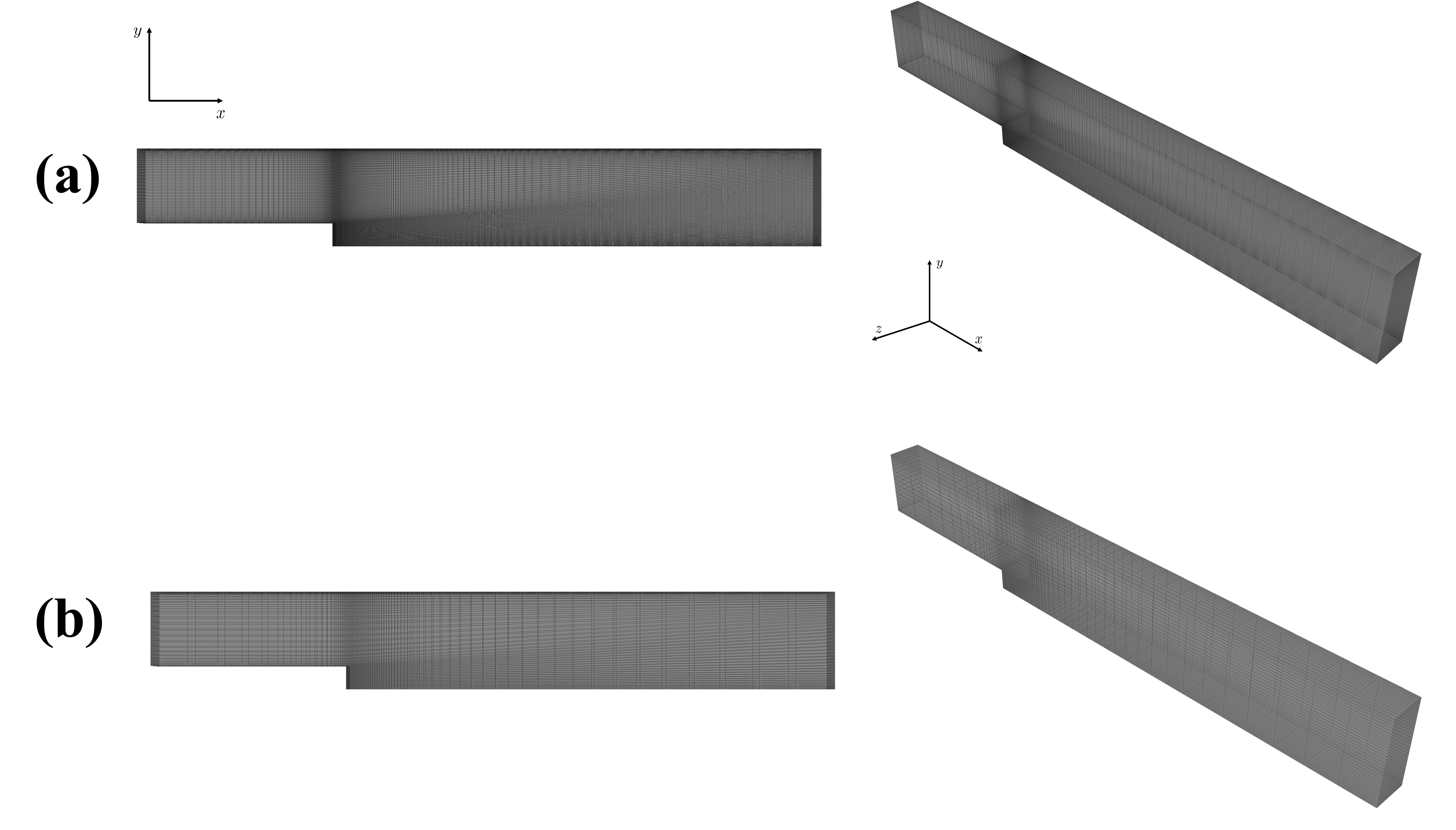}
  \caption{\textbf{(a)} Image of the fine grid for the three-dimensional BFS with the increased step height \textbf{(b)} Image of the coarse grid for the three-dimensional BFS with the increased step height} 
  \label{fig:Grid_BFS_1.2h_3d}
\end{figure}

Figures \ref{fig:Grid_BFS_3d} \&  \ref{fig:Grid_BFS_1.2h_3d} show the images of the fine and coarse grids around the original three-dimensional BFS case and the BFS with the increased step height, respectively.


\section{Grid convergence test}
\label{app:Grid convergence test}
\begin{figure}
  \includegraphics[width=0.7\linewidth]{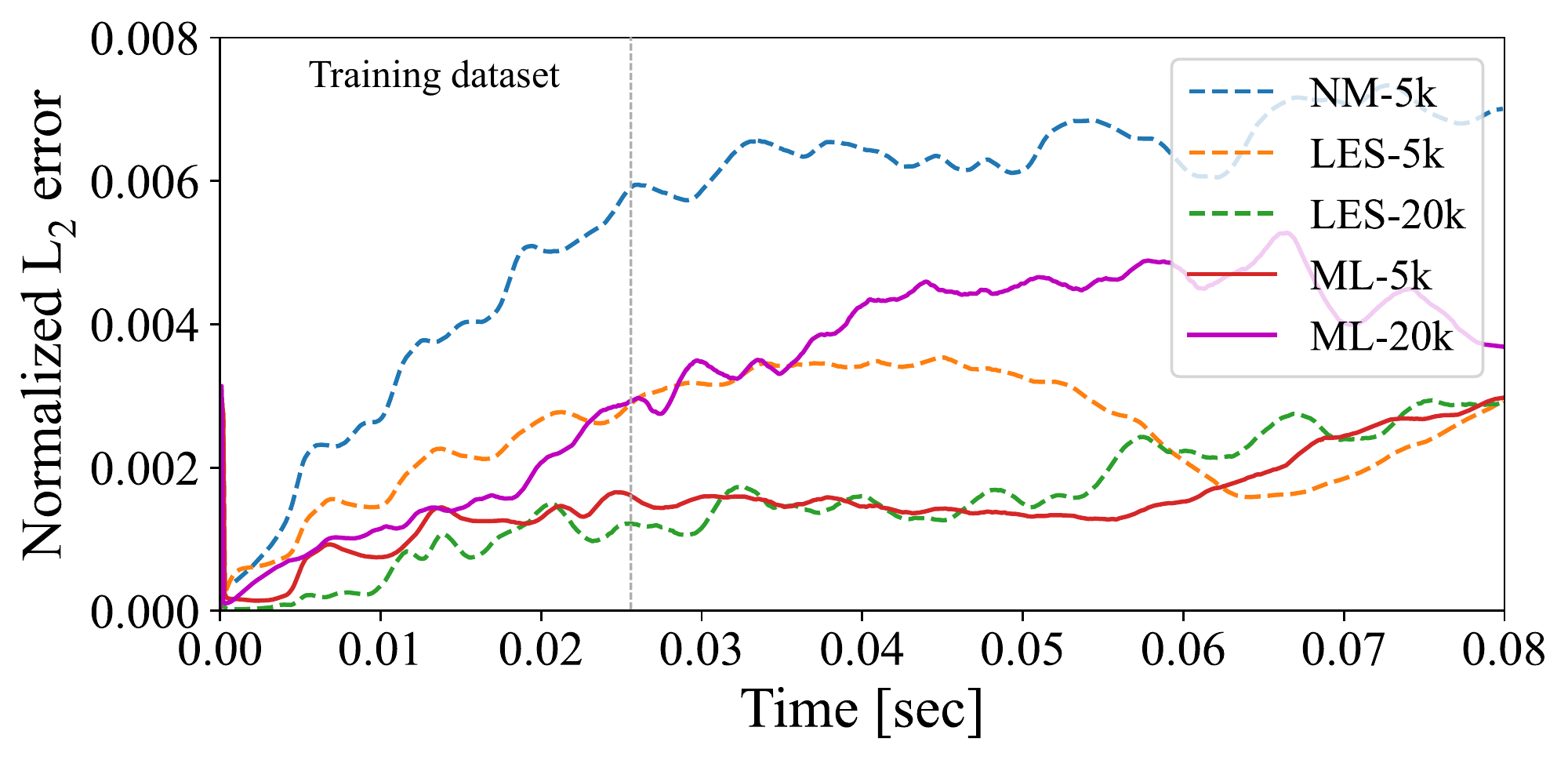}
  \caption{Normalized L2 error on the BFS case} 
  \label{fig:BFS_error_gridconvergence}
\end{figure}

\begin{figure}
  \includegraphics[width=\linewidth]{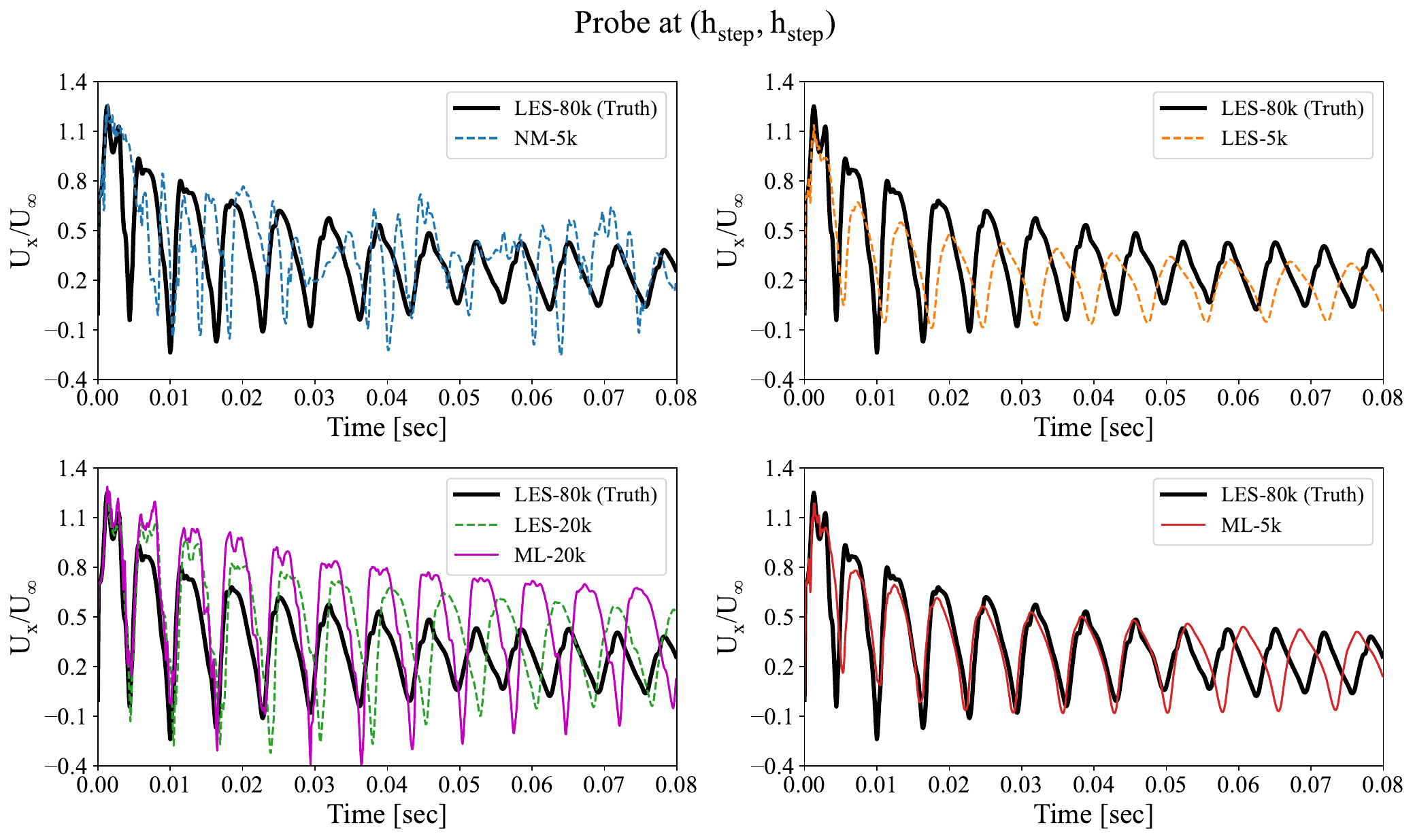}
  \caption{$x$-direction Velocity measurement from a probe in front of the step at $(x, y)=(h, h)$} 
  \label{fig:BFS_probe_gridconvergence}
\end{figure}

\begin{figure}
  \includegraphics[width=0.5\linewidth]{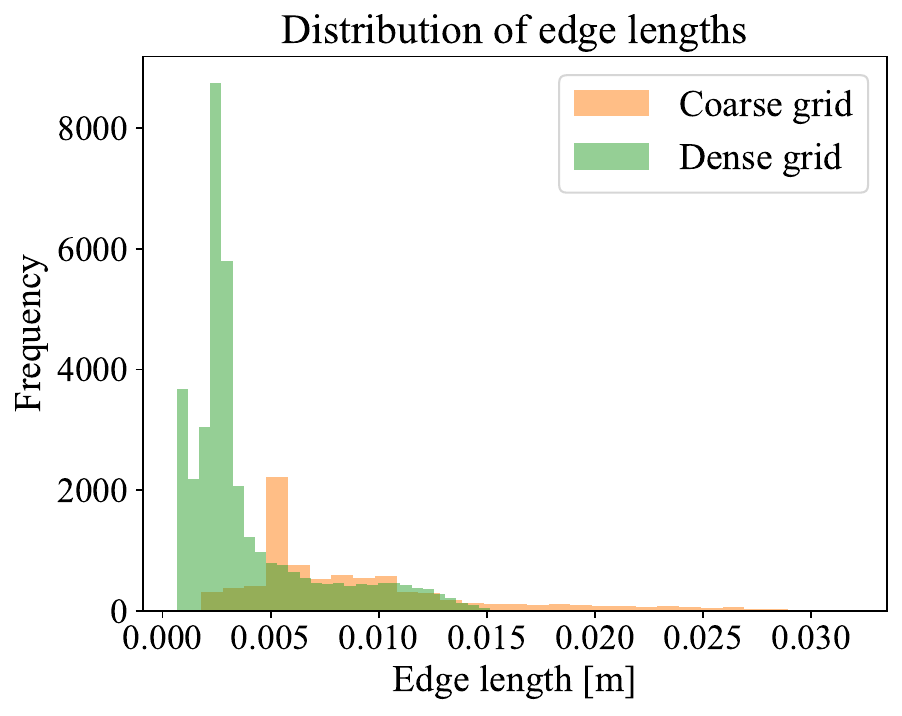}
  \caption{Edge length distributions for dense and coarse grids} 
  \label{fig:edge_distributions}
\end{figure}

The ML-20k model is a model where a subgrid GNN-based closure, trained on a coarse mesh of the backward-facing step case, is applied to a denser grid. According to Fig. \ref{fig:BFS_error_gridconvergence}, the ML-20k model significantly increases the prediction error compared to the LES-20k model. Specifically, as shown in Fig. \ref{fig:BFS_probe_gridconvergence}, the ML-20k model overpredicts peaks of velocity fluctuations. This is because the smallest edge length in the dense grid is half the size of the smallest edge length in the coarse grid, meaning that the model was never exposed to edge lengths as small as those in the dense grid during training. Fig. \ref{fig:edge_distributions} illustrates the distributions of edge lengths on dense and coarse grids.

\section{Learning SGS boundary conditions}
\label{app:Learning SGS boundary conditions}
To explore the viability of our framework for solely learning boundary conditions for closure models, we train our GNN-based subgrid-scale closure model (for the 2D BFS case) to adaptively predict $C_s$ only at the wall boundaries, while assigning a fixed value of $C_s=0.06$ to the interior fluid cells. The fixed value is chosen to match that used in the baseline ``LES-5k`` simulation as in the Sec. \ref{subsubsec:Dataset generation for 2D flows}. During inference, the GNN encoder and processor in the Sec. \ref{sec:Graph neural networks} utilize all node and edge features across the entire domain. However, the outputs at interior fluid nodes are overridden to the fixed value $C_s=0.06$, while the wall nodes retain the adaptive values predicted by the GNN, which depend on local and neighboring flow features. The training dataset and the loss function are the same as in Sec. \ref{subsubsec:Dataset generation for 2D flows}. This formulation enables the model to learn the optimal subgrid-scale contribution at the wall boundaries to match the target flowfield in wall-bounded turbulent flows, while retaining the baseline LES model in the interior fluid region.

\begin{figure}
  \begin{center}
      \includegraphics[width=0.7\linewidth]{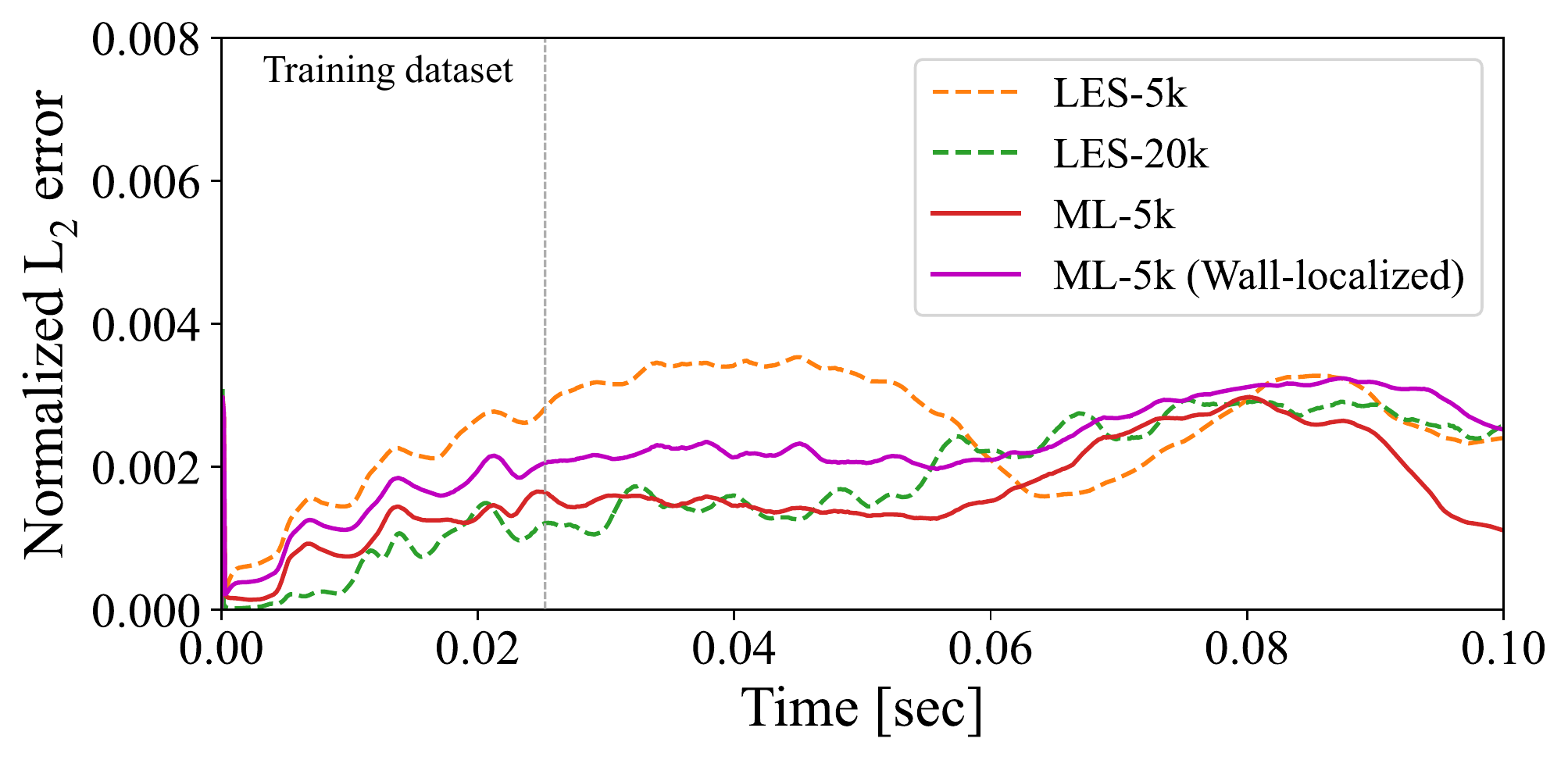}
  \end{center}
  \caption{Normalized $L_2$ error for the BFS case. The gray dashed line represents the extent of training data collected. ``ML-5k`` uses adaptive $C_s$ across the entire domain. ``ML-5k (wall-localized)`` applies adaptive $C_s$ only at the wall boundaries, while using a fixed value of $C_s=0.06$ in the interior fluid cells.}
  \label{fig:BFS_error_wall}
\end{figure}

\begin{figure}
  \begin{center}
      \includegraphics[width=\linewidth]{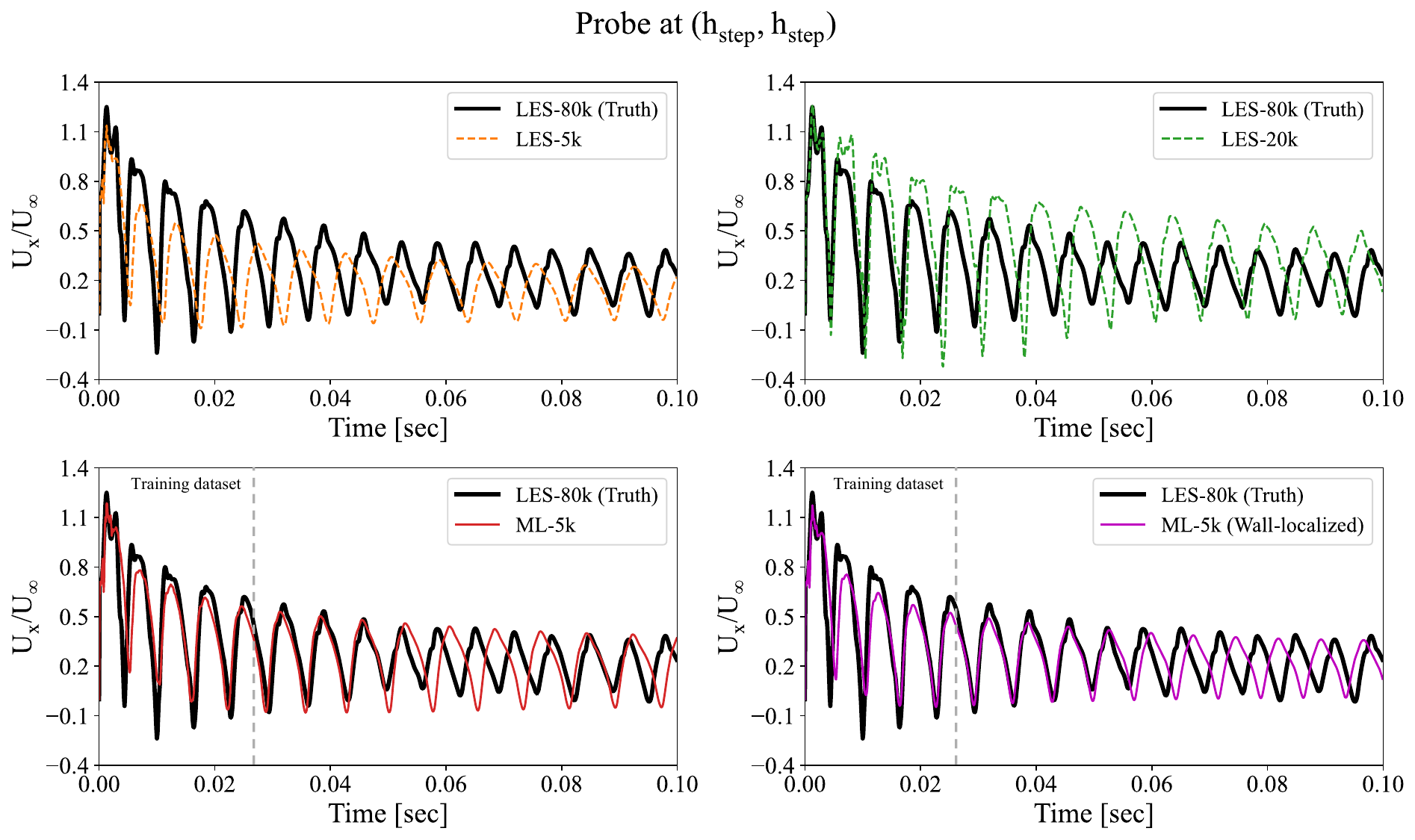}
  \end{center}
  \caption{$x$-direction Velocity measurement from a probe in front of the step at $(x, y)=(h, h)$. Here we assess the performance of the SGS model that is trained to predict $C_s$ adaptively only at the wall.}
  \label{fig:BFS_probe_split_wall}
\end{figure}

\begin{figure*}
  \begin{center}
      \includegraphics[width=\linewidth]{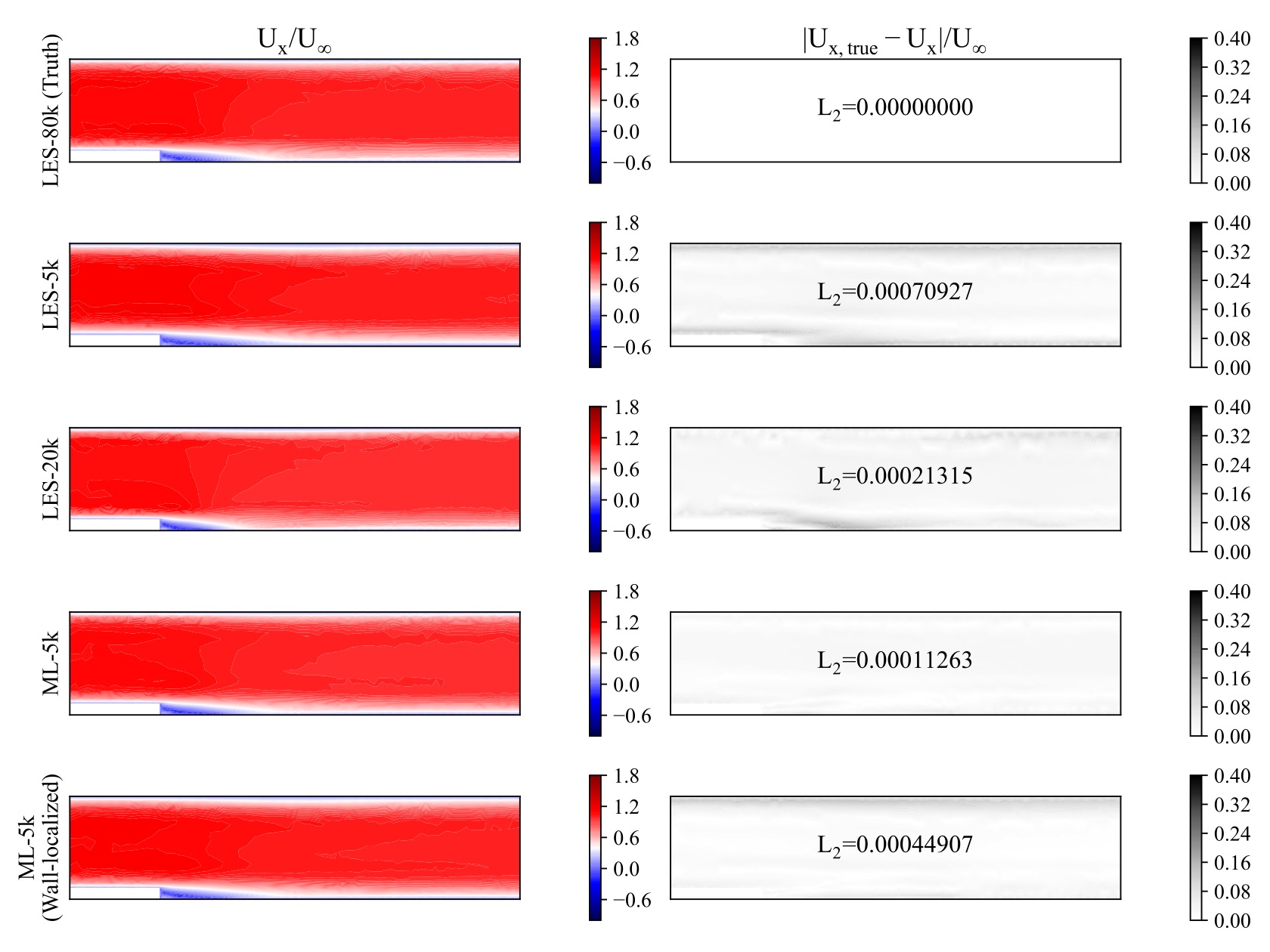}
  \end{center}
  \caption{Plots of the mean flow from the BFS case for various models. \textbf{(Left column)} The normalized mean x-velocity and \textbf{(Right column)} The error with respect to the truth in space, as well as the integrated total.}
  \label{fig:Flowfield_BFS_wall}
\end{figure*}

\begin{figure*}
  \begin{center}
      \includegraphics[width=\linewidth]{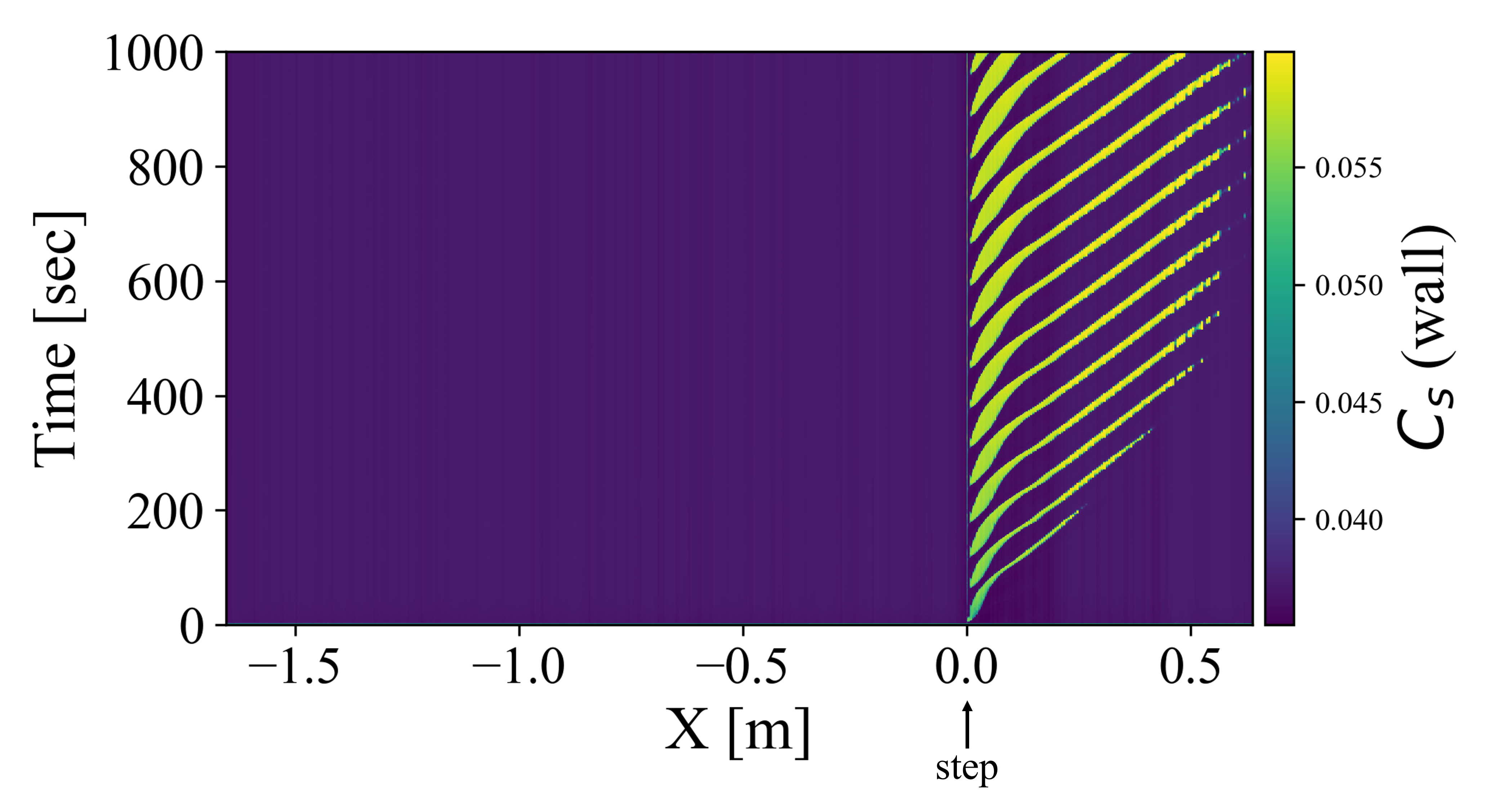}
  \end{center}
  \caption{Space-time map of the wall coefficient $C_s$ for the two-dimensional BFS. The colormap indicates the local value of $C_s$ along the bottom wall over time, showing downstream convection of high-intensity regions associated with vortex passage.}
  \label{fig:Cs_wall_xt}
\end{figure*}

Figure \ref{fig:BFS_error_wall} shows the normalized $L^2$ error over time for the BFS case, comparing the performance of the wall-localized SGS model with the full-domain adaptive SGS model introduced in Sec. \ref{subsubsec:Training results for 2D flows}, which applies adaptive $C_s$ values across the entire domain, and two baseline models.  The wall-localized ML-5k model exhibits a larger $L^2$ error than the full-domain ML-5k model, as it restricts the use of adaptive $C_s$ to the wall boundaries, whereas the full-domain model applies adaptive values at all nodes. Nevertheless, the wall-localized ML-5k model outperforms the baseline LES-5k model, achieving lower overall $L^2$ error despite using adaptive $C_s$ only at the wall boundaries. Figure \ref{fig:BFS_probe_split_wall} compares the predictive accuracy of normalized $x$-direction velocity at the probe located at $(x,y)=(h,h)$. The wall-localized ML-5k model more accurately captures the peaks of periodic velocity fluctuations than both the LES-5k and LES-20k models. Figure \ref{fig:Flowfield_BFS_wall} illustrates the normalized mean $x$-direction velocity field and the corresponding error distribution around the BFS. The wall-localized ML-5k model reduces errors near the downstream region of the step and improves the prediction of the mean flow field compared to the baseline LES-5k model. Figure \ref{fig:Cs_wall_xt} shows the space–time distribution of the wall coefficient $C_s$ along the bottom wall for the backward-facing step flow. The bright oblique streaks represent regions of elevated $C_s$, which are convected downstream over time following the trajectory of the vortical structures detached from the step edge. The step location is indicated by the black arrow. This pattern demonstrates that the wall response in $C_s$ is strongly correlated with the passage of large-scale vortices and propagates downstream at approximately the same convection velocity as the near-wall coherent structures.

\begin{figure}
  \begin{center}
      \includegraphics[width=0.7\linewidth]{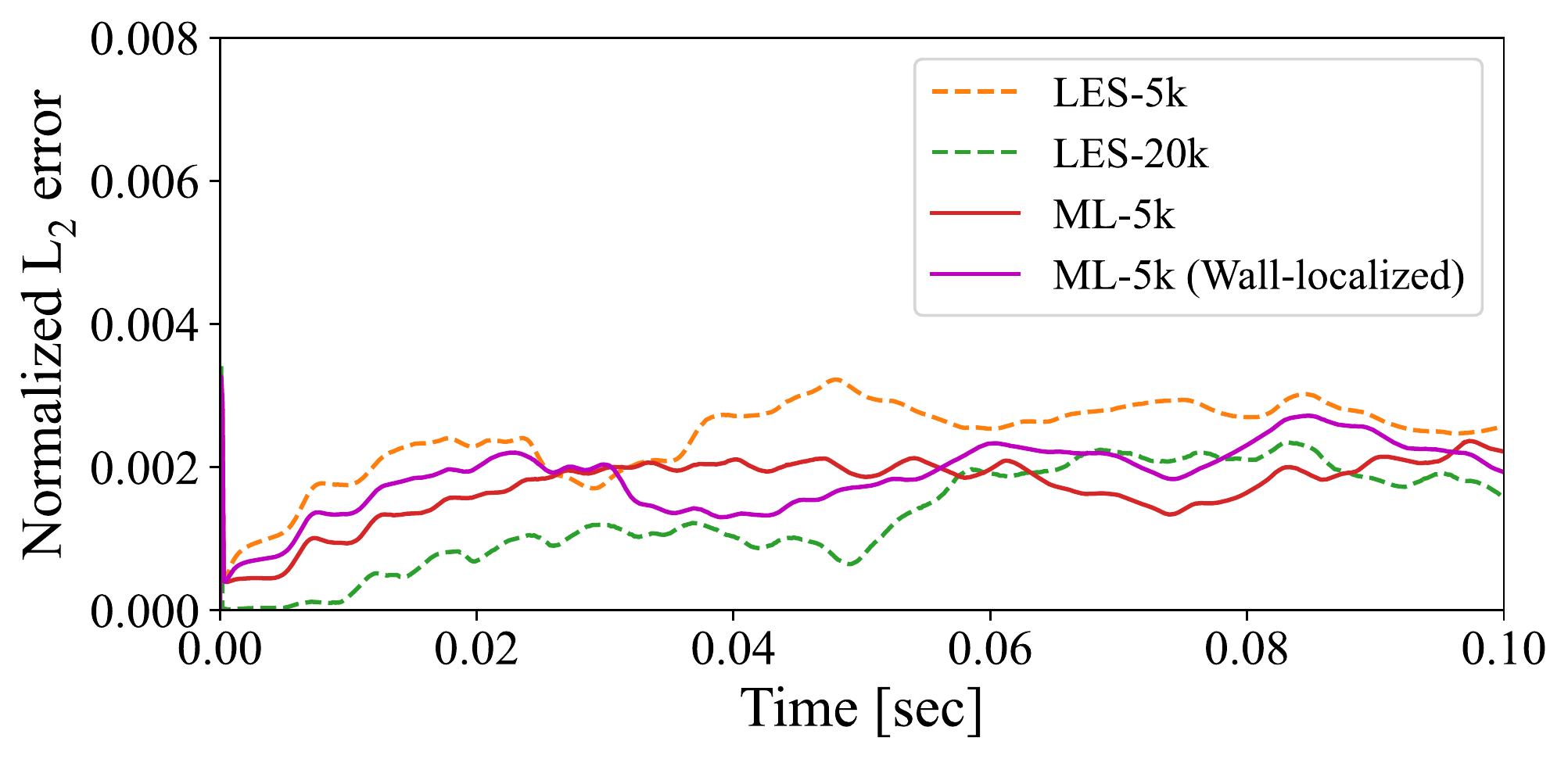}
  \end{center}
  \caption{Normalized $L_2$ error for the ramp case. This geometry was unseen during training.}
  \label{fig:BFR_error_wall}
\end{figure}

\begin{figure}
  \begin{center}
      \includegraphics[width=\linewidth]{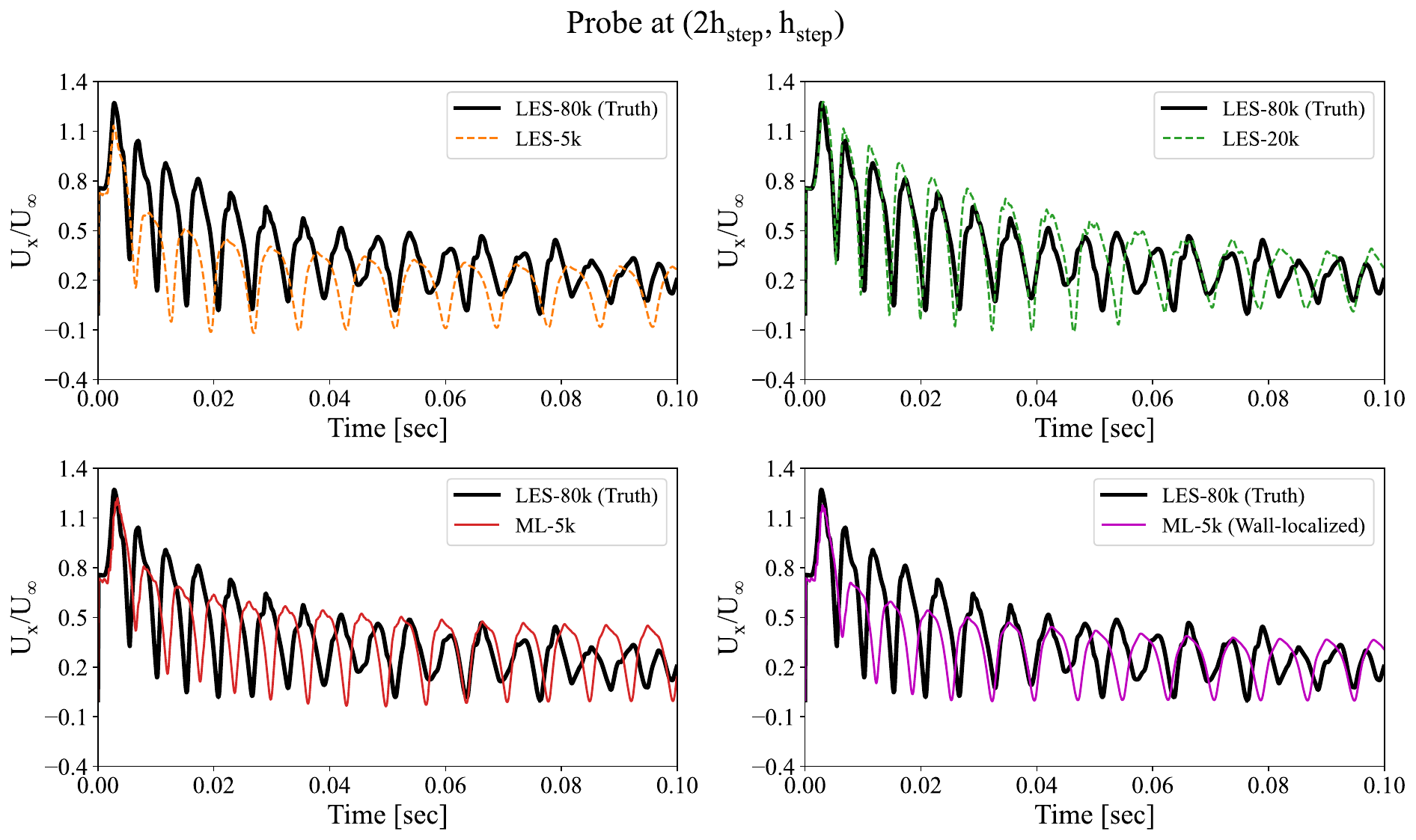}
  \end{center}
  \caption{$x$-direction Velocity measurement from a probe in front of the ramp at $(x, y)=(2h, h)$.}
  \label{fig:BFR_probe_split_wall}
\end{figure}

\begin{figure*}
  \begin{center}
      \includegraphics[width=\linewidth]{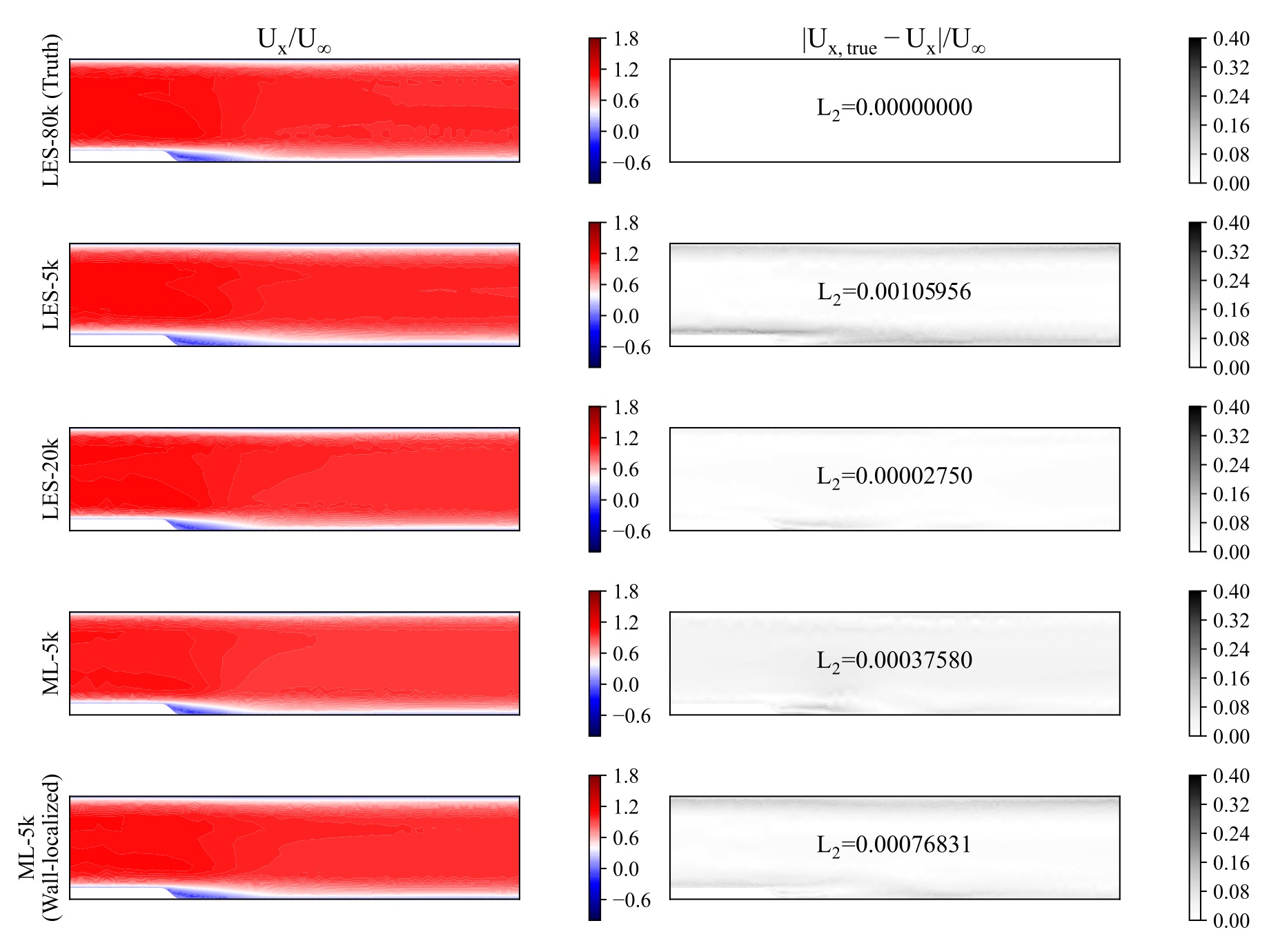}
  \end{center}
  \caption{Plots of the mean flow from the ramp case for various models. \textbf{(Left column)} The normalized mean x-velocity and \textbf{(Right column)} The error with respect to the truth in space, as well as the integrated total.}
  \label{fig:Flowfield_BFR_wall}
\end{figure*}

We also apply our single wall-localized SGS closure model trained with the BFS data to the ramp and wall-mounted cube to evaluate its generalizability. Fig. \ref{fig:BFR_error_wall} shows the normalized $L^2$
error over time for the ramp case, comparing the wall-localized SGS model with three other models. The wall-localized ML-5k model achieves lower error than LES-5k.
Fig. \ref{fig:BFR_probe_split_wall} shows the pointwise accuracy based on normalized x-velocity at the probe located at $(x,y)=(2h,h)$. The wall-localized model improves over LES-5k and remains stable beyond the training horizon. Fig. \ref{fig:Flowfield_BFR_wall} displays the normalized mean x-velocity field and the corresponding error contours, compared to the LES-80k reference. Integrated error values are also reported. The wall-localized ML-5k model reduces errors downstream of the ramp and better captures the mean flow behavior than LES-5k.

\begin{figure}
  \begin{center}
      \includegraphics[width=0.7\linewidth]{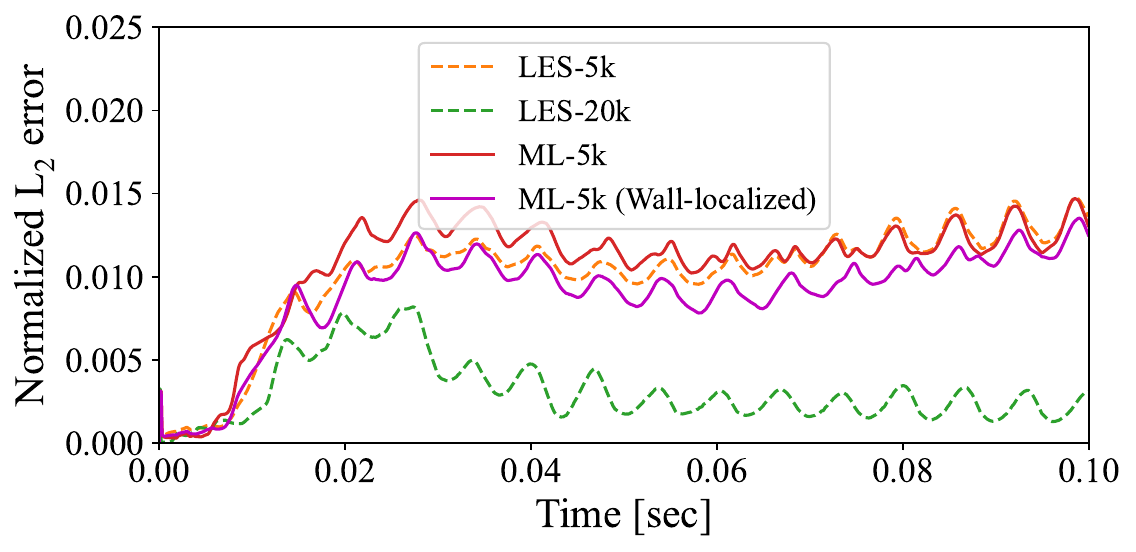}
  \end{center}
  \caption{Normalized $L_2$ error for the wall-mounted cube case. This geometry was unseen during training.}
  \label{fig:WMC_error_wall}
\end{figure}

\begin{figure}
  \begin{center}
      \includegraphics[width=\linewidth]{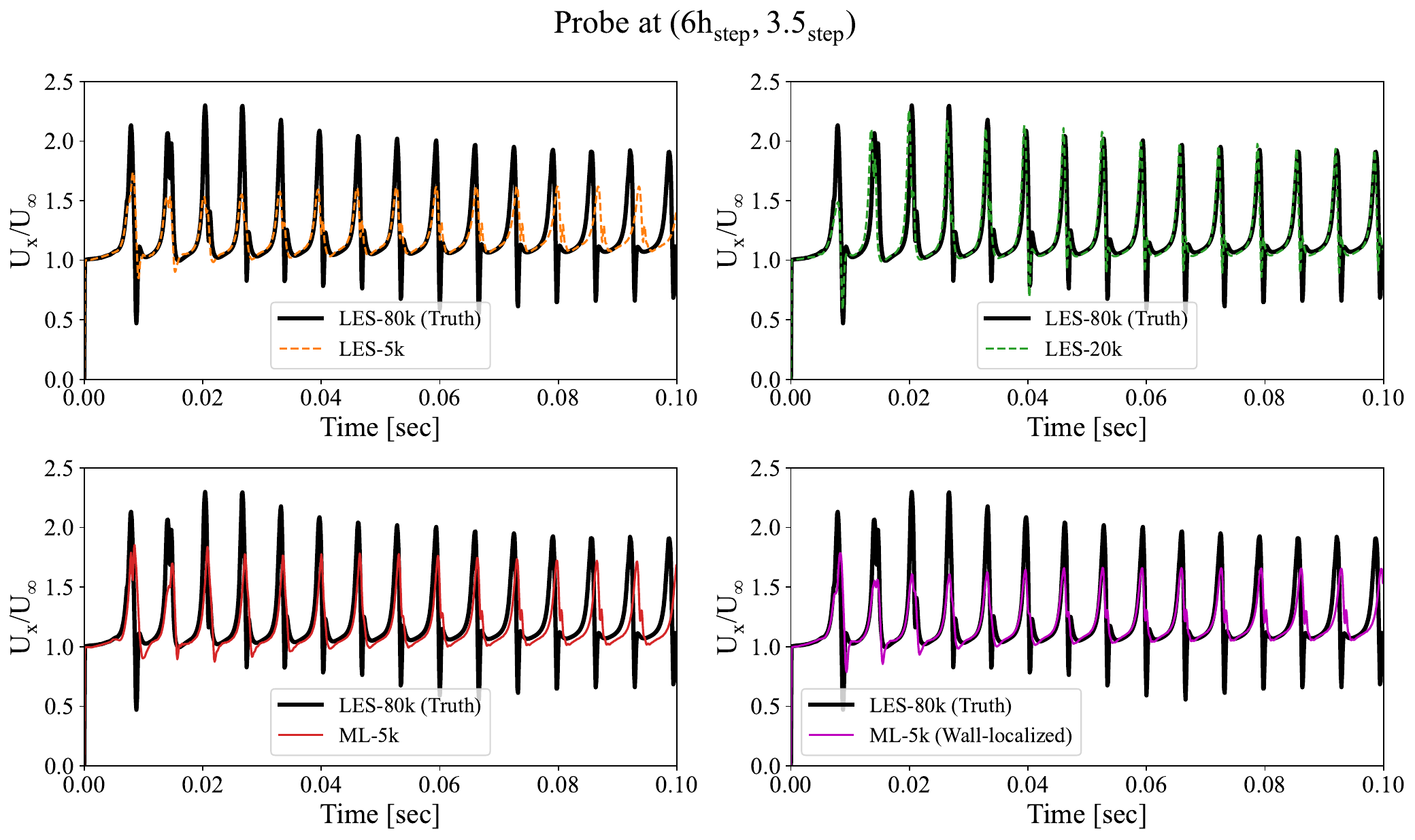}
  \end{center}
  \caption{$x$-direction Velocity measurement from a probe in front of the cube at $(x, y)=(6h, 3.5h)$.}
  \label{fig:WMC_probe_split_wall}
\end{figure}

\begin{figure*}
  \begin{center}
      \includegraphics[width=\linewidth]{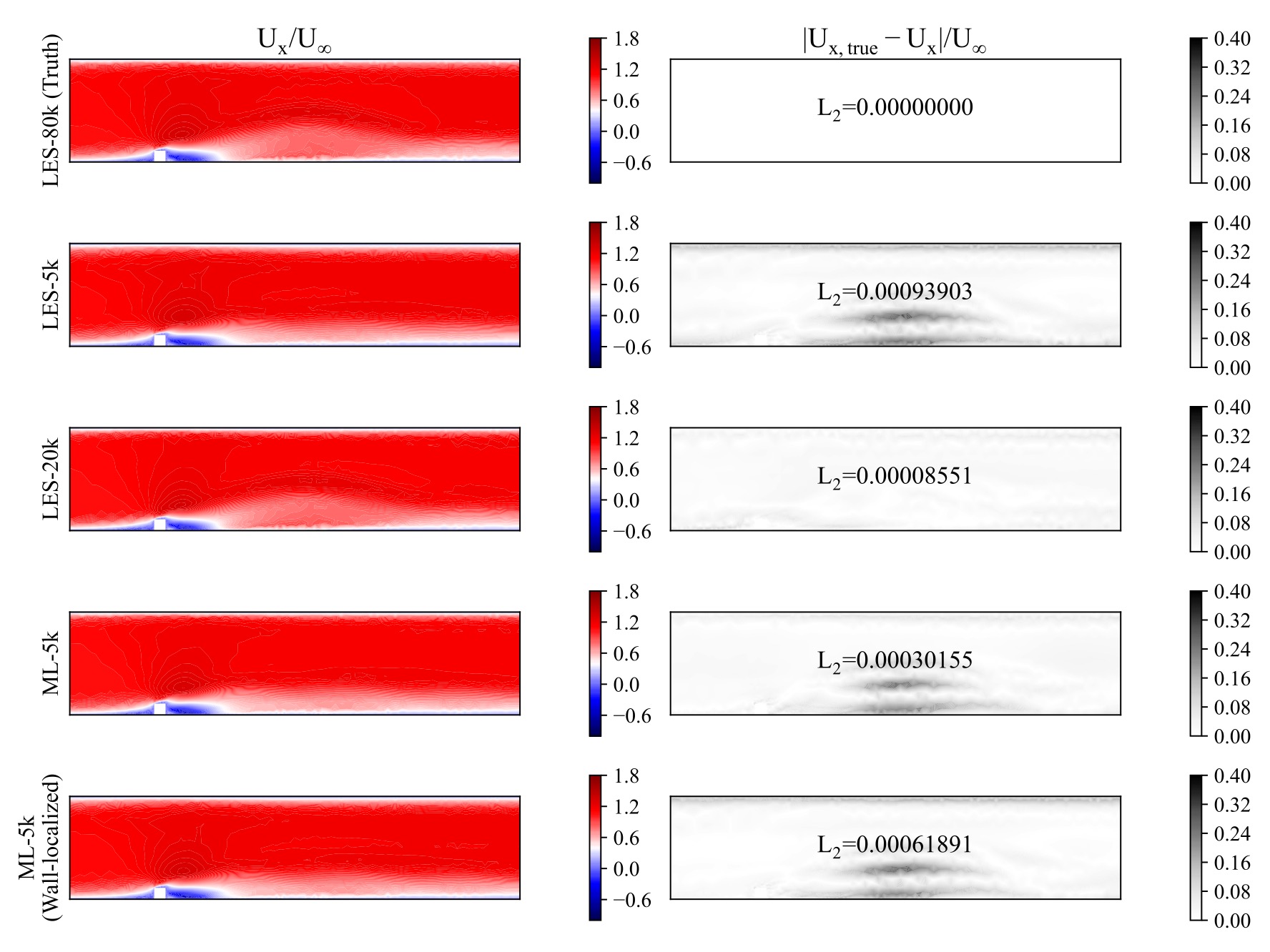}
  \end{center}
  \caption{Plots of the mean flow from BFS case for various models. \textbf{(Left column)} The normalized mean x-velocity and \textbf{(Right column)} The error with respect to the truth in space, as well as the integrated total.}
  \label{fig:Flowfield_WMC_wall}
\end{figure*}

Fig. \ref{fig:WMC_error_wall} shows the normalized $L^2$ 
error over time for the wall-mounted cube case, comparing the wall-localized SGS model with three other models. Similar to the ramp case, the wall-localized ML-5k model improves LES-5k by achieving lower error throughout the prediction horizon. Fig. \ref{fig:WMC_probe_split_wall} presents the pointwise accuracy at the probe located at
$(x,y)=(6h,3.5h)$, showing that the wall-localized model captures velocity fluctuations more accurately and remains stable beyond the training window. Fig. \ref{fig:Flowfield_WMC_wall} shows the normalized mean x-velocity field and the corresponding error contours, compared to the LES-80k reference. Integrated error values confirm that the wall-localized ML-5k model improves prediction accuracy downstream of the cube and better represents the mean flow structure than LES-5k.

These results indicate that our framework can learn optimal subgrid-scale closure at wall boundaries to match the ground truth flowfield, in conjunction with a baseline LES model. The learned closure demonstrates improved and stable \textit{a-posteriori} performance. Moreover, the wall-localized SGS model generalizes well to unseen geometries.

Although the current formulation focuses on learning adaptive subgrid-scale coefficients near the wall, the underlying differentiable physics framework, combined with GNNs, naturally extends to classical flow control tasks. For instance, wall-normal velocity control for drag reduction or inlet perturbation design for transition delay can be cast within the same optimization structure by parameterizing control inputs through GNNs.

\section{Implementation strategies for scaling differentiable FEM-based GNN closure models for three-dimensional flows}
\label{app:Implementation strategies for scaling differentiable FEM-based GNN closure models for three-dimensional flows}

Training neural network models over multiple rollouts is a common strategy to enhance temporal stability in autoregressive predictions. This approach minimizes accumulated prediction errors across several timesteps rather than focusing solely on the next-step prediction. Consequently, during training, gradients from the loss function are backpropagated through all intermediate timesteps—a process known as backpropagation through time (BPTT). While effective in improving stability, this method significantly increases GPU memory consumption since all intermediate states must be stored for gradient computation. In our differentiable FEM-based framework, this requirement translates to saving not only the stiffness matrix at each timestep but also the matrices associated with the IPCS algorithm within each step. For the two-dimensional cases, we employ 128 rollout steps per trajectory; given the relatively small number of nodes, these simulations can be trained via BPTT without GPU memory issues. However, applying the same rollout length to the three-dimensional cases would exceed the 40 GB memory capacity of an NVIDIA A100 GPU due to the substantially larger number of nodes. To mitigate this limitation, we reduce the rollout length to three timesteps for the three-dimensional turbulent cases, effectively controlling the memory growth associated with long-horizon gradient propagation. This decision is inspired by previous studies \cite{lippe2023pde, bengio2015scheduled} that investigated the trade-off between rollout length and training stability for neural network–based dynamical models. Despite using shorter rollouts, our trained closure model remains accurate and stable over more than 1,000 timesteps, as demonstrated in Sec. \ref{subsub:Training results for 3D turbulence}.

\begin{figure}[t]
  \centering
  \includegraphics[width=0.5\linewidth]{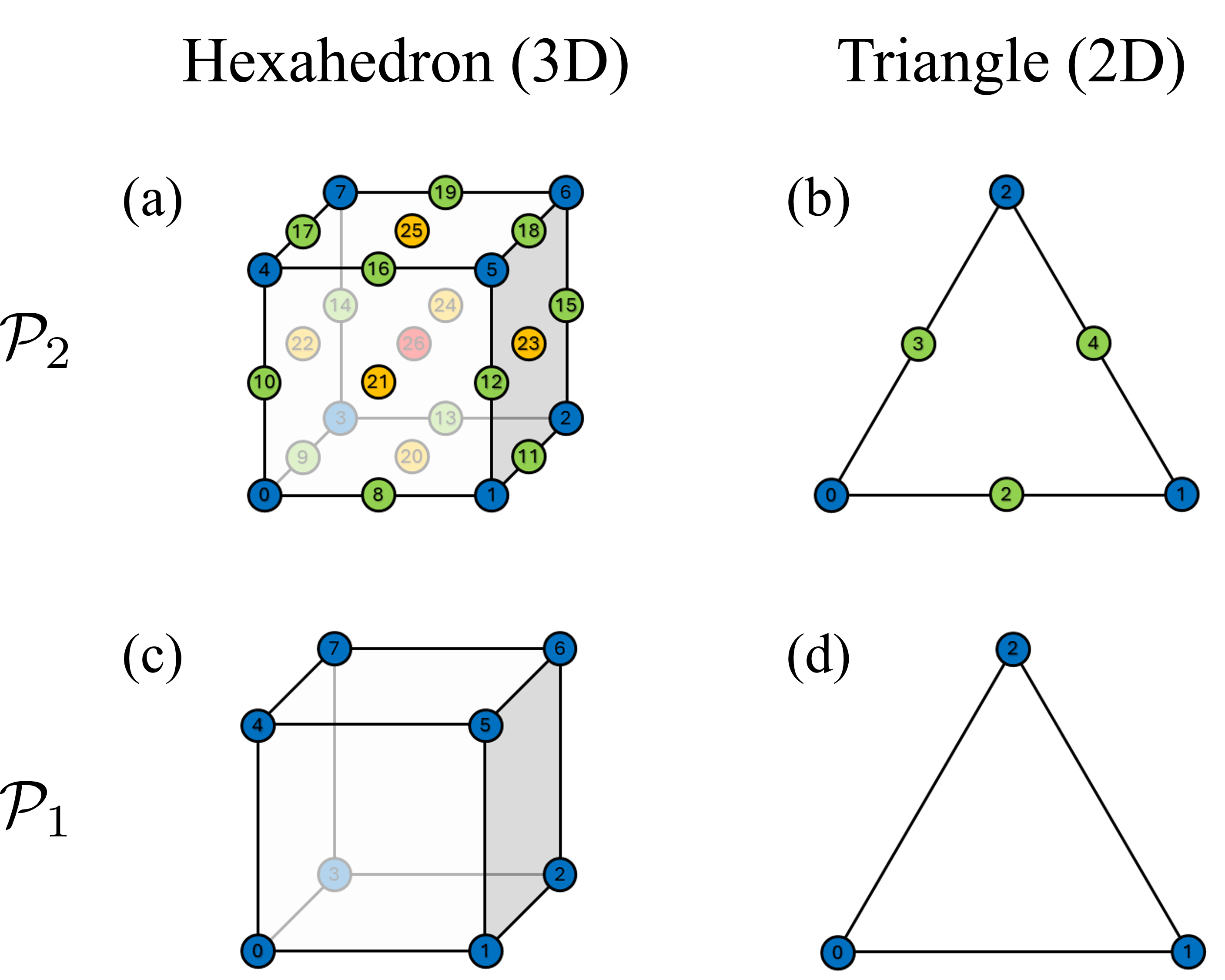}
\captionsetup{singlelinecheck=false} 
  \caption{\justifying
  Comparison of degrees of freedom (DOFs) for second-order and first-order Lagrange elements used in the Taylor–Hood formulation.
(a) Second-order Lagrange element ($\mathcal{P}_2$) on a hexahedral cell (3D) showing 26 DOFs for the velocity field. (b) Second-order Lagrange element ($\mathcal{P}_2$) on a triangular cell (2D) showing 6 DOFs for the velocity field. (c) First-order Lagrange element ($\mathcal{P}_1$) on a hexahedral cell (3D) showing 8 DOFs for the pressure field. (d) First-order Lagrange element ($\mathcal{P}_1$) on a triangular cell (2D) showing 3 DOFs for the pressure field. Note that the Taylor–Hood element pair ($\mathcal{P}_2$-$\mathcal{P}_1$) satisfies the inf–sup (LBB) condition and provides a stable discretization for incompressible Navier–Stokes equations.} 
  \label{fig:TaylorHood}
\end{figure}

Another strategy adopted to scale our differentiable FEM–GNN framework to three-dimensional turbulent flows is to use the nodes associated with the first-order Lagrange element for each cell in the computational domain. In the finite-element discretization of the incompressible Navier–Stokes equations, it is common to employ different interpolation orders for velocity and pressure to satisfy the stability requirement. Specifically, the Taylor–Hood element pair uses $\mathcal{P}_k$ for the velocity field and $\mathcal{P}_{k-1}$ for the pressure field, where $k\geq 2$. This element combination satisfies the inf–sup (or Ladyzhenskaya-Babu\v{s}ka-Brezzi, LBB) condition, ensuring numerical stability and optimal convergence rates for incompressible flow problems \cite{girault2012finite}. By employing a richer function space for velocity, Taylor-Hood elements also suppress the well-known pressure-checkerboard pattern that can appear in equal-order formulations. For computational efficiency, the most widely used configuration corresponds to $k=2$, i.e., second-order Lagrange elements for velocity and first-order elements for pressure. The resulting number of degrees of freedom (DOFs) for pressure is therefore much smaller than that for velocity. Figure \ref{fig:TaylorHood} compares the DOFs per element for both two-dimensional and three-dimensional cases. For a hexahedral element in three dimensions, the first-order Lagrange basis for pressure yields eight DOFs, whereas the second-order basis for velocity yields twenty-six. This difference leads to a substantial reduction in the total number of DOFs across the computational domain. For instance, in the coarse three-dimensional backward-facing-step (BFS) mesh used in this study, the total DOF for velocity is $390,361$, while that for pressure is $51,156$. When constructing the graph for our GNN-based closure model, we therefore define the graph nodes based on the pressure DOFs rather than the velocity DOFs. Although this choice results in a slightly less expressive spatial variation of the modeled Smagorinsky coefficient $C_s$ within each element, it significantly reduces GPU memory consumption—approximately by a factor of eight—thereby making the training of three-dimensional cases computationally feasible.

The final strategy employed to scale our framework for three-dimensional turbulent flows is to allocate multiple MPI ranks across multiple GPUs. When training machine-learning-based closure models in an \textit{a-posteriori} manner using the differentiable FEM-based solver, the governing equations must be iteratively solved for each training trajectory. Consequently, multiple CPUs are required to perform the numerical solution of the governing equations, as is standard in large-scale CFD simulations. In this parallel setting, the computational domain is decomposed into several subdomains, each assigned to a distinct MPI rank. Every rank constructs its own graph representation based on its local subdomain, enabling parallel evaluation of the GNN-based closure model. By mapping ranks to multiple GPUs, the computational workload and memory usage per GPU are reduced, allowing the framework to efficiently handle large-scale three-dimensional turbulent flow simulations.

Overall, these three strategies collectively reduce the memory load per GPU and enable the extension of our framework to three-dimensional turbulent flows.


\clearpage
\bibliography{bib}

\end{document}